\documentclass{article}

\usepackage{arxiv}

\usepackage[utf8]{inputenc} 
\usepackage[T1]{fontenc}    
\usepackage{hyperref}       
\usepackage{url}            
\usepackage{booktabs}       
\usepackage{amsmath, amsfonts}       
\usepackage{nicefrac}       
\usepackage{graphicx}

\usepackage[backend=biber, 
style=numeric-comp, sorting=none, isbn=false, url=false, doi=false, date=year,
firstinits=true]{biblatex}
 
\bibliography{references.bib}

\newcommand{\Vcphi}{\ensuremath{1.81}}
\newcommand{\cphi}{\ensuremath{C_\varphi=\Vcphi}}
\newcommand{\VcphiAl}{\ensuremath{1.75}}

\newcommand{\VcphiTi}{\ensuremath{1.69}}

\newcommand{\VcphiSic}{\ensuremath{2.97}}

\newcommand{\VcphiSi}{\ensuremath{0.61}}

\newcommand{\VcphiCuO}{\ensuremath{1.45}}

\newcommand{\VcphiCu}{\ensuremath{7.46}}

\newcommand{\Vcs}{\ensuremath{0.092}}
\newcommand{\cs}{\ensuremath{C_\text{S}=\Vcs}}
\newcommand{\VcsAl}{\ensuremath{0.16}}

\newcommand{\VcsTi}{\ensuremath{-0.19}}

\newcommand{\VcsSic}{\ensuremath{-1.25}}
\newcommand{\csSic}{\ensuremath{C_\text{S}=\VcsSic}}
\newcommand{\VcsSi}{\ensuremath{0.42}}
\newcommand{\csSi}{\ensuremath{C_\text{S}=\VcsSi}}
\newcommand{\VcsCuO}{\ensuremath{-0.47}}

\newcommand{\VcsCu}{\ensuremath{0.48}}

\newcommand{\Vct}{\ensuremath{0.51}}
\newcommand{\ct}{\ensuremath{C_\text{T}=\Vct}}
\newcommand{\VctAl}{\ensuremath{0.31}}

\newcommand{\VctTi}{\ensuremath{0.63}}

\newcommand{\VctSic}{\ensuremath{-0.018}}

\newcommand{\VctSi}{\ensuremath{0.10}}

\newcommand{\VctCuO}{\ensuremath{0.27}}

\newcommand{\VctCu}{\ensuremath{-0.24}}

\newcommand{\Vcgen}{\ensuremath{1.031}}
\newcommand{\cgen}{\ensuremath{C_0=\Vcgen}}
\newcommand{\VcAl}{\ensuremath{1.025}}

\newcommand{\VcTi}{\ensuremath{1.018}}

\newcommand{\VcSic}{\ensuremath{1.082}}

\newcommand{\VcSi}{\ensuremath{0.994}}

\newcommand{\VcCuO}{\ensuremath{1.051}}

\newcommand{\VcCu}{\ensuremath{1.059}}

\newcommand{\Vbphi}{\ensuremath{0.39}}
\newcommand{\bphi}{\ensuremath{\beta_\varphi=\Vbphi}}
\newcommand{\VbphiAl}{\ensuremath{0.72}}

\newcommand{\VbphiTi}{\ensuremath{0.58}}

\newcommand{\VbphiSic}{\ensuremath{0.70}}

\newcommand{\VbphiSi}{\ensuremath{0.35}}

\newcommand{\VbphiCuO}{\ensuremath{0.54}}

\newcommand{\VbphiCu}{\ensuremath{0.72}}

\newcommand{\Vbs}{\ensuremath{0.34}}
\newcommand{\bs}{\ensuremath{\beta_\text{S}=\Vbs}}
\newcommand{\VbsAl}{\ensuremath{0.12}}

\newcommand{\VbsTi}{\ensuremath{-0.097}}

\newcommand{\VbsSic}{\ensuremath{-0.25}}

\newcommand{\VbsSi}{\ensuremath{0.47}}
\newcommand{\bsSi}{\ensuremath{\beta_\text{S}=\VbsSi}}
\newcommand{\VbsCuO}{\ensuremath{-0.057}}

\newcommand{\VbsCu}{\ensuremath{0.073}}

\newcommand{\Vbt}{\ensuremath{0.27}}
\newcommand{\bt}{\ensuremath{\beta_\text{T}=\Vbt}}
\newcommand{\VbtAl}{\ensuremath{0.21}}

\newcommand{\VbtTi}{\ensuremath{0.51}}

\newcommand{\VbtSic}{\ensuremath{-0.017}}

\newcommand{\VbtSi}{\ensuremath{0.31}}

\newcommand{\VbtCuO}{\ensuremath{0.20}}

\newcommand{\VbtCu}{\ensuremath{-0.086}}

\newcommand{\Vrgen}{\ensuremath{0.29}}
\newcommand{\rgen}{\ensuremath{R^2=\Vrgen}}
\newcommand{\VrAl}{\ensuremath{0.53}}
\newcommand{\rAl}{\ensuremath{R^2=\VrAl}}
\newcommand{\VrTi}{\ensuremath{0.75}}
\newcommand{\rTi}{\ensuremath{R^2=\VrTi}}
\newcommand{\VrSic}{\ensuremath{0.68}}
\newcommand{\rSic}{\ensuremath{R^2=\VrSic}}
\newcommand{\VrSi}{\ensuremath{0.27}}
\newcommand{\rSi}{\ensuremath{R^2=\VrSi}}
\newcommand{\VrCuO}{\ensuremath{0.23}}
\newcommand{\rCuO}{\ensuremath{R^2=\VrCuO}}
\newcommand{\VrCu}{\ensuremath{0.43}}
\newcommand{\rCu}{\ensuremath{R^2=\VrCu}}

\newcommand{\Vn}{\ensuremath{1656}}
\newcommand{\n}{\ensuremath{N=\Vn}}
\newcommand{\VnAl}{\ensuremath{470}}
\newcommand{\nAl}{\ensuremath{N=\VnAl}}
\newcommand{\VnTi}{\ensuremath{188}}
\newcommand{\nTi}{\ensuremath{N=\VnTi}}
\newcommand{\VnSic}{\ensuremath{53}}
\newcommand{\nSic}{\ensuremath{N=\VnSic}}
\newcommand{\VnSi}{\ensuremath{86}}
\newcommand{\nSi}{\ensuremath{N=\VnSi}}
\newcommand{\VnCuO}{\ensuremath{106}}
\newcommand{\nCuO}{\ensuremath{N=\VnCuO}}
\newcommand{\VnCu}{\ensuremath{94}}
\newcommand{\nCu}{\ensuremath{N=\VnCu}}
\newcommand{\nAg}{\ensuremath{N=34}}
\newcommand{\nAlm}{\ensuremath{N=20}}
\newcommand{\nAu}{\ensuremath{N=4}}
\newcommand{\nCNT}{\ensuremath{N=188}}
\newcommand{\nFe}{\ensuremath{N=15}}
\newcommand{\nFeII}{\ensuremath{N=86}}
\newcommand{\nFeIII}{\ensuremath{N=98}}
\newcommand{\nG}{\ensuremath{N=145}}
\newcommand{\nGO}{\ensuremath{N=20}}
\newcommand{\nND}{\ensuremath{N=38}}
\newcommand{\nZn}{\ensuremath{N=3}}

\title{Statistical analysis of thermal conductivity experimentally measured in water-based nanofluids}

\author{
J. Tielke$^{1}$, M. Maas$^{2,3}$, M. Castillo $^{1}$, K. Rezwan$^{2,3}$ and M. Avila\thanks{correspondence to: marc.avila@zarm.uni-bremen.de} ~$^{1,3}$\\
$^{1}$University of Bremen, Center of Applied Space Technology and Microgravity (ZARM),\\ Am Fallturm 2, 28359 Bremen, Germany\\
$^{2}$University of Bremen, Advanced Ceramics, \\ Am Biologischen Garten 2, 28359 Bremen, Germany\\
$^{3}$MAPEX Center for Materials and Processes, University of Bremen,\\ 28359 Bremen, Germany}

\begin{document}
\maketitle

\begin{abstract}
Nanofluids are suspensions of nanoparticles in a base heat-transfer liquid. They have been widely investigated to boost heat transfer since they were proposed in the 1990's. We present a statistical correlation analysis of experimentally measured thermal conductivity of water-based nanofluids available in the literature. The influences of particle concentration, particle size, temperature and surfactants are investigated. For specific particle materials (alumina, titania, copper oxide, copper, silica and silicon carbide), separate analyses are performed.  The conductivity increases with the concentration in qualitative agreement with Maxwell’s theory of homogeneous media. The conductivity also increases with the temperature (in addition to the improvement due to the increased conductivity of water). Surprisingly, only silica nanofluids exhibit a statistically significant effect of the particle size, whereby smaller particles lead to faster heat transfer. Overall, the large scatter in the experimental data prevents a compelling, unambiguous assessment of these effects. Taken together, the results of our analysis suggest that more comprehensive experimental characterizations of nanofluids are necessary to estimate their practical potential.
\end{abstract}

\section{Introduction}
The efficient removal of heat with circulating fluids is pivotal to applications in mechatronics, mechanical, aerospace and chemical engineering. The main limiting material property in heat transfer is the thermal conductivity ($k$) of the fluid, followed by the viscosity. Conventional liquids used for heat transfer, such as water or ethylene glycol, are inexpensive but exhibit low $k$-values. In an influential paper published in 1995, Stephen U.S.\ Choi and Jeffrey A.\ Eastman \cite{Choi1995} proposed that `\emph{an innovative class of heat transfer fluids can be engineered by suspending metallic nanoparticles in conventional heat transfer fluids. The resulting nanofluids are expected to exhibit high thermal conductivities}'. Their experimental measurements  with ethylene glycol exhibited an increase in thermal conductivity up to 20\% at a volume fraction of 4\% through the addition of copper oxide particles ($d\approx20$~nm) and spurred many theoretical and experimental investigations \cite{Lee1999}. Several models were proposed to explain what was historically termed as anomalous heat transfer, meaning that the effective material properties of the suspension (e.g.\ effective viscosity and thermal conductivity of the nanofluid) could not solely account for the enhanced heat transfer \cite{Pak1998,Eastman2004}. In an influential paper, Buongiorno \cite{Buongiorno2005} considered many possible physical mechanisms behind convective heat transfer enhancement in nanofluids and assessed their plausibility and relative importance. He argued that Brownian motion and thermophoresis (i.e. the motion of particles along gradients of temperature) were the two only plausible mechanisms for heat transfer enhancement. Keblinski \textit{et al}. \cite{Keblinski2008} critically analysed some experimental data sets and concluded that effective medium theories are capable of explaining the data.

In 2009, Buongiorno \textit{et al}.\ \cite{Buongiorno2009} performed an experimental benchmark study to determine the influence of the measurement technique on the experimental thermal conductivity of nanofluids. This study showed that the data scattered in a range of at least \( \pm 5\% \) about the median, which implied that a total enhancement of a few percent cannot be detected. Subsequently, Khanafer \textit{et al}.\ \cite{Khanafer2011} developed correlations for thermal conductivity and viscosity based on experimental data. There have been no additional benchmark studies or statistical analysis since their study. Characterization efforts have mainly focused on the analysis of the nanoparticle sizes and particle concentrations \cite{Gowda2010, Fan2011, Lee2014, Gangadevi2018, Mikkola2018}. 

Recently, the state of the art in this field was reviewed by Buschmann \textit{et al}. \cite{Buschmann2018}, who analyzed several experiments on convective heat transfer with nanofluids. Their analysis supported the conclusion of Buongiorno \cite{Buongiorno2005} that there are no anomalies in the convective or conductive heat transfer of nanofluids. Buschmann \textit{et al}. \ \cite{Buschmann2018} argued that nanofluids can be treated as homogeneous fluids by considering their effective properties. Hence, their heat transfer can be correctly predicted by well-established correlations for pure fluids, provided that the effective thermal conductivity and the effective viscosity of the nanofluid are known. Finally, Buschmann \textit{et al}. \ \cite{Buschmann2018} pointed out that there is currently a lack of knowledge of how and why nanoparticles change the thermal conductivity of a fluid.

The observation that thermophysical properties are modified by additions of particles to a fluid dates back to the theoretical work of Maxwell in 1881 \cite{Maxwell1881}. According to his Effective Medium Theory, the effective thermal conductivity of a nanofluid $k_\text{eff}$ depends on the thermal conductivities of the nanoparticles $k_\text{p}$, thermal conductivity of the base fluid $k_\text{f}$, and particle fraction $\varphi$,
\begin{equation}\label{eq:Maxwell}
\frac{k_\text{eff}}{k_\text{f}} = 1 + \frac{3\varphi (k_\text{p} - k_\text{f} ) }{3 k_\text{f} + ( 1 - \varphi) (k_\text{p} - k_\text{f})}.
\end{equation}

Eapen \textit{et al}.\ \cite{Eapen2010} argued that there may be different dispersion states in nanofluids that influence thermal conductivity, which are not considered in Maxwell's theory. For example, particles can form percolating structures at moderately low concentrations \cite{Eapen2010, Carson2005}. Through the use of a theory developed by Hashin and Shtrikman \cite{Hashin1962}, Eapen \textit{et al}.\ \cite{Eapen2010} derived the following HS-bounds for the effective thermal conductivity of nanofluids
\begin{equation}\label{eq:HS}
    k_\text{f} \biggl \lbrack 1 + \frac{3 \varphi (k_\text{p} - k_\text{f})}{3 k_\text{f} + (1 - \varphi) (k_\text{p} - k_\text{f})} \biggr \rbrack \leq k_\text{eff} \leq k_\text{p} \biggl \lbrack 1 - \frac{3 (1 - \varphi) (k_\text{p} - k_\text{f})}{3 k_\text{p} - \varphi (k_\text{p} - k_\text{f})} \biggr \rbrack
\end{equation}
The lower HS-bound represents the well-dispersed state, where the particles are the disperse phase and the base fluid is the continuous phase, which was already described by Maxwell, see eq.~\eqref{eq:Maxwell}. Hence, in the well-dispersed state, heat is mainly transferred through the fluid. The upper HS-bound represents a state in which heat is mainly transferred between particles \cite{Eapen2010, Hashin1962, Mugica2020}. Specifically, at high concentrations ($\varphi \approx 1$), the base fluid becomes the dispersed phase and particles the continuous phase. Even if this limit is not realistic for e.g.\ spherical particles (due to close-packing), it is still useful in describing particle configurations (chains, percolation networks), which can arise even at low concentrations and result in strongly enhanced heat transfer \cite{Bouguerra2018}.

Maxwell's and Hashin and Shtrikman's theories of suspensions do not account for the particle size, whereas the distinct feature of nanofluids is that the particles are less than $100$~nm in one dimension. Vadasz \textit{et al}. \cite{Vadasz2005} considered heat conduction in nanofluids and reviewed the transient-hot-wire method. He showed that by accounting for the dependence of the heat transfer coefficient on the particle size (through the particle's specific area $S$), the experimental data became consistent with classical theories of suspensions. However, he stated that his analysis was inconclusive and that more experimental data were required. Further experimental studies \cite{Lee2014, Das2003,Chon2005, Li2006, Li2007} showed that thermal conductivity increases linearly with increasing temperature and suggested an effect from the particle size. 

Bouguerra \textit{et al}.\ \cite{Bouguerra2018} recently showed that the effective thermal conductivity and the effective viscosity of water-based alumina nanofluids strongly depend on pH. They were able to distinguish between different dispersion states that included well-dispersed particles, the formation of percolation networks, and fully agglomerated particles, through measurements with volume concentrations between 0.2\% and 2\% at different pH levels. Variations of the pH modify the suspension stability through repulsive forces of electrostatic origin. An alternative to this is the use of surfactants. Depending on the choice and concentration of the surfactant, the thermal conductivity of the nanofluid may be increased or decreased \cite{Gangadevi2018, Yang2012, Li2008, Kim2018, Nasiri2011,Cao2015}, and to date there is no general, coherent picture as to what the effect of surfactants on the nanofluid is.

In this paper, we present a statistical correlation analysis of thermal conductivity measurements of water-based nanofluids available in the scientific literature. The aim of our analysis is to assess nanofluids for potential applications and to identify suitable nanomaterials for heat transfer. To this avail, we analyze whether the influences of particle concentration, size,  temperature and surfactants on the thermal conductivity of nanofluids are statistically significant, and we quantify their relative importance. 

\section{Material and Methods}

We compiled a database with \n \ data points (experimental measurements of the thermal conductivity) from $73$ publications concerned with water-based nanofluids. The data points were taken from publications in which temperature, volume concentration, and particle size were fully specified. The database is given in the supplementary materials and contains data for 17 different nanoparticle materials. In the first part of our analysis, all data points were considered. Subsequently, six individual materials were analyzed for which there are at least $4$ different publications available comprising  $N\ge 50$ data points altogether. This was done to avoid spurious results due to small samples. The six individual materials are:
\begin{itemize}
    \item Alumina (\(Al_2O_3\), \nAl): \cite{Lee1999, Gowda2010, Gangadevi2018, Mikkola2018, Bouguerra2018, Das2003, Chon2005, Li2006, Beck2009, Bowers2018, Chandrasekar2011, ElBrolossy2013, HemmatEsfe2014, Heyhat2012, Ho2014, Hong2012, Iacobazzi2016, Kayhani2012, Kim2012, Kumar2018,  Murshed2008, Nair2018, Oh2008, Ruan2011, Rudyak2016, Said2014_2, Teng2010, Topuz2018, Zhu2009, Mintsa2009, Modi2020, Gavili2019, Khurana2019, Patel2010, Vakilinejad2018}
    \item Titania (\(TiO_2\), \nTi) \cite{Nair2018, Topuz2018,  Vakilinejad2018, Das2018, Duangthongsuk2009, Oliveira2017, Said2014, Anbu2019,Fedele2012}
     \item Copper oxide (CuO, \nCuO) \cite{Gangadevi2018, Das2003, Li2006, Nair2018, Mintsa2009,  Patel2010, SinghSokhal2018}
      \item Copper (Cu, \nCu) \cite{Yang2012, Li2008, Patel2010, Liu2006, Saterlie2011}
     \item Silica (\(SiO_2\), \nSi) \cite{Mikkola2018, Bowers2018, Guo2018, Kang2006, Ajeel2019, Ardekani2019, Manay2019, Rejvani2019}
     \item Silicon carbide (SiC, \nSic)  \cite{Singh2009, Chen2017, MANNA2012, Ponnada2019, Xie2002}
\end{itemize}
Descriptive statistics for these data sets are shown in table~1 of the supplementary materials. We note that for the analysis of copper nanofluids we excluded the study of Liu \cite{Liu2006}. This study features results which are very different from those of all other works for the same material. The decision to exclude them is further justified later. Here it suffices to say that as a result of excluding this, the regression significantly improves. For completeness, a comparison to the regressions with all data points (without exclusion) can be found in the supplementary materials.

For the  materials below there are less than $4$ publications and/or less than $50$ data points. Thus we did not analyze them individually, because the analysis would have little statistical significance: 
\begin{itemize}
    \item Iron(III)oxide ($Fe_2O_3$, \nFeII) \cite{Zouli2019, Agarwal2019} 
    \item Iron(II,III)oxide (\(Fe_3O_4\), \nFeIII) \cite{Abareshi2010, Ebrahimi2018, SyamSundar2013}
    \item Zinc oxide (ZnO, \nZn) \cite{Topuz2018}
    \item Graphene (G, \nG) \cite{Kim2018, Gao2018, Akram2019}
    \item Graphene oxide (GO, \nGO) \cite{Hajjar2014}
    \item Carbon-Nanotubes (CNT, \nCNT) \cite{Kim2018, Nasiri2011}
    \item Nanodiamond (ND, \nND) \cite{Sundar2016, Yeganeh2010}
    \item Silver (Ag, \nAg) \cite{Kang2006, Ardekani2019, Pourhoseini2018, Bakhshan2019}
    \item Iron (Fe, \nFe) \cite{Li2005}
    \item Aluminum (Al, \nAlm) \cite{Patel2010}
    \item Gold (Au, \nAu) \cite{JinKim2009, Paul2010}
    \end{itemize} 
We caution that the published data for metals have to be interpreted critically because metallic particles easily oxidize in water \cite{Park2007, Cushing2004}.    

\subsection{Linear statistical model}

We employed a linear model to statistically quantify the effect of the volume concentration $\varphi$, the temperature $T$ and the particle size (through the specific surface $S$) on the normalized thermal conductivity
\begin{equation}
k^*(\varphi,T,S) =\dfrac{k_\text{eff}(\varphi,T,S)}{k_\text{f}(T)},
\end{equation}
where $k_\text{f}(T)$ is the thermal conductivity of the base fluid (pure water) as a function of the temperature. Changes in the thermal conductivity of the base fluid $k_f$ due to the addition of surfactants or changes in the pH were taken account of, if the data were specified in the respective studies. Nevertheless, most of the data are given normalized or are normalized on pure water.

Our linear statistical regression model reads
\begin{equation}\label{eq:model}
    k^*(\varphi,T,S)= C_0 + C_\varphi \,\varphi + C_T\, T^* + C_S\, S^*, 
\end{equation}
where the coefficients $C_i$ (with $i=\{0,\varphi,T,S\}$) were determined from linear regressions of the data sets. These coefficients are dimensionless, due to the definitions of \(T^* = (T - T_\text{ref})/T_\text{ref}\) and $S^* =(S-S_0)/S_\text{ref}$. We chose \( T_\text{ref} = 293\)~K because most measurements found in the literature were taken at room temperature. For large particles (with specific surface $S\approx 0$), we do not expect an increase in thermal conductivity beyond Maxwell's theory, therefore we chose $S_0=0$ as reference. Finally, we chose $S_\text{ref}=6/d_\text{ref}$ with $d_\text{ref}=1$~nm for simplicity (with these choices the last term simplifies to $S^*=S/S_\text{ref}=d_\text{ref}/d$).

\subsection{Physical interpretation of the linear statistical model}

Starting with Maxwell’s equation \eqref{eq:Maxwell} and linearizing about zero concentration (\(\varphi\)=0), one obtains a linear prediction of the normalized effective thermal conductivity for small concentrations
\begin{equation}\label{eq:linMaxwell}
    k^*(\varphi) = 1 + C_{\varphi,\text{Maxwell}}\, \varphi,
\end{equation}
where
\begin{equation}\label{eq:CvarphiMaxwel}
C_{\varphi,\text{Maxwell}} = 3\frac{(k_\text{p} - k_\text{f})}{(k_\text{p} + 2 k_\text{f})}.
\end{equation}
The coefficient $C_{\varphi,\text{Maxwell}}$ increases monotonously as the thermal conductivity of the particles \(k_\text{p}\) increases. In particular, in the low conductivity limit (\(k_\text{p} \ll k_\text{f}\)), $C_{\varphi,\text{Maxwell}}=-1.5$, whereas in the opposite limit (\(k_\text{p} \gg k_\text{f}\)), $C_{\varphi,\text{Maxwell}}=3$. Hence in the framework of the linear Maxwell’s equation \eqref{eq:linMaxwell}, $C_{\varphi,\text{Maxwell}} \in [{-1.5},3]$. This theoretical prediction can be compared with the coefficients obtained form the linear regression of \eqref{eq:model} to the compiled data sets, whereby regression values with $C_\varphi>3$ indicate higher performance than admissible in Maxwell's theory. For example, in the aforementioned experiments of Lee \emph{et al}.\ \cite{Lee1999}, the heat transfer rates increased up to 20\% at a volume fraction of 4\% ($\varphi=0.04$). This would imply $C_\varphi=5$, which cannot be explained with Maxwell's theory. Furthermore, if there are no particles ($\varphi=0$), the thermal conductivity must remain unchanged and it would be expected that the linear correlation analysis yielded values of \(C_0\approx1\). The deviation of \(C_0\) from unity serves as a proxy for the level of quality of the linear model and/or for the scatter of the data.

\subsection{Data processing and statistical methods}

Analyses were performed with IBM SPSS Statistics Version 25 using regressions with the normalized thermal conductivity ($k^*$) as a dependent variable and \(\varphi\),  \(T^*\) and \(S^*\) as independent variables. All variables were entered in a single step and confidence intervals were calculated to 95\%. We evaluated the corrected correlation coefficient \(R^2\), the coefficients $(C_0,C_\varphi,C_\text{T},C_\text{S})$, the standard deviations of the coefficients and the standardized regression coefficients $(\beta_\varphi,\beta_\text{T},\beta_\text{S})$. The values of $(\beta_\varphi,\beta_\text{T},\beta_\text{S})$ were calculated by subtracting the mean from the variable and dividing by its standard deviation. The larger the standardized regression coefficient of a parameter, the higher the influence of that parameter on $k^*$. Our analysis demonstrated that when \(|\beta_i|< 0.1\) the influence of parameter $i$ was insignificant. 

Additional analyses were also carried out with subsets of the data or variations of the model, which allowed an assessment of the robustness of the data and model. For example, to verify the use of the linear regression in terms of the concentration, we restricted the data to $\varphi \leq 0.02$. The influence of surfactants was tested by performing separate regressions for the data without surfactants. Furthermore, we fitted nonlinear regressions with quadratic and cubic terms on the concentration and also with \(C_0=1\) fixed. Detailed results for all the linear and nonlinear regressions can be found in the supplementary materials.

\section{Results}

\begin{figure}
\centering
   \includegraphics [width=3.34in]{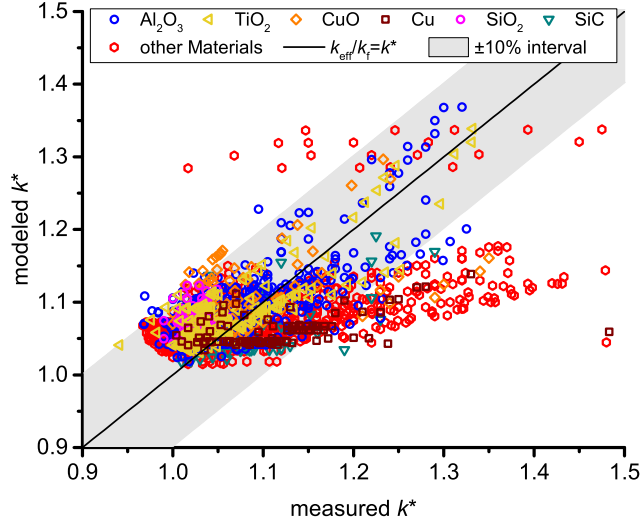}
\caption{Experimentally measured versus modeled normalized thermal conductivity (\(k^*\)) for water-based nanofluids.  The ideal values of \(k^*\) and a \(\pm10\%\) interval are displayed as a solid line and a grey area, respectively. Colors are used to distinguish single materials (analysed separately) from other materials (see the legend). The model is \(k^*=\Vcgen + \Vcphi \ \varphi + \Vct \ T^* + \Vcs \ S^*\) (\rgen, \ \n).}
\label{fig:1}
\end{figure}

The result of using a linear regression to fit all data points of our database with model eq.~(\ref{eq:model}) is shown in the first row of table~\ref{tab:1}. In figure~\ref{fig:1} the model prediction (vertical axis) is plotted against the experimentally measured normalized thermal conductivity (horizontal axis). Data points lying along the black line exhibit perfect agreement with our linear model, eq.~\eqref{eq:model}, whereas the discrepancy is larger the further the data points are from the line. The shaded region depicts a \( \pm 10\% \) interval about the model prediction. The correlation coefficient is \rgen \ and \cgen. The most significant parameter is the concentration, with  \cphi \ (\bphi), which is well within the bounds from the linearized Maxwell equation, $C_{\varphi,\text{Maxwell}} \in [-1.5,3]$. The thermal conductivity also increases with increasing temperature and specific surface, with  \ct\ (\bt) and \cs \ (\bs), respectively. These results allow a first estimation of the performance of water-based nanofluids. For example, if $d=30$~nm at a 4\% concentration and $T=303$~K is substituted into eq.~\eqref{eq:model}
$$
k^*= C_0 + C_\varphi \,\varphi + C_\text{T}\, T^* + C_\text{S}\, S^*= 1.031+0.072+0.017+0.003=1.124,
$$
a $12.4$\% increase in thermal conductivity is obtained (when compared to pure water at $T=303$~K). Noteworthy, the contribution directly from the particle size (last term) is negligible.

\begin{table}[!h]
\caption{Results of the linear regressions fitted to the entire database and for each individual material, with the number of data points ($N$), corrected correlation coefficient ($R^2$),  model coefficients (\(C_\varphi, C_\text{T}, C_\text{S}\)), and their corresponding standardized correlation coefficients (\(\beta_\varphi, \beta_\text{T}, \beta_\text{S}\))}
\label{tab:1}
\begin{tabular} {|c|c|c|c|c c|c c|c c|}
\hline 
Material & $N$ & $R^2$ & \(C_0\) & \(C_\varphi\) & \(\beta_\varphi\) & \(C_T\) & \(\beta_T\) & \(C_S\) & \(\beta_S\) \\ \hline
General & \Vn & \Vrgen & \Vcgen & \Vcphi & \Vbphi & \Vct & \Vbt & \Vcs & \Vbs \\ \hline
Alumina & \VnAl & \VrAl & \VcAl & \VcphiAl & \VbphiAl & \VctAl & \VbtAl & \VcsAl & \VbsAl \\ 
Titania & \VnTi & \VrTi & \VcTi & \VcphiTi & \VbphiTi & \VctTi & \VbtTi & \VcsTi & \VbsTi \\
Copper oxide & \VnCuO & \VrCuO & \VcCuO & \VcphiCuO & \VbphiCuO & \VctCuO & \VbtCuO & \VcsCuO & \VbsCuO \\
Copper & \VnCu & \VrCu & \VcCu & \VcphiCu & \VbphiCu & \VctCu & \VbtCu & \VcsCu & \VbsCu \\
Silica & \VnSi & \VrSi & \VcSi & \VcphiSi & \VbphiSi & \VctSi & \VbtSi & \VcsSi & \VbsSi \\
Silicon carbide & \VnSic & \VrSic & \VcSic & \VcphiSic & \VbphiSic & \VctSic & \VbtSic & \VcsSic & \VbsSic \\ \hline

\end{tabular}
\vspace*{-4pt}
\end{table}

\begin{figure}
    \centering
     \begin{tabular} {cc}
    a & b\\
    \includegraphics[width=2in]{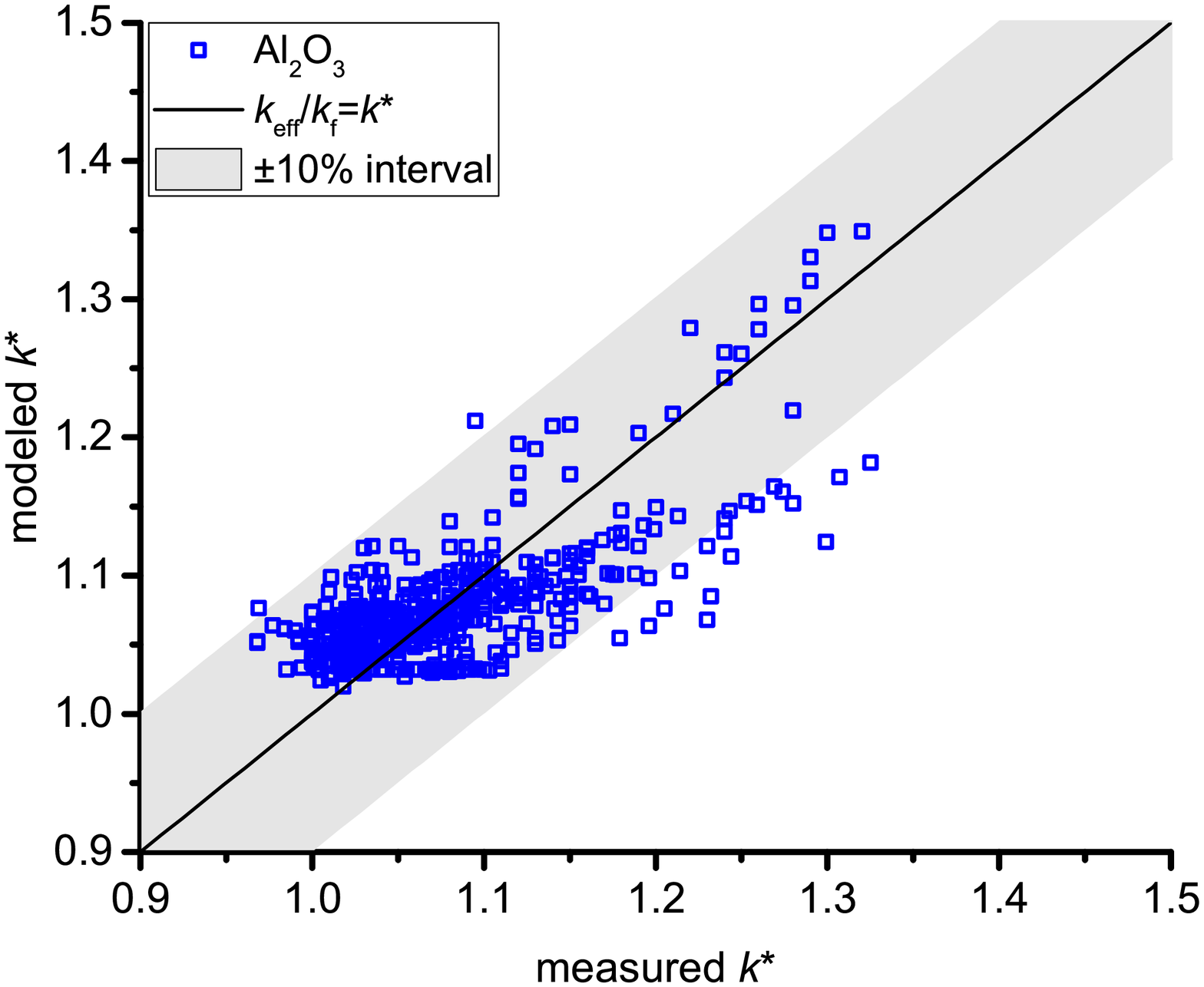} & \includegraphics[width=2in]{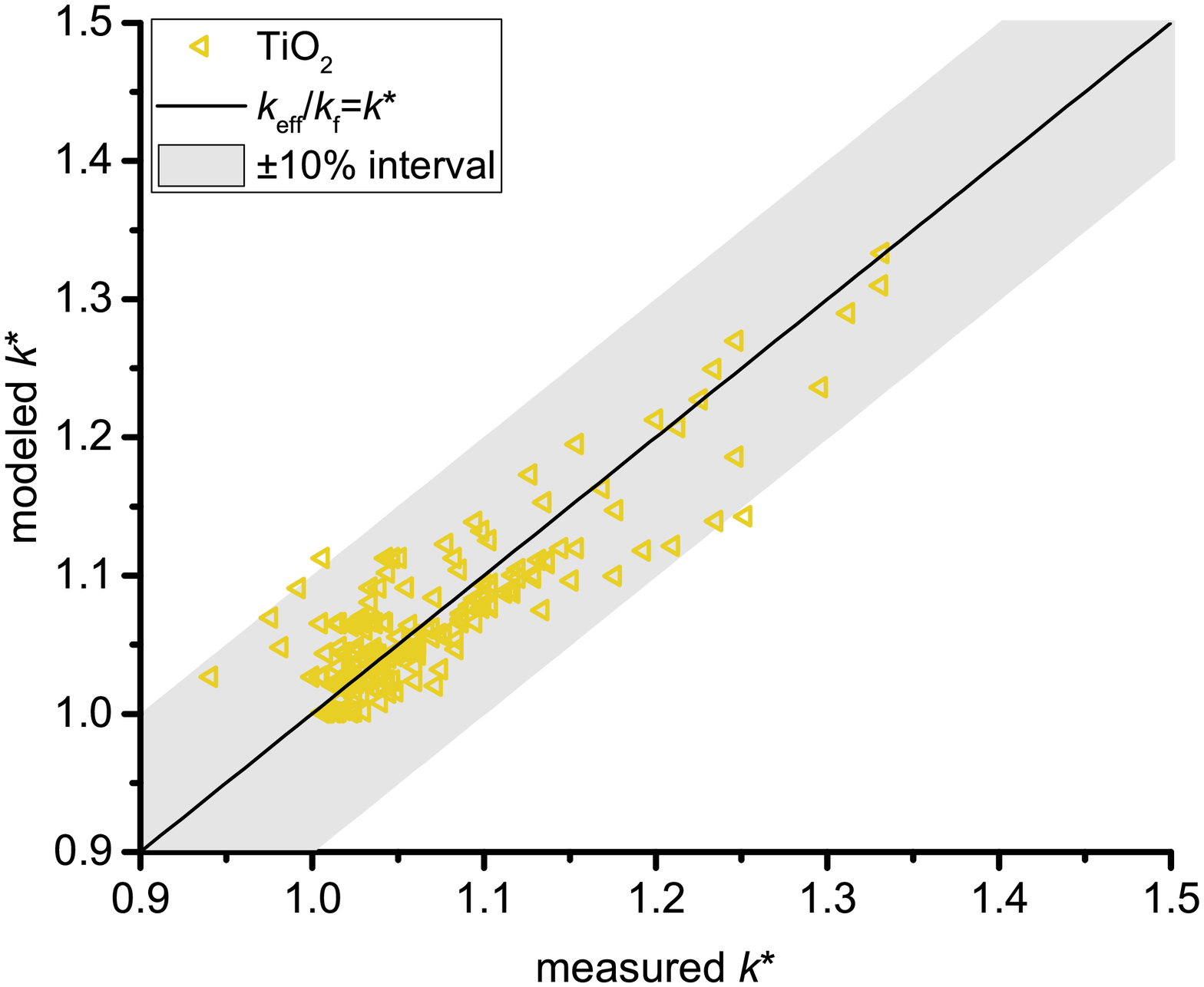}\\
     c & d\\
     \includegraphics[width=2in]{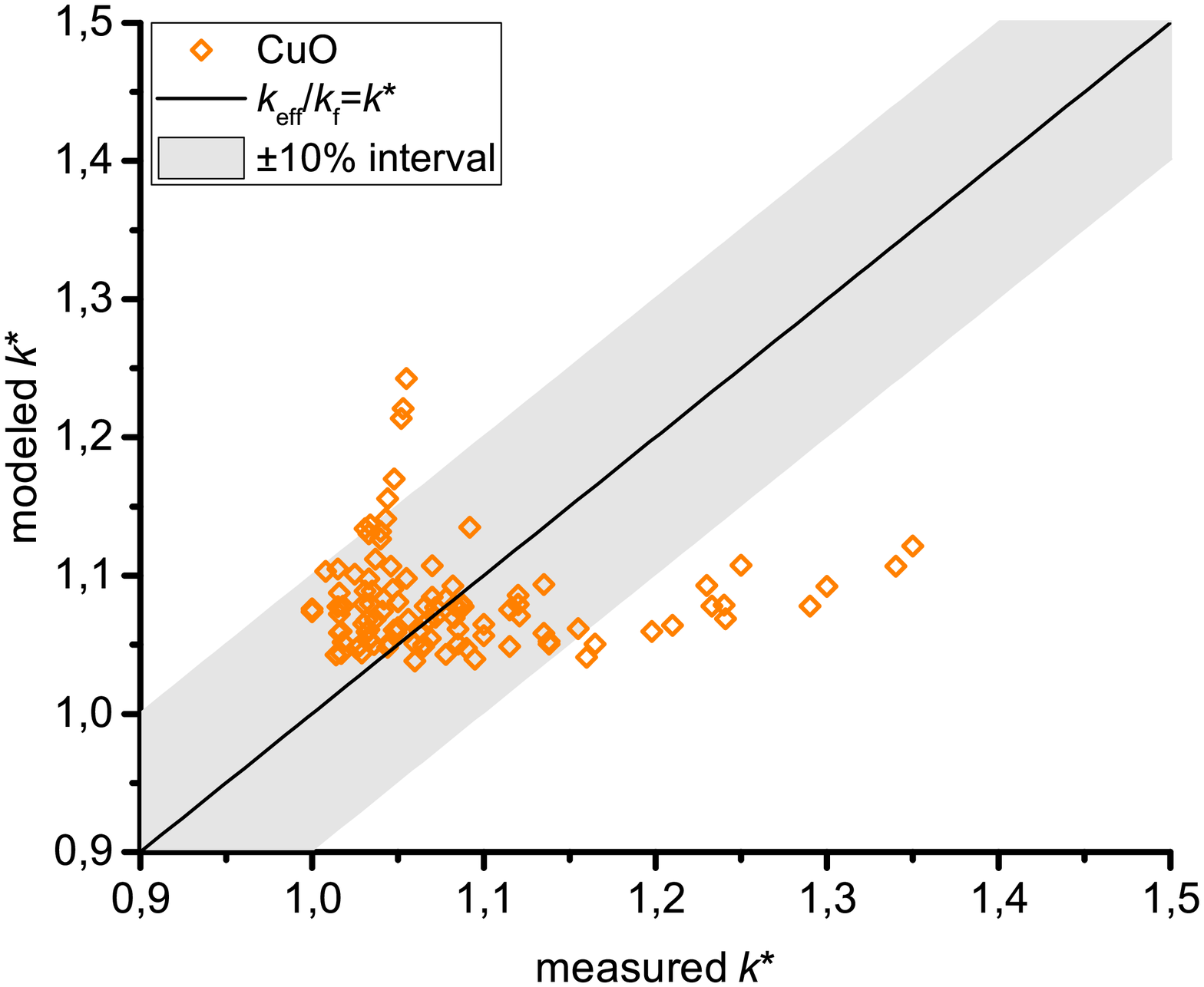} & \includegraphics[width=2in]{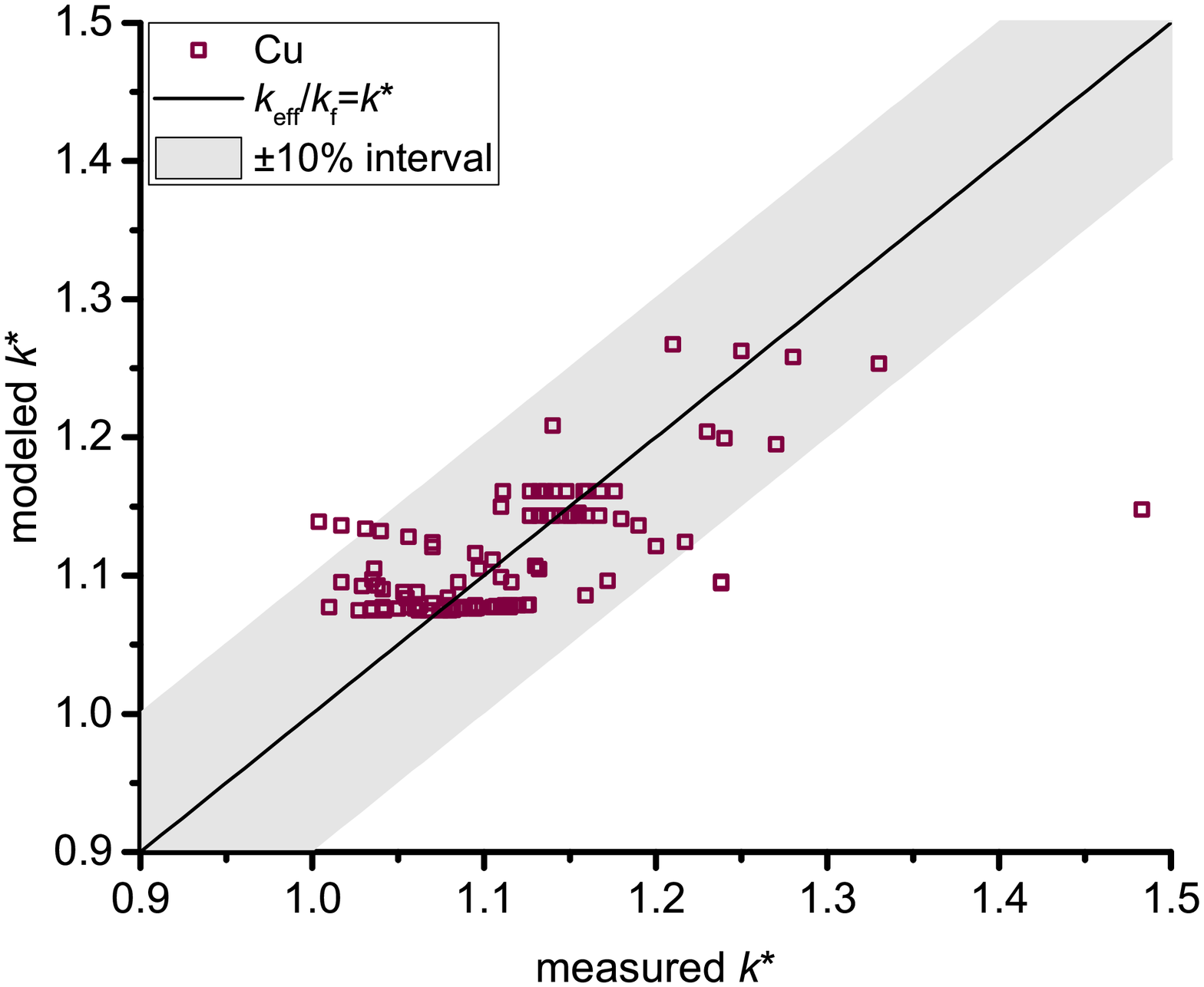} \\
     e & f\\
    \includegraphics[width=2in]{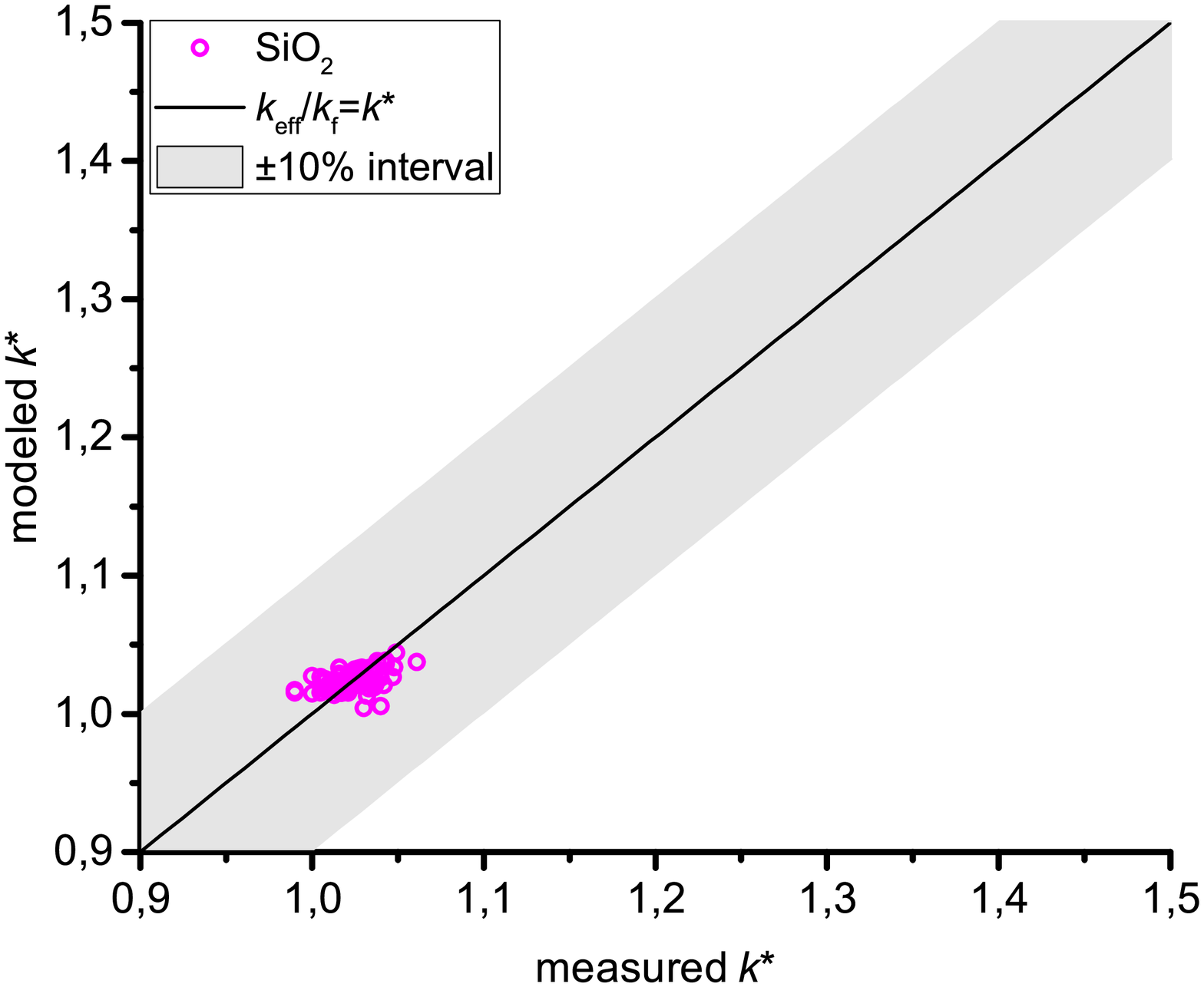} & \includegraphics[width=2in]{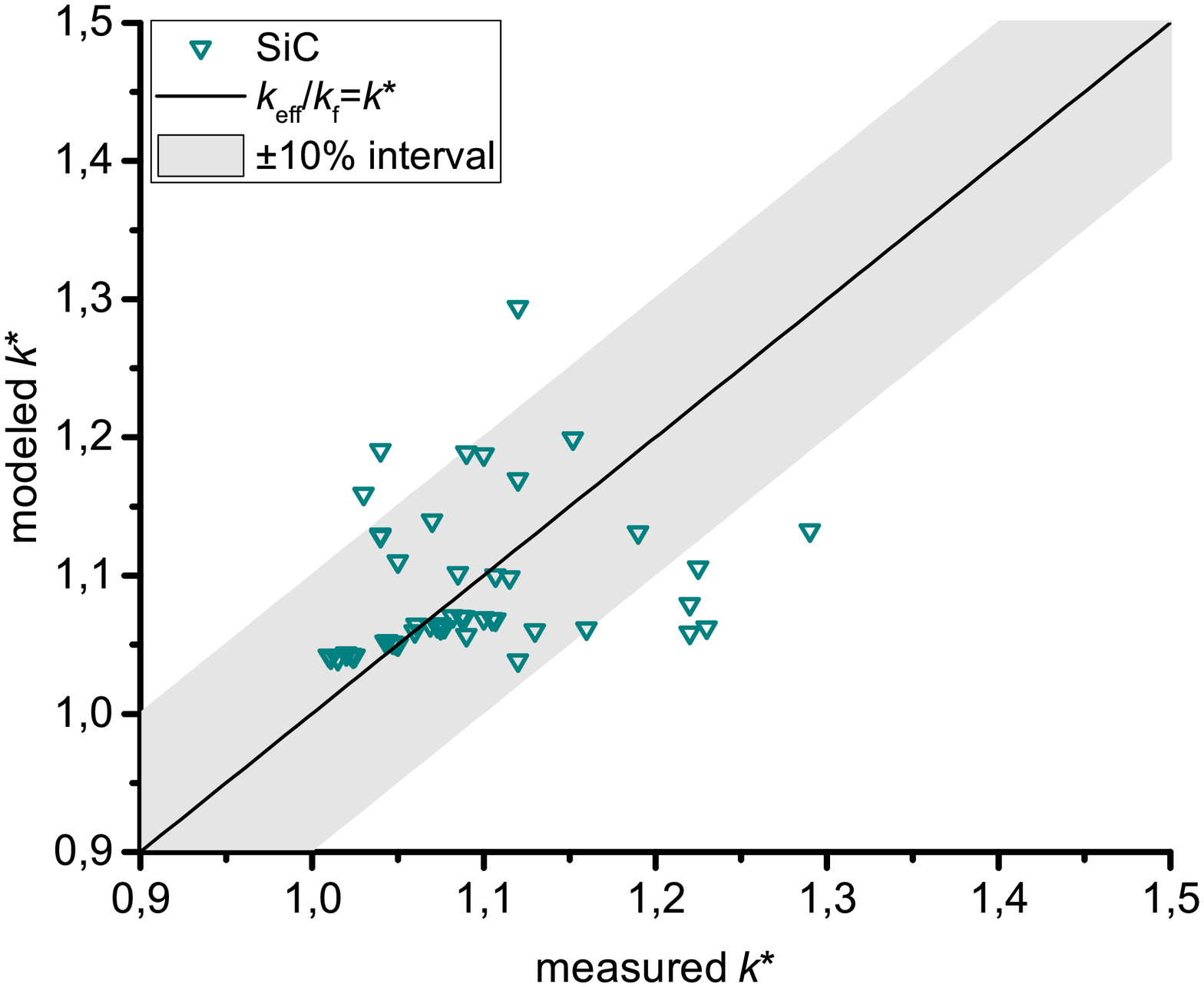}
     \end{tabular}
    \caption{Experimentally measured versus modeled normalized thermal conductivity (\(k^*\)). The ideal values of \(k^*\) and a \(\pm10\%\) interval are displayed as a solid line and a grey area, respectively. (a) Alumina with \nAl\ and \(k^*=\VcAl + \VcphiAl \ \varphi + \VctAl \ T^* + \VcsAl \ S^*\) (\rAl). (b) Titania with \nTi\ and \(k^*=\VcTi + \VcphiTi \ \varphi + \VctTi \ T^* \VcsTi \ S^*\) (\rTi). (c)  Copper oxide with \nCuO\ and \(k^*=\VcCuO + \VcphiCuO \ \varphi + \VctCuO \ T^* \VcsCu \ S^*\) (\rCuO). (d) Copper with \nCu\ and \(k^*=\VcCu + \VcphiCu \ \varphi + \VctCu \ T^* \VcsCu \ S^*\) (\rCu) (e) Silica with \nSi\ and \(k^*=\VcsSi + \VcphiSi \ \varphi + \VctSi \ T^* + \VcsSi \ S^*\) (\rSi). (f) Silicon carbide with \nSic\ and \(k^*=\VcSic + \VcphiSic \ \varphi + \VctSic \ T^* \VcsSic \ S^*\) (\rSic).}
    \label{fig:2}
\end{figure}

The linear regressions to the data sets for single materials are shown in figure~\ref{fig:2}. Clearly, the scatter in the data and the goodness of the fit depend strongly on the material, which is quantified by the respective $R^2$ (given in the second column of table~\ref{tab:1}). Overall, the values of $\beta_\varphi$ given in the fourth column of table~\ref{tab:1} confirm that increasing the particle concentration leads to a statistically significant enhancement of the thermal conductivity for all materials analyzed. In addition, there is also a significant influence of the temperature and/or surface  for specific materials (see the sixth and eight columns of table~\ref{tab:1}). In what follows, we discuss the influence of each of these three factors separately.

\subsection{Effect of particle concentration}

The computed values of $C_\varphi$ are shown in figure~\ref{fig:3}a as a function of the thermal conductivity of the particle material. The corresponding 95\% confidence intervals indicate that silicon carbide nanofluids are in excellent agreement with Maxwell's prediction (eq.~\eqref{eq:linMaxwell} and lower solid line in the figure), whereas silica, alumina, titania and copper oxide nanofluids are below it. Only copper nanofluids appear to exceed Maxwell's prediction. Overall, the discrepancy with Maxwell's prediction is not large in view of the large scatter in the data sets.  By linearizing the upper HS-bound of eq.~\eqref{eq:HS}, we obtained an upper bound for $C_\varphi$ (upper solid line in figure~\ref{fig:3}a). This upper bound embraces all possible dispersion states and is satisfied also by copper.

\subsection{Effect of nanofluid temperature}

The computed values of $C_\text{T}$ for single materials are shown in figure~\ref{fig:3}b. Increasing the temperature in titania nanofluids enhances the thermal conductivity significantly and strongly. For alumina and copper oxide nanofluids the increase in thermal conductivity is less pronounced and also less significant. For silica the effect is significant, but the performance increase is even weaker.  Finally, there is no significant influence of temperature on the thermal conductivity for copper and silicon carbide nanofluids, which is due to a lack of experimental measurements for a sufficiently large range of temperatures (see table 1, figs. 5 and 7 in the supplementary material).

\subsection{Effect of the particle size}

\begin{figure}
    \centering
    \begin{tabular}{c}
   a \\
    \includegraphics[width=2.7in]{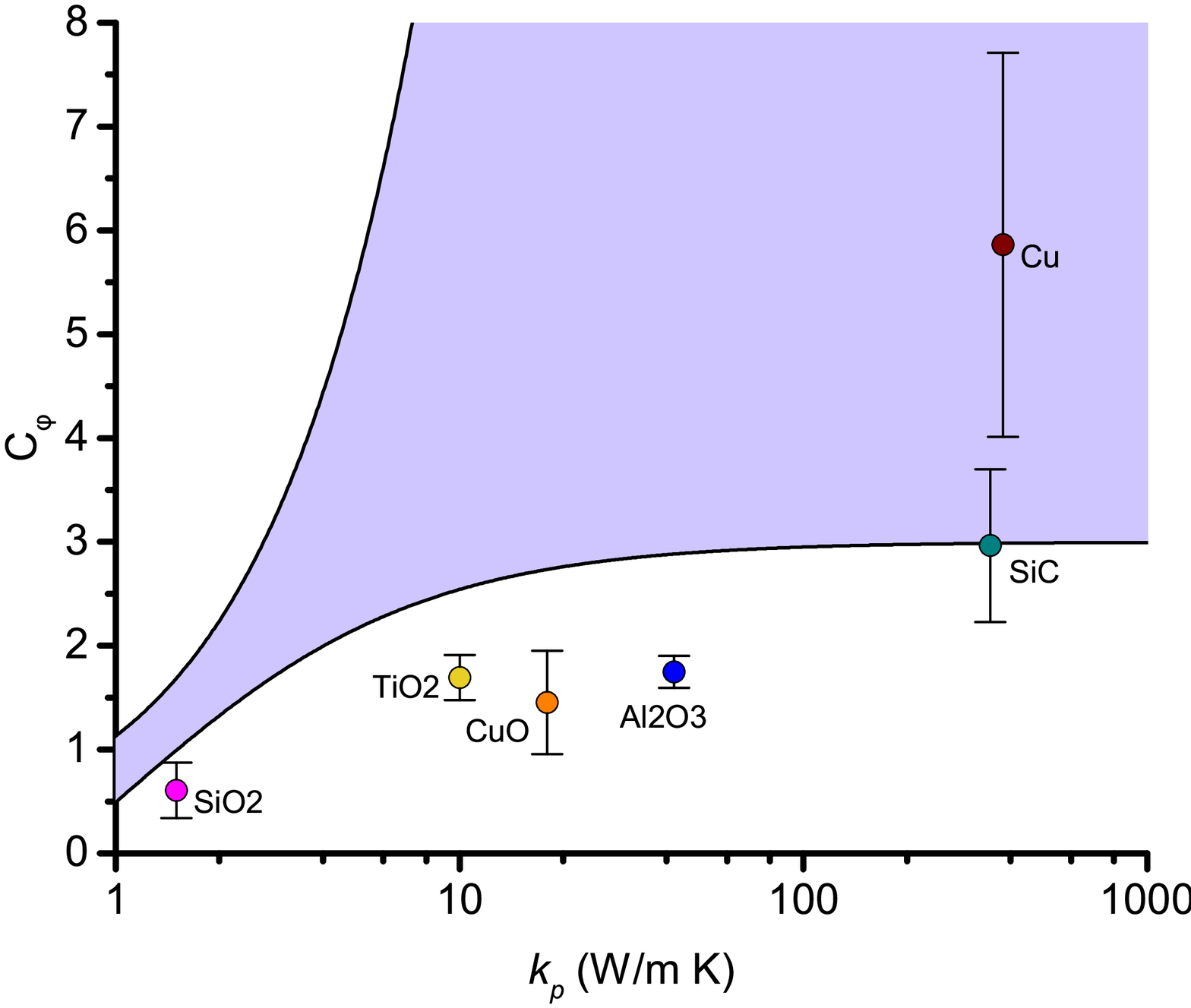}\\
     b\\
     \includegraphics[width=2.7in]{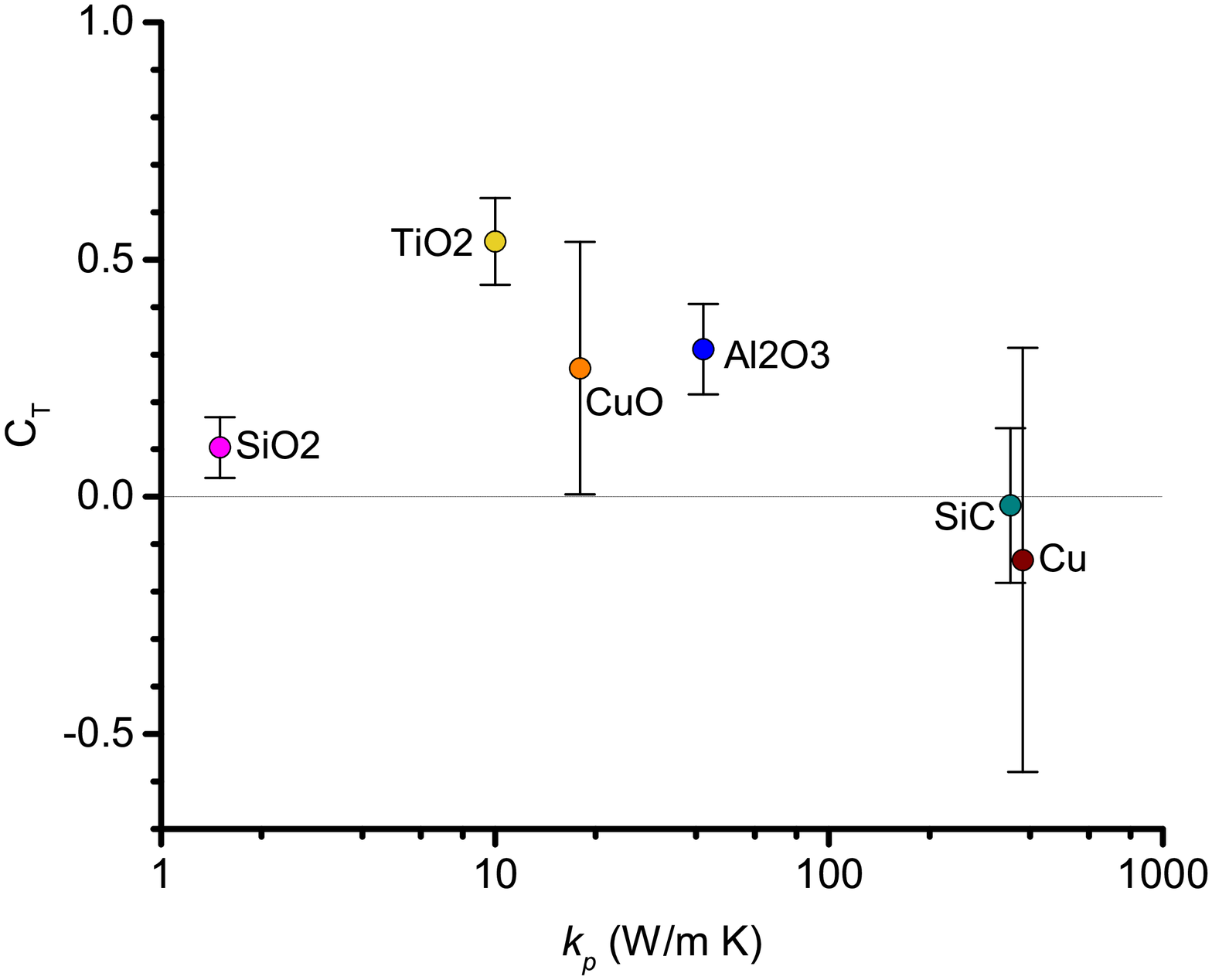}\\
     c\\
      \includegraphics[width=2.7in]{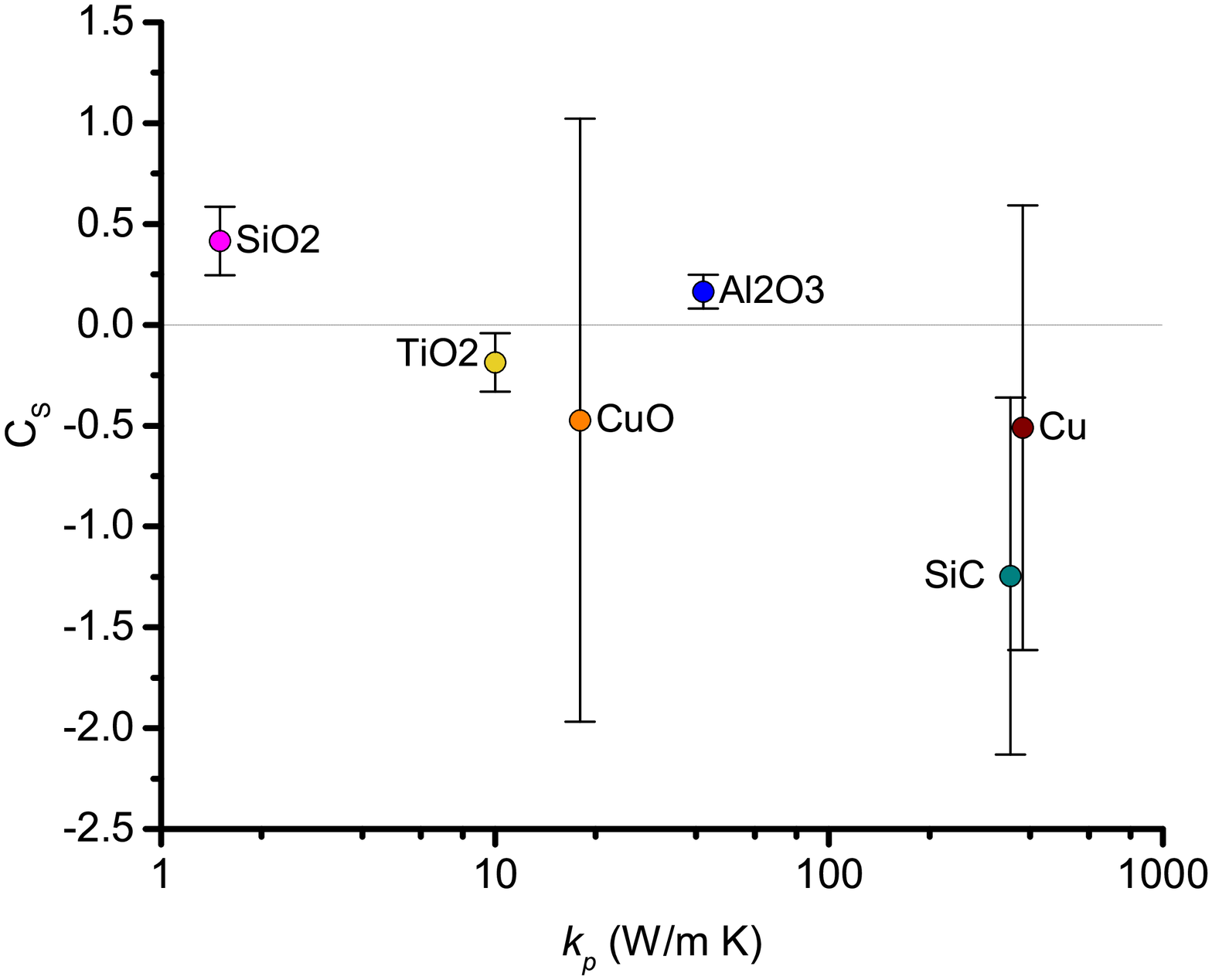}\\
    \end{tabular}
\caption{The colored symbols show the values of the model coefficient $C_i$ (with corresponding 95\% confidence-interval, shown as error-bars) as a function of the thermal conductivity \(k_\text{p}\) of each material.  (a) Concentration coefficients $C_\varphi$. The linearized HS-bounds are displayed as solid black line. The lower bound is given by the linearized Maxwell equation \eqref{eq:linMaxwell}, whereas the upper HS-bound was calculated by linearizing the right-hand-side of eq.~\eqref{eq:HS} (with $\varphi \ k_p/k_f=0$). (b) Temperature coefficients $C_\text{T}$. (c) Surface coefficients $C_\text{S}$.}
    \label{fig:3}
\end{figure}

The computed values of $C_\text{S}$ are shown in figure~\ref{fig:3}c. A strong effect of the particle size is found only in silica nanofluids (\csSi, \bsSi). For alumina nanofluids the particle size has a mildly significant and weak effect. In both cases, reducing the particle size appears to enhance the thermal conductivity. By contrast, for silicon carbide nanofluids the computed coefficient is negative (\csSic), suggesting that increasing the particle size leads to higher thermal conductivity. However, the large scatter in the data for this material (see the corresponding confidence intervals in figure~\ref{fig:3}c) does not allow  a conclusive statement. For  copper oxide and copper nanofluids the effect is statistically insignificant, which may be  attributed to the absence of variation in particle size in the experimental measurements available (see table 1, fig. 2 in the supplementary material).

The case of titania is more complicated to interpret. Here the experimental measurements span a wide range of particle sizes, yet the particle-size effect is insignificant if all data are used. If only the measurements without surfactant are considered in the statistical analysis, the particle size appears to have a very strong and significant influence, with $C_\text{S}=1.931$ and $\beta_\text{S}=0.367$.  The descriptive statistics and boxplot diagram in the supplementary material (see figs.~2 and ~4 therein) show an irregular distribution of the measured particle sizes, with most measurements for $d=21$~nm and a few for $100$~nm. This may result in the significant regression coefficients.

\subsection{Effect of surfactants}

Out of \n~data points, $1038$~data points ($62\%$) were obtained in suspensions without surfactants, whereas $618$ ($38\%$) data points were obtained with surfactants. Half of these data points ($309$) were from surfactant concentrations of less than $1.67 wt\%$, the rest with concentrations higher than $10 wt\%$ ($225$~data points) or provided no information about the concentration ($84$~data points). Descriptive statistics and corresponding boxplots can be found in section 2.2 in the supplementary materials. Depending on the type of the surfactant and concentration, the thermal conductivity of both, the base fluid and the nanofluid, may increase or decrease \cite{Gangadevi2018, Yang2012, Li2008,  Kim2018, Nasiri2011}. Because of the heterogeneity of the measurements with surfactants (in type and concentration), it would be meaningful to analyze data with similar surfactant configurations together. However, this would lead to small data subsets and would prevent statistically significant analyses.  Hence, in this work we statistically investigated the effect of surfactants by performing regressions of the data points obtained without surfactants (separately for each material), and then comparing the results to those presented in the previous sections (with and without surfactants).

The data for copper were all obtained with surfactants except for the study of Liu \emph{et al.}~\cite{Liu2006}; this is the reason why their work was excluded from the analysis for copper. For $SiO_2$ and $SiC$ nanofluids no surfactants were used. Hence the effect of surfactants cannot be investigated for these materials, as no comparisons (with/without surfactants) are possible. Still, we stress that the results for copper reveal a $C_\varphi$ beyond Maxwell's theory in contrast to the other materials and insignificant $C_\text{T}$ and $C_\text{S}$ coefficients with large error-bars.

Comparisons were possible for $Al_2O_3$ and $CuO$, with  $(405/470)$ and $(94/106)$ data points obtained without surfactants, respectively. The changes in the regressions of alumina and copper oxide without surfactants lie within the 95\% intervals of the regression for the whole data sets. Major changes occurred in the case of $TiO_2$. Out of  \nTi~data points, $83$ were obtained without surfactants. The concentration coefficient $C_\varphi$ increases and nearly approaches Maxwell's prediction. The coefficients $C_\text{T}$ and $C_\text{S}$ increase beyond the 95\% interval of the whole data set, whereas $C_0$ decreases.

Overall it can be concluded that while surfactants are expected to change the stability and the thermal conductivity of nanofluids, no conclusive statements can be made with the data available in the literature.

\section{Discussion}

Buongiorno \emph{et al.\ } \cite{Buongiorno2009} pointed out already one decade ago that nanofluid data in the literature exhibit large scatter (\( \pm 5\% \) about the median). A first critical statistical analysis of the scatter in the experimental data was done by Khanafer \emph{et al.\ } \cite{Khanafer2011}. Our statistical analysis extends and updates their analyses and confirms that the scatter in the different data sets is very large, as illustrated by the regression coefficients $R^2\in[0.23,0.75]$. In line with Buongiorno \emph{et al.}~\cite{Buongiorno2009}, we attribute the large scatter in the data to two factors. First, the experimental determination of thermophysical properties (e.g.\ thermal conductivity) of fluids is known to be susceptible of errors (even for pure fluids). Chirico \textit{et al}.~\cite{Chirico2013} reviewed  thermophysical data from five major journals and found errors in nearly one third of all publications. The most common problems were data with omitted uncertainties and the use of volume-based units \cite{Chirico2013}. Furthermore, in many publications the particle parameters (e.g.\ the size) were directly taken from the manufacturers, which is another source of error. Second, the characterizations of the nanofluids performed in the literature are often insufficient and additional variables need to be taken into account. Nieto de Castro \emph{et al.\ } \cite{NietodeCastro2017}  considered the different factors influencing the thermal conductivity of nanofluids and emphasized the need for accurate, comparable measurements including stability characterizations (see \cite{NietodeCastro2020, Bobbo2021} also).

\subsection{Dispersion state}

The stability of nanofluids depends on their preparation, i.e.\ on the use of surfactants and/or the pH value of the fluid. For example, the colloidal stabilization of titania, alumina and other oxidic particles highly depends on pH \cite{Bouguerra2018}, whereas hydrophobic particles (e.g.\ silicon carbide) agglomerate without the addition of dispersing agents \cite{Brunelli2013, Singh2007,Keblinski2008}. Hence the surfactant itself and the concentration must be selected very carefully to maximize the thermal conductivity. Due to the heterogeneity in surfactant type and concentration (and particle material and size) used across studies, it is  difficult to compare data and draw general conclusions from them. Hence, despite the widespread use of surfactants, the results are ambiguous at best. Many studies suggest improved thermal conductivity, whereas others suggest the opposite \cite{Gangadevi2018, Yang2012, Li2008, Kim2018, Nasiri2011}. The copper nanofluids in the literature were synthesized with the use of surfactants. In most studies, the concentration of surfactants was above the critical micelle concentration, possibly generating networks of micelles in the nanofluid. Hence the  high value of $C_\varphi$ obtained for copper may arise from a dispersion state different than well-dispersed particles. We note however that micellar aggregates, worm-like micelles and network structures increase the viscosity of the nanofluid \cite{Cao2015, Mingzheng2012}. Thus, the use of high surfactant concentrations may lead to high thermal conductivities, but may be rather unfavorable for the application of nanofluids.

Our analysis employs a linear regression to assess the importance of concentration, temperature and specific surface. This precludes an investigation of combined effects, whereas surface (particle-size) effects may depend on the concentration. Unfortunately, the quality and quantity of the published data do not currently allow for a statistically significant analysis of such combined effects via the inclusion of nonlinear terms (e.g.\ of the form $C_{\varphi,\text{S}}\,\varphi\,S^*$). We note however that we did not find any significant cross-correlations between variables. The aging and stability of the nanofluid, which change the dispersion state, could not be included in our statistical analysis, because these factors are seldom described in sufficient detail in published data \cite{Buongiorno2009,  Ali_2018, Sezer2018}. For the same reason, we could not analyze the dependence of the effective viscosity on concentration, temperature and size of the nanoparticles. 

The viscosity is crucial in heat transfer applications and complex structures, such as particle clusters or percolation networks, dramatically increase the viscosity of suspensions. Bouguerra \emph{et al.}~\cite{Bouguerra2018} have shown that for alumina nanofluids the well-dispersed regime (when available), is the preferred one for heat transfer applications. Studies employing molecular dynamics simulations have also shown that clustering influences the thermal conductivity of nanofluids \cite{Prasher2006, Jabbari2017, Tahmooressi2020}, and increases  their viscosity \cite{Prasher2006, Jabbari2017}. Tahmooressi \emph{et al.}~\cite{Tahmooressi2020} showed for nanoparticles at high concentrations ($\varphi = 0.1$) that the agglomeration of nanoparticles into small clusters may be more efficient towards high thermal conductivities than well-dispersed particles or percolating networks. This is in line with the experimental observations of Bouguerra \emph{et al.}~\cite{Bouguerra2018} measuring higher thermal conductivities for percolation than for dispersed particles. Clearly, simultaneous measurements of thermal conductivity and viscosity, and direct measurements of the state of dispersion of the nanofluid would be very useful. Here the determination of the diffusion coefficient and the size of the hydration layer would help \cite{Mugica2020, Gerardi2009, Thoma2019}.

\subsection{Regressions with $C_0=1$}

We also fitted all data sets discussed above, but imposing $C_0=1$ in eq.~\eqref{eq:model}. The results are shown in table~6a--g and figs.~5--7 in the supplementary materials. For silica nanofluids the results do not change much when $C_0=1$ is imposed. For all other particle materials, imposing $C_0=1$ leads to an increase in the values of $C_\varphi$, $C_\text{T}$ and $C_\text{S}$. This is because for all these materials $C_0>1$ in the linear regressions. In fact, the larger the value of $C_0$ in the linear regression, the larger the increase of the other coefficients when $C_0=1$ is set. For silicon carbide, copper oxide and copper nanofluids, the value of $C_\text{S}$ even changes from negative to positive. Hence nothing can be said about particle size for these nanofluids. In both linear and nonlinear analyses the 95\% confidence intervals for the surface coefficient $C_S$ reflect the large scatter in the data regarding particle size effects. Overall, it can be concluded that setting $C_0=1$ does not qualitatively change the results of our analysis, but rather fortifies our findings and our assessment of the data quality. 

\subsection{Analysis for small concentrations $\varphi \le 0.02$} 

We analysed the data for concentrations below 2\% separately to assess the validity of the linear assumption for the concentration in our statistical regression model. We found statistically identical results for silica, silicon carbide and copper. For copper oxide, titania and alumina (see section 2.1 and figure 3 in the supplementary materials) the (relative) performance is better at low concentrations. For alumina, there is then perfect agreement with Maxwell, whereas titania and copper oxide are above the Maxwell curve, but with confidence intervals nearly touching it. It can be concluded that higher concentrations decrease the (relative) performance, possibly due to unfavourable agglomeration effects (see e.g.\ Bouguerra \emph{et al}.~\cite{Bouguerra2018}).

\section{Conclusions and outlook}

 Our statistical analysis shows that the experimentally measured thermal conductivity of water-based nanofluids increases approximately linearly with the particle concentration. Within statistical uncertainty, this increase can be accounted for by Maxwell's Effective Medium Theory for all materials. A possible exception may be copper nanofluids, although more experiments are necessary to assess their performance and the role of surfactants. A linear increase of the conductivity with the temperature was observed for all materials in which measurements covering sufficiently large temperature ranges are available. Finally, only  silica nanofluids exhibit statistically significant, strong particle-size effects, which makes this scarcely investigated material interesting for fundamental investigations. In particular, in silica nanofluids, reducing the particle size (at constant concentration) leads to higher conductivity. Their performance is however low (compared to other nanofluids) because of their low value of $C_\varphi$.
 
In view of the detrimental effect of the increased viscosity and of stability issues, we suggest that the potential of nanofluids in engineering practice is limited. The main advantage of nanofluids, when compared to suspensions of larger particles, is their reduced sedimentation speed. We conclude that the large scatter found in the experimental measurements makes it difficult to test and compare theories for the effective thermal conductivity of water-based nanofluids. More comprehensive and precise characterizations, including the analysis of the dispersion state and of the stability on nanofluids (e.g.\ depending on the surfactant), are needed to quantify the sources of the data scatter. We believe that an improvement of the state of the art can only be achieved by ensuring the reproducibility of results with a priori identical conditions in different research groups. Silica nanoparticles are usually homogeneous in their size, are easy to handle and are sufficiently stable in dispersion. Hence, silica nanofluids are good candidates to precisely quantify the effects of particle size and dispersion stability on nanofluids. Copper nanofluids exhibit high thermal conductivities, but in order to compare their performance to other materials, it would be necessary to stabilize them with small concentrations of surfactant (below the critical micelle concentration). Unfortunately, this is challenging because metal nanoparticles rapidly oxidize and agglomerate without surfactants \cite{Gawande2016}.

Finally, we suggest that more sophisticated statistical analyses \cite{HLM62007, GelmData2007} could be employed to shed light on the source of the variability in the measured thermal conductivity of nanofluids. We will investigate possible factors in future work.

\section{Competing Interests}
The authors declare no competing interests.

\section{Funding}
his research did not receive any specific grant from funding agencies in the public, commercial, or not-for-profit sectors.

\section{Acknowledgements}
We thank Dr. Scharpenberg (Faculty Mathematics/Computer Science, University of Bremen) for critically reading an earlier version of the manuscript and providing helpful comments on the statistical analysis.

\printbibliography

@article{Keblinski2008,
author = {Keblinski, Pawel and Prasher, Ravi and Eapen, Jacob},
doi = {10.1007/s11051-007-9352-1},
issn = {1572-896X},
journal = {Journal of Nanoparticle Research},
number = {7},
pages = {1089--1097},
title = {{Thermal conductance of nanofluids: is the controversy over?}},
volume = {10},
year = {2008}
}

@article{Lee2014,
author = {Lee, Ji-Hwan and Lee, Seung-Hyun and {Pil Jang}, Seok},
doi = {10.1063/1.4872164},
issn = {0003-6951},
journal = {Applied Physics Letters},
number = {16},
pages = {161908},
publisher = {American Institute of Physics},
title = {{Do temperature and nanoparticle size affect the thermal conductivity of alumina nanofluids?}},
volume = {104},
year = {2014}
}

@article{Fan2011,
author = {Fan, Jing and Wang, Liqiu},
doi = {10.1016/j.ijheatmasstransfer.2011.05.009},
issn = {0017-9310},
journal = {International Journal of Heat and Mass Transfer},
number = {19},
pages = {4349--4359},
title = {{Heat conduction in nanofluids: Structure–property correlation}},
volume = {54},
year = {2011}
}

@article{MANNA2012,
author = {Manna, O and Singh, S and Paul, Gayatri},
doi = {10.1007/s12034-012-0366-7},
journal = {Bulletin of Materials Science},
title = {{Enhanced thermal conductivity of nano-SiC dispersed water based nanofluid}},
volume = {35},
year = {2012}
}

@article{Singh2007,
author = {Singh, Bimal P and Jena, Jayadev and Besra, Laxmidhar and Bhattacharjee, Sarama},
doi = {10.1007/s11051-006-9121-6},
issn = {1572-896X},
journal = {Journal of Nanoparticle Research},
number = {5},
pages = {797--806},
title = {{Dispersion of nano-silicon carbide (SiC) powder in aqueous suspensions}},
volume = {9},
year = {2007}
}

@article{Brunelli2013,
author = {Brunelli, Andrea and Pojana, Giulio and Callegaro, Sarah and Marcomini, Antonio},
doi = {10.1007/s11051-013-1684-4},
issn = {1572-896X},
journal = {Journal of Nanoparticle Research},
number = {6},
pages = {1684},
title = {{Agglomeration and sedimentation of titanium dioxide nanoparticles (n-TiO2) in synthetic and real waters}},
volume = {15},
year = {2013}
}

@article{Murshed2008,
author = {Murshed, S M S and Leong, K C and Yang, C},
doi = {10.1016/j.ijthermalsci.2007.05.004},
issn = {1290-0729},
journal = {International Journal of Thermal Sciences},
number = {5},
pages = {560--568},
title = {{Investigations of thermal conductivity and viscosity of nanofluids}},
volume = {47},
year = {2008}
}

@article{Chirico2013,
author = {Chirico, Robert D and Frenkel, Michael and Magee, Joseph W and Diky, Vladimir and Muzny, Chris D and Kazakov, Andrei F and Kroenlein, Kenneth and Abdulagatov, Ilmutdin and Hardin, Gary R and Acree, William E and Brenneke, Joan F and Brown, Paul L and Cummings, Peter T and de Loos, Theo W and Friend, Daniel G and Goodwin, Anthony R H and Hansen, Lee D and Haynes, William M and Koga, Nobuyoshi and Mandelis, Andreas and Marsh, Kenneth N and Mathias, Paul M and McCabe, Clare and O'Connell, John P and P{\'{a}}dua, Agilio and Rives, Vicente and Schick, Christoph and Trusler, J P Martin and Vyazovkin, Sergey and Weir, Ron D and Wu, Jiangtao},
doi = {10.1021/je400569s},
issn = {0021-9568},
journal = {Journal of Chemical {\&} Engineering Data},
number = {10},
pages = {2699--2716},
publisher = {American Chemical Society},
title = {{Improvement of Quality in Publication of Experimental Thermophysical Property Data: Challenges, Assessment Tools, Global Implementation, and Online Support}},
volume = {58},
year = {2013}
}

@article{Hashin1962,
author = {Hashin, Z and Shtrikman, S},
doi = {10.1063/1.1728579},
issn = {0021-8979},
journal = {Journal of Applied Physics},
number = {10},
pages = {3125--3131},
publisher = {American Institute of Physics},
title = {{A Variational Approach to the Theory of the Effective Magnetic Permeability of Multiphase Materials}},
volume = {33},
year = {1962}
}

@article{Eapen2010,
author = {Eapen, Jacob and Rusconi, Roberto and Piazza, Roberto and Yip, Sidney},
doi = {10.1115/1.4001304},
issn = {0022-1481},
journal = {Journal of Heat Transfer},
number = {10},
pages = {102402--102414},
publisher = {ASME},
title = {{The Classical Nature of Thermal Conduction in Nanofluids}},
volume = {132},
year = {2010}
}

@article{Li2007,
author = {Li, Calvin H and Peterson, G P},
doi = {10.1063/1.2436472},
issn = {0021-8979},
journal = {Journal of Applied Physics},
number = {4},
pages = {44312},
publisher = {American Institute of Physics},
title = {{The effect of particle size on the effective thermal conductivity of Al2O3-water nanofluids}},
volume = {101},
year = {2007}
}

@article{Chen2017,
author = {Chen, Wenjing and Zou, Changjun and Li, Xiaoke and Li, Lu},
journal = {International Journal of Heat and Mass Transfer},
doi = {10.1016/j.ijheatmasstransfer.2016.11.048},
pages = {264--270},
title = {{Experimental investigation of SiC nanofluids for solar distillation system: Stability, optical properties and thermal conductivity with saline water-based fluid}},
volume = {107},
year = {2017}
}

@article{Vadasz2005,
doi = {10.1115/1.2175149},
author = {Vadasz, Peter},
issn = {0022-1481},
journal = {Journal of Heat Transfer},
pages = {465--477},
publisher = {ASME},
title = {{Heat Conduction in Nanofluid Suspensions}},
volume = {128},
year = {2006}
}

@book{Maxwell1881,
author = {Maxwell, J. C.},
edition = {2},
pages = {435},
publisher = {Clarendon Press},
title = {{A Treatise on Electricity and Magnetism}},
year = {1881}
}

@article{Pak1998,
author = {Pak, Bock Choon and Cho, Young I},
doi = {10.1080/08916159808946559},
issn = {0891-6152},
journal = {Experimental Heat Transfer},
number = {2},
pages = {151--170},
publisher = {Taylor {\&} Francis},
title = {{Hydrodynamic And Heat Transfer Study Of Dispersed Fluids With Submicron Metallic Oxide Particles}},
volume = {11},
year = {1998}
}

@article{Rudyak2016,
author = {Rudyak, V Ya. and Minakov, A V and Krasnolutskii, S L},
doi = {10.1134/S1029959916030085},
issn = {1029-9599},
journal = {Physical Mesomechanics},
number = {3},
pages = {298--306},
title = {{Physics and mechanics of heat exchange processes in nanofluid flows}},
volume = {19},
year = {2016}
}

@article{Hong2012,
author = {Hong, Jonggan and Kim, Dongsik},
doi = {10.1016/j.tca.2011.12.019},
issn = {0040-6031},
journal = {Thermochimica Acta},
number = {SI},
pages = {28--32},
title = {{Effects of aggregation on the thermal conductivity of alumina/water nanofluids}},
volume = {542},
year = {2012}
}

@article{Ho2014,
author = {Ho, C J and Lin, Y J},
doi = {10.1016/j.icheatmasstransfer.2014.08.017},
issn = {0735-1933},
journal = {International Communications In Heat And Mass Transfer},
pages = {247--253},
title = {{Turbulent forced convection effectiveness of alumina-water nanofluid in a circular tube with elevated inlet fluid temperatures: An experimental study}},
volume = {57},
year = {2014}
}

@article{Gowda2010,
author = {Gowda, Raghu and Sun, Hongwei and Wang, Pengtao and Charmchi, Majid and Gao, Fan and Gu, Zhiyong and Budhlall, Bridgette},
doi = {10.1155/2010/807610},
issn = {1687-8132},
journal = {Advances In Mechanical Engineering},
title = {{Effects of Particle Surface Charge, Species, Concentration, and Dispersion Method on the Thermal Conductivity of Nanofluids}},
year = {2010}
}

@article{Ruan2011,
author = {Ruan, Binglu and Jacobi, Anthony M},
doi = {10.1115/1.4002980},
issn = {0022-1481},
journal = {Journal Of Heat Transfer-Transactions of the ASME},
number = {5},
title = {{Investigation on Intertube Falling-Film Heat Transfer and Mode Transitions of Aqueous-Alumina Nanofluids}},
volume = {133},
year = {2011}
}

@article{ElBrolossy2013,
author = {El-Brolossy, T A and Saber, O},
doi = {10.1016/j.expthermflusci.2012.08.011},
issn = {0894-1777},
journal = {Experimental Thermal And Fluid Science},
pages = {498--503},
title = {{Non-intrusive method for thermal properties measurement of nanofluids}},
volume = {44},
year = {2013}
}

@article{Carson2005,
author = {Carson, James K and Lovatt, Simon J and Tanner, David J and Cleland, Andrew C},
doi = {10.1016/j.ijheatmasstransfer.2004.12.032},
issn = {0017-9310},
journal = {International Journal of Heat and Mass Transfer},
number = {11},
pages = {2150--2158},
title = {{Thermal conductivity bounds for isotropic, porous materials}},
volume = {48},
year = {2005}
}

@article{Said2014,
author = {Said, Z and Saidur, R and Hepbasli, A and Rahim, N A},
doi = {10.1016/j.icheatmasstransfer.2014.08.034},
issn = {0735-1933},
journal = {International Communications in Heat and Mass Transfer},
pages = {85--95},
title = {{New thermophysical properties of water based TiO2 nanofluid—The hysteresis phenomenon revisited}},
volume = {58},
year = {2014}
}

@article{Teng2010,
author = {Teng, Tun-Ping and Hung, Yi-Hsuan and Teng, Tun-Chien and Mo, Huai-En and Hsu, How-Gao},
doi = {10.1016/j.applthermaleng.2010.05.036},
issn = {1359-4311},
journal = {Applied Thermal Engineering},
number = {14},
pages = {2213--2218},
title = {{The effect of alumina/water nanofluid particle size on thermal conductivity}},
volume = {30},
year = {2010}
}

@article{Zhu2009,
author = {Zhu, Dongsheng and Li, Xinfang and Wang, Nan and Wang, Xianju and Gao, Jinwei and Li, Hua},
doi = {10.1016/j.cap.2007.12.008},
issn = {1567-1739},
journal = {Current Applied Physics},
number = {1},
pages = {131--139},
title = {{Dispersion behavior and thermal conductivity characteristics of Al2O3–H2O nanofluids}},
volume = {9},
year = {2009}
}

@article{Li2008,
author = {Li, X F and Zhu, D S and Wang, X J and Wang, N and Gao, J W and Li, H},
doi = {10.1016/j.tca.2008.01.008},
issn = {0040-6031},
journal = {Thermochimica Acta},
number = {1},
pages = {98--103},
title = {{Thermal conductivity enhancement dependent pH and chemical surfactant for Cu-H2O nanofluids}},
volume = {469},
year = {2008}
}

@article{Duangthongsuk2009,
author = {Duangthongsuk, Weerapun and Wongwises, Somchai},
doi = {10.1016/j.expthermflusci.2009.01.005},
issn = {0894-1777},
journal = {Experimental Thermal and Fluid Science},
number = {4},
pages = {706--714},
title = {{Measurement of temperature-dependent thermal conductivity and viscosity of TiO2-water nanofluids}},
volume = {33},
year = {2009}
}

@article{Mintsa2009,
author = {Mintsa, Honorine Angue and Roy, Gilles and Nguyen, Cong Tam and Doucet, Dominique},
doi = {10.1016/j.ijthermalsci.2008.03.009},
issn = {1290-0729},
journal = {International Journal of Thermal Sciences},
number = {2},
pages = {363--371},
title = {{New temperature dependent thermal conductivity data for water-based nanofluids}},
volume = {48},
year = {2009}
}

@article{Singh2009,
author = {Singh, D and Timofeeva, E and Yu, W and Routbort, J and France, D and Smith, D and Lopez-Cepero, J M},
doi = {10.1063/1.3082094},
issn = {0021-8979},
journal = {Journal of Applied Physics},
number = {6},
pages = {64306},
publisher = {American Institute of Physics},
title = {{An investigation of silicon carbide-water nanofluid for heat transfer applications}},
volume = {105},
year = {2009}
}

@article{Chon2005,
author = {Chon, Chan Hee and Kihm, Kenneth D and Lee, Shin Pyo and Choi, Stephen U S},
doi = {10.1063/1.2093936},
issn = {0003-6951},
journal = {Applied Physics Letters},
number = {15},
pages = {153107},
publisher = {American Institute of Physics},
title = {{Empirical correlation finding the role of temperature and particle size for nanofluid (Al2O3) thermal conductivity enhancement}},
volume = {87},
year = {2005}
}

@article{Beck2009,
author = {Beck, Michael P and Yuan, Yanhui and Warrier, Pramod and Teja, Amyn S},
doi = {10.1007/s11051-008-9500-2},
issn = {1572-896X},
journal = {Journal of Nanoparticle Research},
number = {5},
pages = {1129--1136},
title = {{The effect of particle size on the thermal conductivity of alumina nanofluids}},
volume = {11},
year = {2009}
}

@article{Abareshi2010,
author = {Abareshi, Maryam and Goharshadi, Elaheh K and {Mojtaba Zebarjad}, Seyed and {Khandan Fadafan}, Hassan and Youssefi, Abbas},
doi = {10.1016/j.jmmm.2010.08.016},
issn = {0304-8853},
journal = {Journal of Magnetism and Magnetic Materials},
number = {24},
pages = {3895--3901},
title = {{Fabrication, characterization and measurement of thermal conductivity of Fe3O4 nanofluids}},
volume = {322},
year = {2010}
}

@article{Kang2006,
author = {Kang, Hyun Uk and Kim, Sung Hyun and Oh, Je Myung},
doi = {10.1080/08916150600619281},
issn = {0891-6152},
journal = {Experimental Heat Transfer},
number = {3},
pages = {181--191},
publisher = {Taylor {\&} Francis},
title = {{Estimation of Thermal Conductivity of Nanofluid Using Experimental Effective Particle Volume}},
volume = {19},
year = {2006}
}

@article{Li2005,
author = {Li, Qiang and Xuan, Yimin and Wang, Jian},
doi = {10.1016/j.expthermflusci.2005.03.021},
issn = {0894-1777},
journal = {Experimental Thermal and Fluid Science},
number = {2},
pages = {109--116},
title = {{Experimental investigations on transport properties of magnetic fluids}},
volume = {30},
year = {2005}
}

@article{JinKim2009,
author = {{Jin Kim}, Ho and Bang, In Cheol and Onoe, Jun},
journal = {Optics and Lasers in Engineering},
doi = {10.1016/j.optlaseng.2008.10.011},
pages = {532--538},
title = {{Characteristic stability of bare Au-water nanofluids fabricated by pulsed laser ablation in liquids}},
volume = {47},
year = {2009}
}

@article{Liu2006,
author = {Liu, Min-Sheng and Lin, Mark Ching-Cheng and Tsai, C Y and Wang, Chi-Chuan},
doi = {10.1016/j.ijheatmasstransfer.2006.02.012},
issn = {0017-9310},
journal = {International Journal of Heat and Mass Transfer},
number = {17},
pages = {3028--3033},
title = {{Enhancement of thermal conductivity with Cu for nanofluids using chemical reduction method}},
volume = {49},
year = {2006}
}

@article{Yang2012,
author = {Yang, Juan-Cheng and Li, Feng-Chen and Zhou, Wen-Wu and He, Yu-Rong and Jiang, Bao-Cheng},
doi = {10.1016/j.ijheatmasstransfer.2012.02.052},
issn = {0017-9310},
journal = {International Journal of Heat and Mass Transfer},
number = {11},
pages = {3160--3166},
title = {{Experimental investigation on the thermal conductivity and shear viscosity of viscoelastic-fluid-based nanofluids}},
volume = {55},
year = {2012}
}

@article{Li2006,
author = {Li, Calvin H and Peterson, G P},
doi = {10.1063/1.2191571},
issn = {0021-8979},
journal = {Journal of Applied Physics},
number = {8},
pages = {84314},
publisher = {American Institute of Physics},
title = {{Experimental investigation of temperature and volume fraction variations on the effective thermal conductivity of nanoparticle suspensions (nanofluids)}},
volume = {99},
year = {2006}
}

@article{Das2003,
author = {Das, Sarit Kumar and Putra, Nandy and Thiesen, Peter and Roetzel, Wilfried},
doi = {10.1115/1.1571080},
issn = {0022-1481},
journal = {Journal of Heat Transfer},
number = {4},
pages = {567--574},
publisher = {ASME},
title = {{Temperature Dependence of Thermal Conductivity Enhancement for Nanofluids}},
volume = {125},
year = {2003}
}

@article{Oliveira2017,
author = {Oliveira, Let{\'{i}}cia Raquel and Silva, Anielle Christine Almeida and Dantas, Noelio Oliveira and {Bandarra Filho}, Enio P},
doi = {10.1016/j.ijheatmasstransfer.2017.07.094},
issn = {0017-9310},
journal = {International Journal of Heat and Mass Transfer},
pages = {795--808},
title = {{Thermophysical properties of TiO2-PVA/water nanofluids}},
volume = {115},
year = {2017}
}

@article{Vakilinejad2018,
author = {Vakilinejad, Ali and Aroon, Mohammad Ali and Al-Abri, Mohammed and Bahmanyar, Hossein and Myint, Myo Tay Zar and Vakili-Nezhaad, G Reza},
doi = {10.1080/00986445.2017.1407922},
issn = {0098-6445},
journal = {Chemical Engineering Communications},
number = {5},
pages = {610--623},
publisher = {Taylor {\&} Francis},
title = {{Experimental and theoretical investigation of thermal conductivity of some water-based nanofluids}},
volume = {205},
year = {2018}
}

@article{Kumar2018,
author = {Kumar, Nishant and Sonawane, Shriram S and Sonawane, Shirish H},
doi = {10.1016/j.icheatmasstransfer.2017.10.001},
issn = {0735-1933},
journal = {International Communications in Heat and Mass Transfer},
pages = {1--10},
title = {{Experimental study of thermal conductivity, heat transfer and friction factor of Al2O3 based nanofluid}},
volume = {90},
year = {2018}
}

@article{Gangadevi2018,
author = {Gangadevi, R and Vinayagam, B K and Senthilraja, S},
doi = {10.1016/j.matpr.2017.12.347},
issn = {2214-7853},
journal = {Materials Today: Proceedings},
number = {2, Part 3},
pages = {9004--9011},
title = {{Effects of sonication time and temperature on thermal conductivity of CuO/water and Al2O3/water nanofluids with and without surfactant}},
volume = {5},
year = {2018}
}

@article{Choi1995,
author = {Choi, S U S and Eastman, Jeffrey},
journal = {Proceedings of the ASME International Mechanical Engineering Congress and Exposition},
title = {{Enhancing thermal conductivity of fluids with nanoparticles}},
volume = {66},
year = {1995}
}

@article{Buongiorno2005,
doi = {10.1115/1.2150834},
author = {Buongiorno, J},
issn = {0022-1481},
journal = {Journal of Heat Transfer},
number = {3},
pages = {240--250},
publisher = {ASME},
title = {{Convective Transport in Nanofluids}},
volume = {128},
year = {2005}
}

@article{Nasiri2011,
author = {Nasiri, Aida and Shariaty-Niasar, Mojtaba and Rashidi, Alimorad and Amrollahi, Azadeh and Khodafarin, Ramin},
doi = {10.1016/j.expthermflusci.2011.01.006},
issn = {0894-1777},
journal = {Experimental Thermal and Fluid Science},
number = {4},
pages = {717--723},
title = {{Effect of dispersion method on thermal conductivity and stability of nanofluid}},
volume = {35},
year = {2011}
}

@article{Bowers2018,
author = {Bowers, James and Cao, Hui and Qiao, Geng and Li, Qi and Zhang, Gan and Mura, Ernesto and Ding, Yulong},
doi = {10.1016/j.pnsc.2018.03.005},
issn = {1002-0071},
journal = {Progress in Natural Science: Materials International},
number = {2},
pages = {225--234},
title = {{Flow and heat transfer behaviour of nanofluids in microchannels}},
volume = {28},
year = {2018}
}

@article{Buschmann2018,
author = {Buschmann, M H and Azizian, R and Kempe, T and Juli{\'{a}}, J E and Mart{\'{i}}nez-Cuenca, R and Sund{\'{e}}n, B and Wu, Z and Sepp{\"{a}}l{\"{a}}, A and Ala-Nissila, T},
doi = {10.1016/j.ijthermalsci.2017.11.003},
issn = {1290-0729},
journal = {International Journal of Thermal Sciences},
pages = {504--531},
title = {{Correct interpretation of nanofluid convective heat transfer}},
volume = {129},
year = {2018}
}

@article{Mikkola2018,
author = {Mikkola, V and Puupponen, S and Granbohm, H and Saari, K and Ala-Nissila, T and Sepp{\"{a}}l{\"{a}}, A},
doi = {10.1016/j.ijthermalsci.2017.10.015},
issn = {1290-0729},
journal = {International Journal of Thermal Sciences},
pages = {187--195},
title = {{Influence of particle properties on convective heat transfer of nanofluids}},
volume = {124},
year = {2018}
}

@article{Buongiorno2009,
author = {Buongiorno, Jacopo and Venerus, David C and Prabhat, Naveen and McKrell, Thomas and Townsend, Jessica and Christianson, Rebecca and Tolmachev, Yuriy V and Keblinski, Pawel and Hu, Lin-wen and Alvarado, Jorge L and Bang, In Cheol and Bishnoi, Sandra W and Bonetti, Marco and Botz, Frank and Cecere, Anselmo and Chang, Yun and Chen, Gang and Chen, Haisheng and Chung, Sung Jae and Chyu, Minking K and Das, Sarit K and {Di Paola}, Roberto and Ding, Yulong and Dubois, Frank and Dzido, Grzegorz and Eapen, Jacob and Escher, Werner and Funfschilling, Denis and Galand, Quentin and Gao, Jinwei and Gharagozloo, Patricia E and Goodson, Kenneth E and Gutierrez, Jorge Gustavo and Hong, Haiping and Horton, Mark and Hwang, Kyo Sik and Iorio, Carlo S and Jang, Seok Pil and Jarzebski, Andrzej B and Jiang, Yiran and Jin, Liwen and Kabelac, Stephan and Kamath, Aravind and Kedzierski, Mark A and Kieng, Lim Geok and Kim, Chongyoup and Kim, Ji-Hyun and Kim, Seokwon and Lee, Seung Hyun and Leong, Kai Choong and Manna, Indranil and Michel, Bruno and Ni, Rui and Patel, Hrishikesh E and Philip, John and Poulikakos, Dimos and Reynaud, Cecile and Savino, Raffaele and Singh, Pawan K and Song, Pengxiang and Sundararajan, Thirumalachari and Timofeeva, Elena and Tritcak, Todd and Turanov, Aleksandr N and {Van Vaerenbergh}, Stefan and Wen, Dongsheng and Witharana, Sanjeeva and Yang, Chun and Yeh, Wei-Hsun and Zhao, Xiao-Zheng and Zhou, Sheng-Qi},
doi = {10.1063/1.3245330},
issn = {0021-8979},
journal = {Journal of Applied Physics},
number = {9},
pages = {94312},
publisher = {American Institute of Physics},
title = {{A benchmark study on the thermal conductivity of nanofluids}},
volume = {106},
year = {2009}
}

@article{Sundar2016,
author = {Sundar, L Syam and Hortiguela, Maria J and Singh, Manoj K and Sousa, Antonio C M},
doi = {10.1016/j.icheatmasstransfer.2016.05.025},
issn = {0735-1933},
journal = {International Communications in Heat and Mass Transfer},
pages = {245--255},
title = {{Thermal conductivity and viscosity of water based nanodiamond (ND) nanofluids: An experimental study}},
volume = {76},
year = {2016}
}

@article{Iacobazzi2016,
author = {Iacobazzi, Fabrizio and Milanese, Marco and Colangelo, Gianpiero and Lomascolo, Mauro and de Risi, Arturo},
doi = {10.1016/j.energy.2016.10.027},
issn = {0360-5442},
journal = {Energy},
number = {1},
pages = {786--794},
title = {{An explanation of the Al2O3 nanofluid thermal conductivity based on the phonon theory of liquid}},
volume = {116},
year = {2016}
}

@article{Kim2012,
author = {Kim, Chang Kyu and Lee, Gyoung-Ja and Rhee, Chang Kyu},
doi = {10.1016/j.tca.2011.12.016},
issn = {0040-6031},
journal = {Thermochimica Acta},
number = {SI},
pages = {33--36},
title = {{A study on heat transfer characteristics of spherical and fibrous alumina nanofluids}},
volume = {542},
year = {2012}
}

@article{Kim2018,
author = {Kim, Sedong and Tserengombo, Baasandulam and Choi, Soon-Ho and Noh, Jungpil and Huh, Sunchul and Choi, Byeongkeun and Chung, Hanshik and Kim, Junhyo and Jeong, Hyomin},
doi = {10.1016/j.icheatmasstransfer.2017.12.011},
issn = {0735-1933},
journal = {International Communications in Heat and Mass Transfer},
pages = {95--102},
title = {{Experimental investigation of dispersion characteristics and thermal conductivity of various surfactants on carbon based nanomaterial}},
volume = {91},
year = {2018}
}

@article{Topuz2018,
author = {Topuz, Adnan and Engin, Tahsin and {Alper {\"{O}}zalp}, A and Erdoğan, Beytullah and Mert, Serdar and Yeter, Alper},
doi = {10.1007/s10973-017-6790-6},
issn = {1588-2926},
journal = {Journal of Thermal Analysis and Calorimetry},
number = {3},
pages = {2843--2863},
title = {{Experimental investigation of optimum thermal performance and pressure drop of water-based Al2O3, TiO2 and ZnO nanofluids flowing inside a circular microchannel}},
volume = {131},
year = {2018}
}

@article{Guo2018,
author = {Guo, Wenwen and Li, Guoneng and Zheng, Youqu and Dong, Cong},
doi = {10.1016/j.tca.2018.01.008},
issn = {0040-6031},
journal = {Thermochimica Acta},
pages = {84--97},
title = {{Measurement of the thermal conductivity of SiO2 nanofluids with an optimized transient hot wire method}},
volume = {661},
year = {2018}
}

@article{Bouguerra2018,
author = {Bouguerra, Nizar and Poncet, S{\'{e}}bastien and Elkoun, Said},
doi = {10.1016/j.icheatmasstransfer.2018.02.015},
issn = {0735-1933},
journal = {International Communications in Heat and Mass Transfer},
pages = {51--55},
title = {{Dispersion regimes in alumina/water-based nanofluids: Simultaneous measurements of thermal conductivity and dynamic viscosity}},
volume = {92},
year = {2018}
}

@article{Das2018,
author = {Das, Pritam Kumar and Mallik, Arnab Kumar and Ganguly, Ranjan and Santra, Apurba Kumar},
doi = {10.1016/j.molliq.2018.01.075},
issn = {0167-7322},
journal = {Journal of Molecular Liquids},
pages = {98--107},
title = {{Stability and thermophysical measurements of TiO2 (anatase) nanofluids with different surfactants}},
volume = {254},
year = {2018}
}

@article{Nair2018,
author = {Nair, Vipin and Parekh, A D and Tailor, P R},
doi = {10.1007/s40430-018-1177-6},
issn = {1806-3691},
journal = {Journal of the Brazilian Society of Mechanical Sciences and Engineering},
number = {5},
pages = {262},
title = {{Water-based Al2O3, CuO and TiO2 nanofluids as secondary fluids for refrigeration systems: a thermal conductivity study}},
volume = {40},
year = {2018}
}

@article{Ebrahimi2018,
author = {Ebrahimi, Samaneh and Saghravani, Seyed Fazlolah},
doi = {10.1007/s00231-017-2188-z},
issn = {1432-1181},
journal = {Heat and Mass Transfer},
number = {4},
pages = {999--1008},
title = {{Experimental study of the thermal conductivity features of the water based Fe3O4/CuO nanofluid}},
volume = {54},
year = {2018}
}

@article{Pourhoseini2018,
author = {Pourhoseini, S H and Naghizadeh, N and Hoseinzadeh, H},
doi = {10.1016/j.powtec.2018.03.058},
issn = {0032-5910},
journal = {Powder Technology},
pages = {279--286},
title = {{Effect of silver-water nanofluid on heat transfer performance of a plate heat exchanger: An experimental and theoretical study}},
volume = {332},
year = {2018}
}

@article{Gao2018,
author = {Gao, Yuguo and Wang, Haochang and Sasmito, Agus P and Mujumdar, Arun S},
doi = {10.1016/j.ijheatmasstransfer.2018.02.089},
issn = {0017-9310},
journal = {International Journal of Heat and Mass Transfer},
pages = {97--109},
title = {{Measurement and modeling of thermal conductivity of graphene nanoplatelet water and ethylene glycol base nanofluids}},
volume = {123},
year = {2018}
}

@article{HemmatEsfe2014,
author = {{Hemmat Esfe}, Mohammad and Saedodin, Seyfolah and Mahian, Omid and Wongwises, Somchai},
doi = {10.1007/s10973-014-3771-x},
issn = {1588-2926},
journal = {Journal of Thermal Analysis and Calorimetry},
number = {2},
pages = {675--681},
title = {{Thermal conductivity of Al2O3/water nanofluids}},
volume = {117},
year = {2014}
}

@article{Patel2010,
author = {Patel, Hrishikesh E and Sundararajan, T and Das, Sarit K},
doi = {10.1007/s11051-009-9658-2},
issn = {1572-896X},
journal = {Journal of Nanoparticle Research},
number = {3},
pages = {1015--1031},
title = {{An experimental investigation into the thermal conductivity enhancement in oxide and metallic nanofluids}},
volume = {12},
year = {2010}
}

@article{Eastman2004,
author = {Eastman, J A and Phillpot, S R and Choi, S U S and Keblinski, P},
doi = {10.1146/annurev.matsci.34.052803.090621},
issn = {1531-7331},
journal = {Annual Review of Materials Research},
number = {1},
pages = {219--246},
publisher = {Annual Reviews},
title = {{Thermal Transport In Nanofluids}},
volume = {34},
year = {2004}
}

@article{Oh2008,
author = {Oh, Dong-Wook and Jain, Ankur and Eaton, John K and Goodson, Kenneth E and Lee, Joon Sik},
doi = {10.1016/j.ijheatfluidflow.2008.04.007},
issn = {0142-727X},
journal = {International Journal of Heat and Fluid Flow},
number = {5},
pages = {1456--1461},
title = {{Thermal conductivity measurement and sedimentation detection of aluminum oxide nanofluids by using the 3$\omega$ method}},
volume = {29},
year = {2008}
}

@article{Lee1999,
doi = {10.1115/1.2825978},
author = {Lee, S and Choi, S U.-S. and Li, S and Eastman, J A},
issn = {0022-1481},
journal = {Journal of Heat Transfer},
number = {2},
pages = {280--289},
publisher = {ASME},
title = {{Measuring Thermal Conductivity of Fluids Containing Oxide Nanoparticles}},
volume = {121},
year = {1999}
}

@article{Heyhat2012,
author = {Heyhat, M M and Kowsary, F and Rashidi, A M and Esfehani, S Alem Varzane and Amrollahi, A},
doi = {10.1016/j.icheatmasstransfer.2012.06.024},
issn = {0735-1933},
journal = {International Communications in Heat And Mass Transfer},
number = {8},
pages = {1272--1278},
title = {{Experimental investigation of turbulent flow and convective heat transfer characteristics of alumina water nanofluids in fully developed flow regime}},
volume = {39},
year = {2012}
}

@article{Chandrasekar2011,
author = {Chandrasekar, M and Suresh, S},
doi = {10.1080/08916152.2010.523809},
issn = {0891-6152},
journal = {Experimental Heat Transfer},
number = {3},
pages = {234--256},
title = {{Experiments To Explore The Mechanisms Of Heat Transfer In Nanocrystalline Alumina/Water Nanofluid Under Laminar And Turbulent Flow Conditions}},
volume = {24},
year = {2011}
}

@article{Said2014_2,
author = {Said, Z and Sajid, M H and Saidur, R and Rahim, N A and Bhuiyan, M H U},
doi = {10.1179/1432891714Z.000000000930},
issn = {1432-8917},
journal = {Material Research Innovations},
number = {S6},
pages = {47--50},
title = {{Rheological behaviour and the hysteresis phenomenon of Al2O3 nanofluids}},
volume = {18},
year = {2014}
}

@article{Kayhani2012,
author = {Kayhani, M H and Nazari, M and Soltanzadeh, H and Heyhat, M M and Kowsary, F},
doi = {10.1049/mnl.2011.0706},
issn = {1750-0443},
journal = {Micro {\&} Nano Letters},
number = {3},
pages = {223--227},
title = {{Experimental analysis of turbulent convective heat transfer and pressure drop of Al2O3/water nanofluid in horizontal tube}},
volume = {7},
year = {2012}
}

@article{Cushing2004,
author = {Cushing, Brian L and Kolesnichenko, Vladimir L and O'Connor, Charles J},
doi = {10.1021/cr030027b},
issn = {0009-2665},
journal = {Chemical Reviews},
number = {9},
pages = {3893--3946},
publisher = {American Chemical Society},
title = {{Recent Advances in the Liquid-Phase Syntheses of Inorganic Nanoparticles}},
volume = {104},
year = {2004}
}

@article{Park2007,
author = {Park, Jongnam and Joo, Jin and Kwon, Soon Gu and Jang, Youngjin and Hyeon, Taeghwan},
doi = {10.1002/ange.200603148},
issn = {0044-8249},
journal = {Angewandte Chemie},
number = {25},
pages = {4714--4745},
publisher = {John Wiley {\&} Sons, Ltd},
title = {{Synthese monodisperser sph{\"{a}}rischer Nanokristalle}},
volume = {119},
year = {2007}
}

@article {Zouli2019,
title = "Enhancement of Thermal Conductivity and Local Heat Transfer Coefficients Using Fe2O3/Water Nanofluid for Improved Thermal Desalination Processes",
journal = "Journal of Nanofluids",
year = "2019",
volume = "8",
number = "5",
pages = "1103-1122",
itemtype = "ARTICLE",
issn = "2169-432X",
doi = "doi:10.1166/jon.2019.1653",
author = "Zouli, Nasser and Said, Ibrahim A. and Al-Dahhan, Muthanna",
}

@article{Gavili2019,
author = {Gavili, Anwar and Isfahani, Taghi},
doi = {10.1007/s00231-019-02752-5},
journal = {Heat and Mass Transfer},
title = {{Experimental investigation of transient heat transfer coefficient in natural convection with Al2O3-nanofluids}},
url = {https://doi.org/10.1007/s00231-019-02752-5},
year = {2019}
}

@article{SinghSokhal2018,
author = {{Singh Sokhal}, Gurpreet and Gangacharyulu, Dasaroju and Bulasara, Vijaya Kumar},
doi = {10.1016/j.vacuum.2018.08.048},
issn = {0042-207X},
journal = {Vacuum},
pages = {268--276},
title = {{Influence of copper oxide nanoparticles on the thermophysical properties and performance of flat tube of vehicle cooling system}},
volume = {157},
year = {2018}
}

@article{Ajeel2019,
author = {Ajeel, Raheem K and Salim, W.S.-I. and Sopian, K and Yusoff, M Z and Hasnan, Khalid and Ibrahim, Adnan and Al-Waeli, Ali H A},
doi = {10.1016/j.ijheatmasstransfer.2019.118806},
issn = {0017-9310},
journal = {International Journal of Heat and Mass Transfer},
pages = {118806},
title = {{Turbulent convective heat transfer of silica oxide nanofluid through corrugated channels: An experimental and numerical study}},
volume = {145},
year = {2019}
}

@article{Modi2020,
author = {Modi, Mihir and Kangude, Prasad and Srivastava, Atul},
doi = {10.1016/j.ijheatmasstransfer.2019.118833},
issn = {0017-9310},
journal = {International Journal of Heat and Mass Transfer},
pages = {118833},
title = {{Performance evaluation of alumina nanofluids and nanoparticles-deposited surface on nucleate pool boiling phenomena}},
volume = {146},
year = {2020}
}

@article{Rejvani2019,
author = {Rejvani, Mousa and Alipour, Ali and Vahedi, Seyed Masoud and Chamkha, Ali J and Wongwises, Somchai},
doi = {10.1002/er.4854},
issn = {0363-907X},
journal = {International Journal of Energy Research},
number = {14},
pages = {8548--8571},
publisher = {John Wiley {\&} Sons, Ltd},
title = {{Optimal characteristics and heat transfer efficiency of SiO2/water nanofluid for application of energy devices: A comprehensive study}},
url = {https://doi.org/10.1002/er.4854},
volume = {43},
year = {2019}
}

@article{Agarwal2019,
author = {Agarwal, Ravi and Verma, Kamalesh and Agrawal, Narendra Kumar and Singh, Ramvir},
doi = {10.1007/s11665-019-04202-z},
issn = {1544-1024},
journal = {Journal of Materials Engineering and Performance},
number = {8},
pages = {4602--4609},
title = {{Comparison of Experimental Measurements of Thermal Conductivity of Fe2O3 Nanofluids Against Standard Theoretical Models and Artificial Neural Network Approach}},
volume = {28},
year = {2019}
}

@article{Khurana2019,
author = {Khurana, Deepak and Subudhi, Sudhakar},
doi = {10.1007/s00231-019-02629-7},
issn = {1432-1181},
journal = {Heat and Mass Transfer},
number = {10},
pages = {2831--2843},
title = {{Forced convection of Al2O3/water nanofluids with simple and modified spiral tape inserts}},
volume = {55},
year = {2019}
}

@article{Akram2019,
author = {Akram, Naveed and Sadri, Rad and Kazi, S N and Ahmed, S M and Zubir, M N M and Ridha, Mohd and Soudagar, Manzoore and Ahmed, Waqar and Arzpeyma, Mazdak and Tong, Goh Boon},
doi = {10.1007/s10973-019-08153-4},
issn = {1588-2926},
journal = {Journal of Thermal Analysis and Calorimetry},
number = {1},
pages = {609--621},
title = {{An experimental investigation on the performance of a flat-plate solar collector using eco-friendly treated graphene nanoplatelets–water nanofluids}},
volume = {138},
year = {2019}
}

@article{Manay2019,
author = {Manay, Eyuphan and Mandev, Emre},
doi = {10.1007/s10973-018-7463-9},
issn = {1588-2926},
journal = {Journal of Thermal Analysis and Calorimetry},
number = {2},
pages = {887--900},
title = {{Experimental investigation of mixed convection heat transfer of nanofluids in a circular microchannel with different inclination angles}},
url = {https://doi.org/10.1007/s10973-018-7463-9},
volume = {135},
year = {2019}
}

@article{Anbu2019,
author = {Anbu, S and Venkatachalapathy, S and Suresh, S},
doi = {10.1007/s10973-019-08008-y},
issn = {1588-2926},
journal = {Journal of Thermal Analysis and Calorimetry},
number = {3},
pages = {849--864},
title = {{Convective heat transfer studies on helically corrugated tubes with spiraled rod inserts using TiO2/DI water nanofluids}},
url = {https://doi.org/10.1007/s10973-019-08008-y},
volume = {137},
year = {2019}
}

@article{Ardekani2019,
author = {Ardekani, A Mokhtari and Kalantar, V and Heyhat, M M},
doi = {10.1007/s10973-018-08001-x},
issn = {1588-2926},
journal = {Journal of Thermal Analysis and Calorimetry},
number = {3},
pages = {779--790},
title = {{Experimental study on the flow and heat transfer characteristics of Ag/water and SiO2/water nanofluids flows in helically coiled tubes}},
volume = {137},
year = {2019}
}

@article{Bakhshan2019,
author = {Bakhshan, Y and Samari, F and Ghaemi, M and Ghafarigousheh, S and Kakoee, A},
doi = {10.1007/s40997-018-0153-1},
issn = {2364-1835},
journal = {Iranian Journal of Science and Technology, Transactions of Mechanical Engineering},
number = {1},
pages = {251--257},
title = {{Experimental Study on the Thermal Conductivity of Silver Nanoparticles Synthesized Using Sargassum Angostifolium}},
volume = {43},
year = {2019}
}

@article{Ponnada2019,
author = {Ponnada, Sneha and Subrahmanyam, T and Naidu, S V},
doi = {10.1016/j.egypro.2019.01.682},
issn = {1876-6102},
journal = {Energy Procedia},
pages = {5156--5161},
title = {{An experimental investigation on heat transfer and friction factor of Silicon Carbide/water nanofluids in a circular tube}},
volume = {158},
year = {2019}
}

@article{Yeganeh2010,
author = {Yeganeh, M and Shahtahmasebi, N and Kompany, A and Goharshadi, E K and Youssefi, A and {\v{S}}iller, L},
doi = {10.1016/j.ijheatmasstransfer.2010.03.008},
issn = {0017-9310},
journal = {International Journal of Heat and Mass Transfer},
number = {15},
pages = {3186--3192},
title = {{Volume fraction and temperature variations of the effective thermal conductivity of nanodiamond fluids in deionized water}},
volume = {53},
year = {2010}
}

@article{Xie2002,
author = {Xie, Huaqing and Wang, J and Xi, T and Liu, Y},
doi = {10.1023/A:1015121805842},
issn = {1572-9567},
journal = {International Journal of Thermophysics},
number = {2},
pages = {571--580},
title = {{Thermal Conductivity of Suspensions Containing Nanosized SiC Particles}},
volume = {23},
year = {2002}
}

@article{Saterlie2011,
author = {Saterlie, Michael and Sahin, Huseyin and Kavlicoglu, Barkan and Liu, Yanming and Graeve, Olivia},
doi = {10.1186/1556-276X-6-217},
issn = {1556-276X},
journal = {Nanoscale Research Letters},
number = {1},
pages = {217},
title = {{Particle size effects in the thermal conductivity enhancement of copper-based nanofluids}},
volume = {6},
year = {2011}
}

@article{Hajjar2014,
author = {Hajjar, Zeinab and morad Rashidi, Ali and Ghozatloo, Ahmad},
doi = {10.1016/j.icheatmasstransfer.2014.07.018},
issn = {0735-1933},
journal = {International Communications in Heat and Mass Transfer},
pages = {128--131},
title = {{Enhanced thermal conductivities of graphene oxide nanofluids}},
volume = {57},
year = {2014}
}

@article{Paul2010,
author = {Paul, G and Pal, T and Manna, I},
doi = {10.1016/j.jcis.2010.05.086},
issn = {0021-9797},
journal = {Journal of Colloid and Interface Science},
number = {1},
pages = {434--437},
title = {{Thermo-physical property measurement of nano-gold dispersed water based nanofluids prepared by chemical precipitation technique}},
volume = {349},
year = {2010}
}

@article{Sezer2018,
	Author = {Sezer, Nurettin and Atieh, Muataz and Ko{\c{c}}, Muammer},
	Journal = {Powder Technology},
	Title = {{A comprehensive review on synthesis, stability, thermophysical properties, and characterization of nanofluids}},
	Volume = {344},
	Year = {2018},
}

@article{Ali_2018,
	Address = {ADAM HOUSE, 3RD FLR, 1 FITZROY SQ, LONDON, W1T 5HF, ENGLAND},
	Author = {Ali, Naser and Teixeira, Joao A and Addali, Abdulmajid},
	Doi = {10.1155/2018/6978130},
	Issn = {1687-4110},
	Journal = {Journal of Nanomaterials},
	Publisher = {HINDAWI LTD},
	Title = {{A Review on Nanofluids: Fabrication, Stability, and Thermophysical Properties}},
	Type = {Review},
	Year = {2018},
	}

@article{SyamSundar2013,
author = {{Syam Sundar}, L and Singh, Manoj K and Sousa, Antonio C M},
doi = {10.1016/j.icheatmasstransfer.2013.02.014},
issn = {0735-1933},
journal = {International Communications in Heat and Mass Transfer},
pages = {7--14},
title = {{Investigation of thermal conductivity and viscosity of Fe3O4 nanofluid for heat transfer applications}},
volume = {44},
year = {2013}
}

@article{Jabbari2017,
title = "Thermal conductivity and viscosity of nanofluids: A review of recent molecular dynamics studies",
journal = "Chemical Engineering Science",
volume = "174",
pages = "67 - 81",
year = "2017",
issn = "0009-2509",
doi = "10.1016/j.ces.2017.08.034",
url = "http://www.sciencedirect.com/science/article/pii/S0009250917305432",
author = "Fatemeh Jabbari and Ali Rajabpour and Seifollah Saedodin",
}

@article{Tahmooressi2020,
title = "Numerical simulation of aggregation effect on nanofluids thermal conductivity using the lattice Boltzmann method",
journal = "International Communications in Heat and Mass Transfer",
volume = "110",
pages = "104408",
year = "2020",
issn = "0735-1933",
doi = "10.1016/j.icheatmasstransfer.2019.104408",
url = "http://www.sciencedirect.com/science/article/pii/S073519331930274X",
author = "Hamed Tahmooressi and Alibakhsh Kasaeian and Ali Tarokh and Roya Rezaei and Mina Hoorfar",
}

@article{Prasher2006,
author = {Prasher,Ravi  and Evans,William  and Meakin,Paul  and Fish,Jacob  and Phelan,Patrick  and Keblinski,Pawel },
title = {Effect of aggregation on thermal conduction in colloidal nanofluids},
journal = {Applied Physics Letters},
volume = {89},
number = {14},
pages = {143119},
year = {2006},
doi = {10.1063/1.2360229},
}

@article{Gerardi2009,
author = {Gerardi,Craig  and Cory,David  and Buongiorno,Jacopo  and Hu,Lin-Wen  and McKrell,Thomas },
title = {Nuclear magnetic resonance-based study of ordered layering on the surface of alumina nanoparticles in water},
journal = {Applied Physics Letters},
volume = {95},
number = {25},
pages = {253104},
year = {2009},
doi = {10.1063/1.3276551},
}

@article{Thoma2019,
author = {Thom{\"{a}}, Sabrina L J and Krauss, Sebastian W and Eckardt, Mirco and Chater, Phil and Zobel, Mirijam},
doi = {10.1038/s41467-019-09007-1},
issn = {2041-1723},
journal = {Nature Communications},
number = {1},
pages = {995},
title = {{Atomic insight into hydration shells around facetted nanoparticles}},
url = {10.1038/s41467-019-09007-1},
volume = {10},
year = {2019}
}

@article{Mugica2020,
author = {Mugica, Ibai and Poncet, S{\'{e}}bastien},
doi = {10.1007/s11051-020-4776-y},
issn = {1572-896X},
journal = {Journal of Nanoparticle Research},
number = {5},
pages = {113},
title = {{A critical review of the most popular mathematical models for nanofluid thermal conductivity}},
url = {10.1007/s11051-020-4776-y},
volume = {22},
year = {2020}
}

@article{Cao2015,
author = {Cao, Fangyu and Liu, Ying and Xu, Jiajun and He, Yadong and Hammouda, B and Qiao, Rui and Yang, Bao},
doi = {10.1038/srep16040},
issn = {2045-2322},
journal = {Scientific Reports},
number = {1},
pages = {16040},
title = {{Probing Nanoscale Thermal Transport in Surfactant Solutions}},
url = {10.1038/srep16040},
volume = {5},
year = {2015}
}

@article{Mingzheng2012,
author = {Mingzheng, Zhou and Guodong, Xia and Jian, Li and Lei, Chai and Lijun, Zhou},
doi = {10.1016/j.expthermflusci.2011.07.014},
issn = {0894-1777},
journal = {Experimental Thermal and Fluid Science},
pages = {22--29},
title = {{Analysis of factors influencing thermal conductivity and viscosity in different kinds of surfactant solutions}},
url = {http://www.sciencedirect.com/science/article/pii/S0894177711001610},
volume = {36},
year = {2012}
}

@article {NietodeCastro2017,
title = {Understanding Stability, Measurements, and Mechanisms of Thermal Conductivity of Nanofluids},
journal = {Journal of Nanofluids},
year = {2017},
volume = {6},
number = {5},
pages = {804-811},
url = "https://www.ingentaconnect.com/content/asp/jon/2017/00000006/00000005/art00002",
doi = "doi:10.1166/jon.2017.1388",
author = {Nieto de Castro, Carlos A. and Vieira, Salom{\’e} I. C. and Louren{\c{c}}o, Maria J. and Murshed, S. M. Sohel},
}

@Article{NietodeCastro2020,
AUTHOR = {Nieto de Castro, Carlos A. and Lourenço, Maria José V.},
TITLE = {Towards the Correct Measurement of Thermal Conductivity of Ionic Melts and Nanofluids},
JOURNAL = {Energies},
VOLUME = {13},
YEAR = {2020},
NUMBER = {1},
ARTICLE-NUMBER = {99},
URL = {https://www.mdpi.com/1996-1073/13/1/99},
ISSN = {1996-1073},
DOI = {10.3390/en13010099}
}

@article{Khanafer2011,
title = {A critical synthesis of thermophysical characteristics of nanofluids},
journal = {International Journal of Heat and Mass Transfer},
volume = {54},
number = {19},
pages = {4410-4428},
year = {2011},
issn = {0017-9310},
doi = {10.1016/j.ijheatmasstransfer.2011.04.048},
url = {https://www.sciencedirect.com/science/article/pii/S0017931011002699},
author = {Khalil Khanafer and Kambiz Vafai},
}

@article{Fedele2012,
title = {Viscosity and thermal conductivity measurements of water-based nanofluids containing titanium oxide nanoparticles},
journal = {International Journal of Refrigeration},
volume = {35},
number = {5},
pages = {1359-1366},
year = {2012},
issn = {0140-7007},
doi = {10.1016/j.ijrefrig.2012.03.012},
url = {https://www.sciencedirect.com/science/article/pii/S0140700712000709},
author = {Laura Fedele and Laura Colla and Sergio Bobbo},
}

@Article{Bobbo2021,
AUTHOR = {Bobbo, Sergio and Buonomo, Bernardo and Manca, Oronzio and Vigna, Silvio and Fedele, Laura},
TITLE = {Analysis of the Parameters Required to Properly Define Nanofluids for Heat Transfer Applications},
JOURNAL = {Fluids},
VOLUME = {6},
YEAR = {2021},
NUMBER = {2},
ARTICLE-NUMBER = {65},
URL = {https://www.mdpi.com/2311-5521/6/2/65},
ISSN = {2311-5521},
DOI = {10.3390/fluids6020065}
}

@article{Gawande2016,
author = {Gawande, Manoj B and Goswami, Anandarup and Felpin, Fran{\c{c}}ois-Xavier and Asefa, Tewodros and Huang, Xiaoxi and Silva, Rafael and Zou, Xiaoxin and Zboril, Radek and Varma, Rajender S},
doi = {10.1021/acs.chemrev.5b00482},
issn = {0009-2665},
journal = {Chemical Reviews},
number = {6},
pages = {3722--3811},
publisher = {American Chemical Society},
title = {{Cu and Cu-Based Nanoparticles: Synthesis and Applications in Catalysis}},
url = {https://doi.org/10.1021/acs.chemrev.5b00482},
volume = {116},
year = {2016}
}

@book{GelmData2007,
title = {Data analysis using regression and multilevel/hierarchical models},
author = {Andrew Gelman and Jennifer Hill},
series = {Analytical methods for social research},
edition = {Repr. with corr., 3. print.},
publisher = {University Press},
address = {Cambridge},
year = {2007},
ISBN = {0521867061 and 052168689X and 9780521867061 and 9780521686891 and 9780521686891},
}

@book{HLM62007,
title = {HLM 6 : hierarchical linear and nonlinear modeling},
author = {Stephen Raudenbush and Stephen W. Raudenbush},
edition = {4th print.},
publisher = {Scientific Software International},
address = {Lincolnwood, Ill.},
year = {2007},
ISBN = {9780894980541},
}

\end{document}


\maketitle
\section{Descriptive statistics}

\begin{figure} [h]
    \centering
    \begin{tabular} {cc}
   \includegraphics[width=0.5\textwidth]{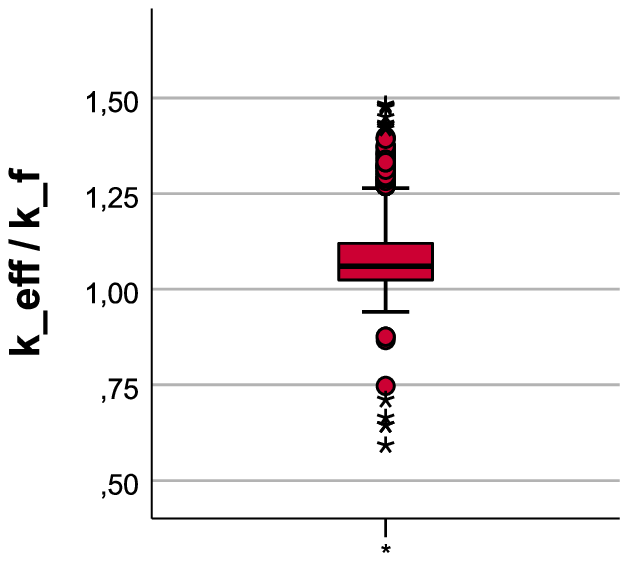} & \includegraphics[width=0.5\textwidth]{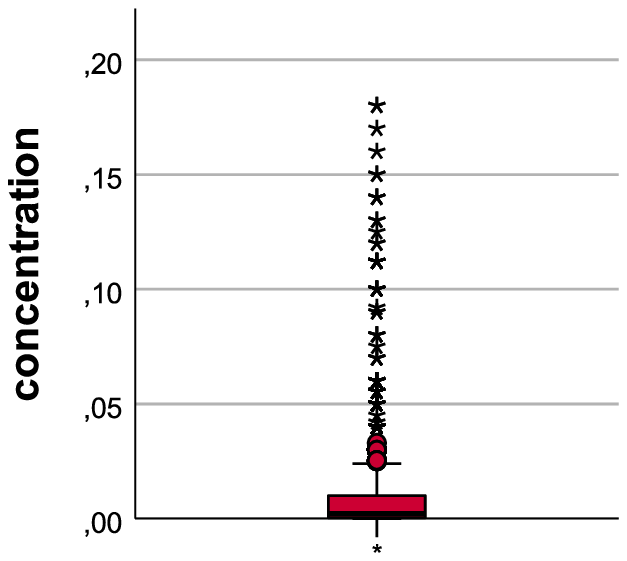} \\
    \includegraphics[width=0.5\textwidth]{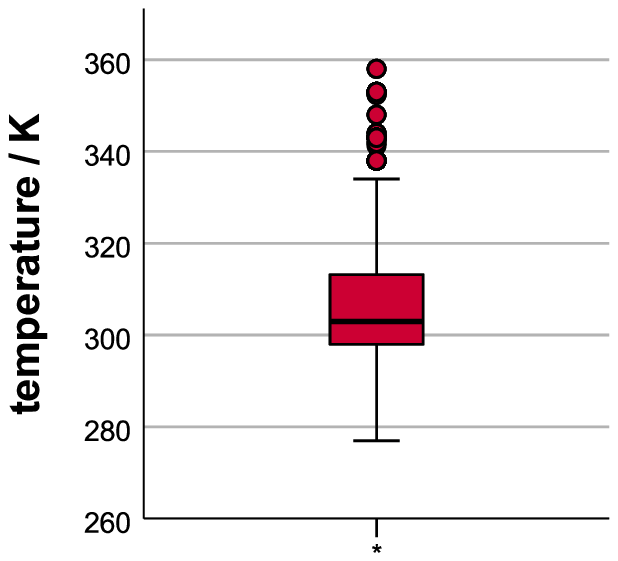} &
    \includegraphics[width=0.5\textwidth]{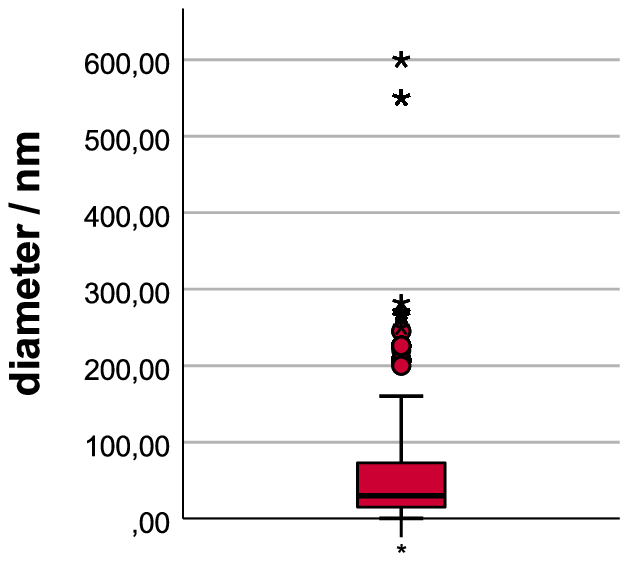} \\
    \end{tabular}
    \caption{Boxplott diagrams showing the distribution of the data for all data points with respect to the thermal conductivity $k_\text{eff}/k_\text{f}$, concentration  $\varphi$, the temperature $T$, and the nanoparticle size $d$.}
    \label{fig:box1}
\end{figure}

\begin{figure} 
    \centering
    \begin{tabular} {c}
   \includegraphics[width=0.67\textwidth]{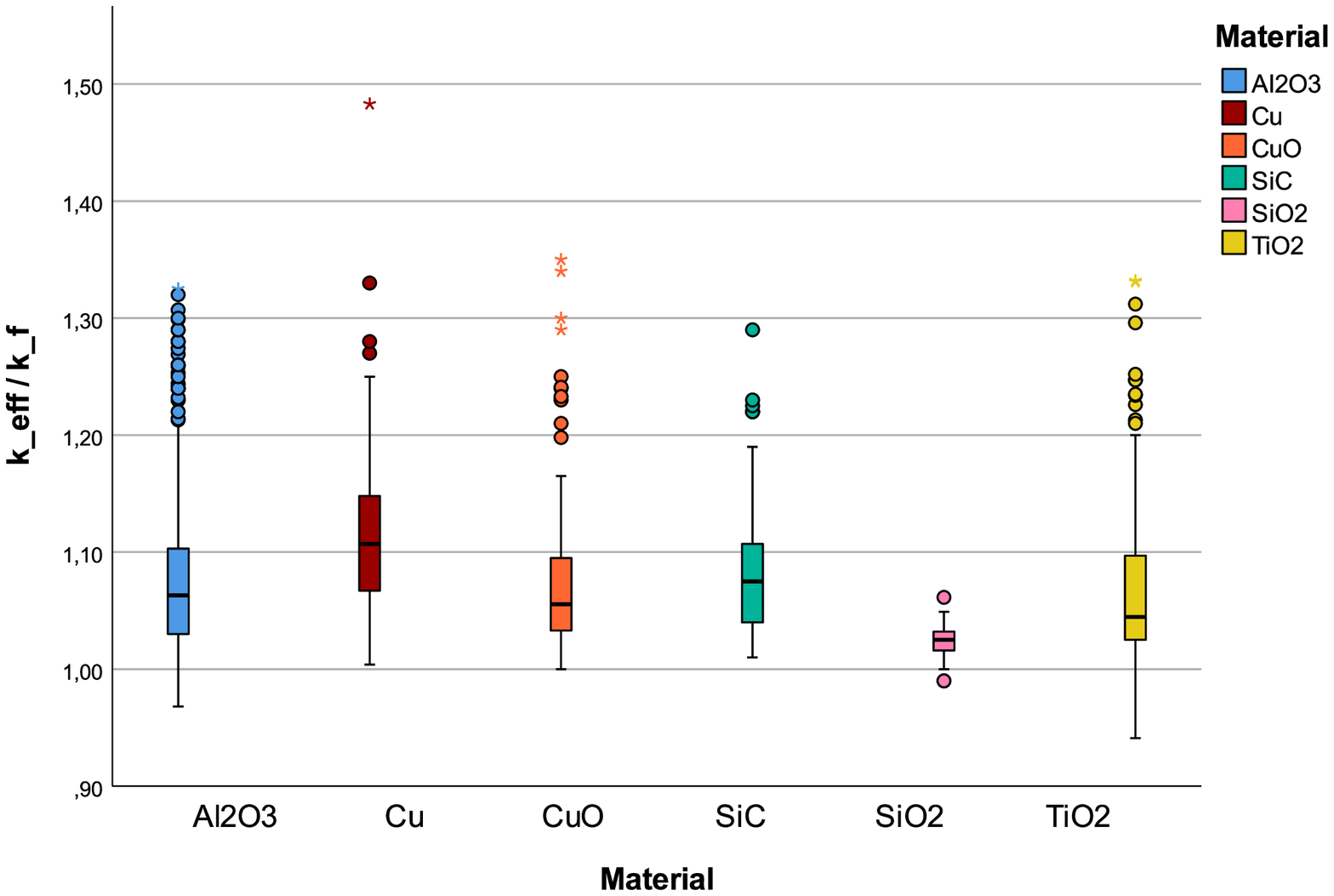} \\
    \includegraphics[width=0.67\textwidth]{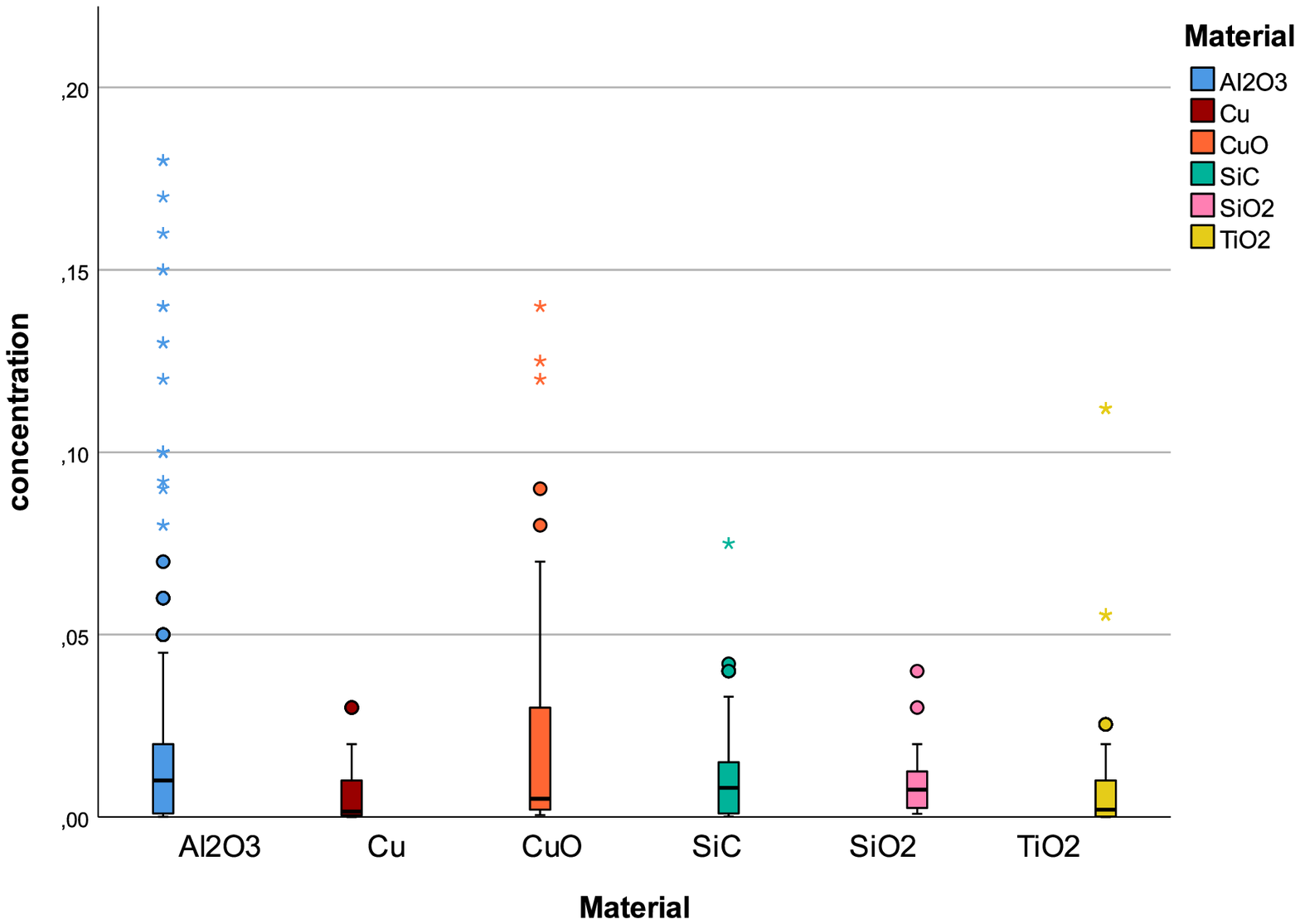} \\
    \includegraphics[width=0.67\textwidth]{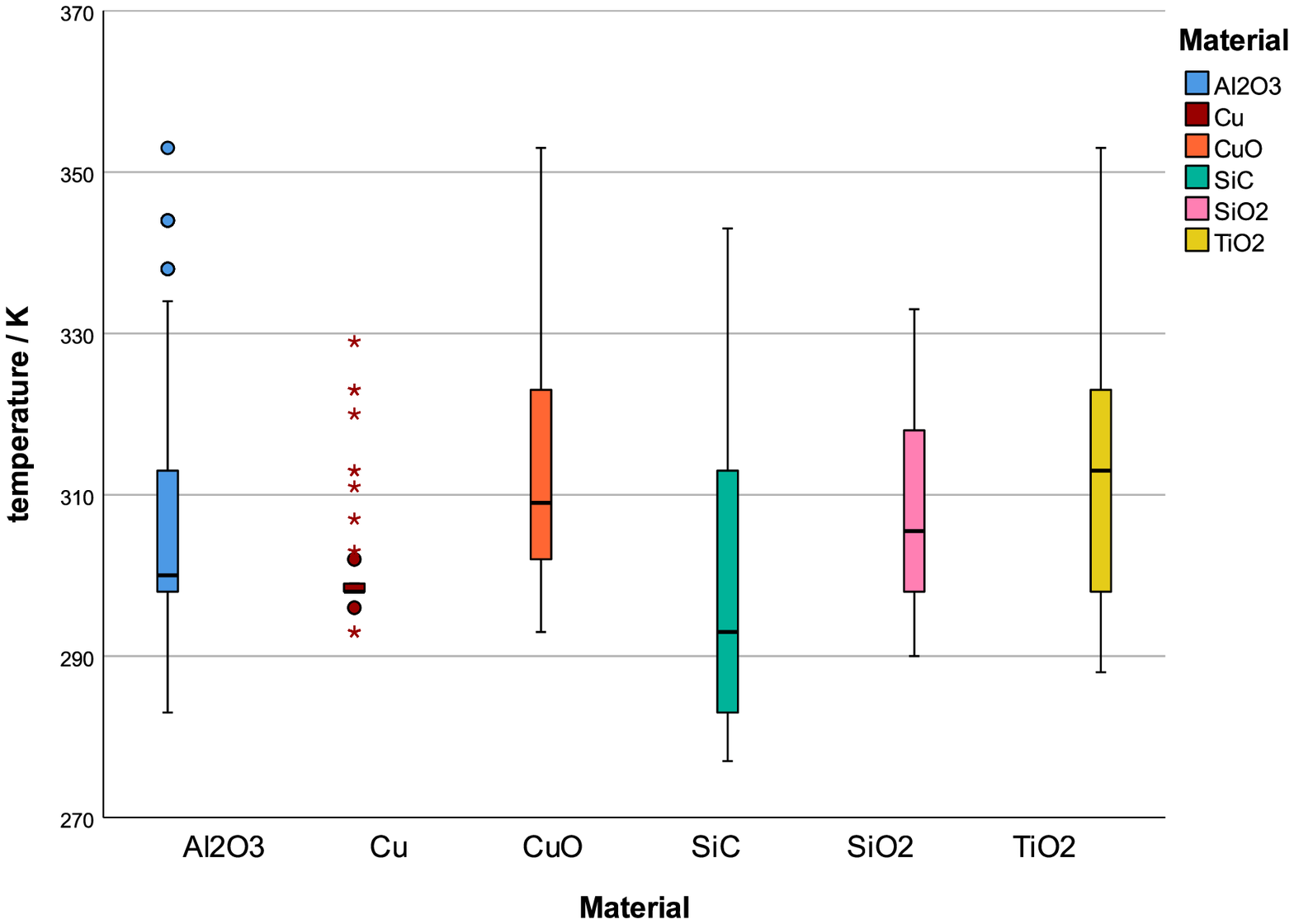} \\  
    \includegraphics[width=0.67\textwidth]{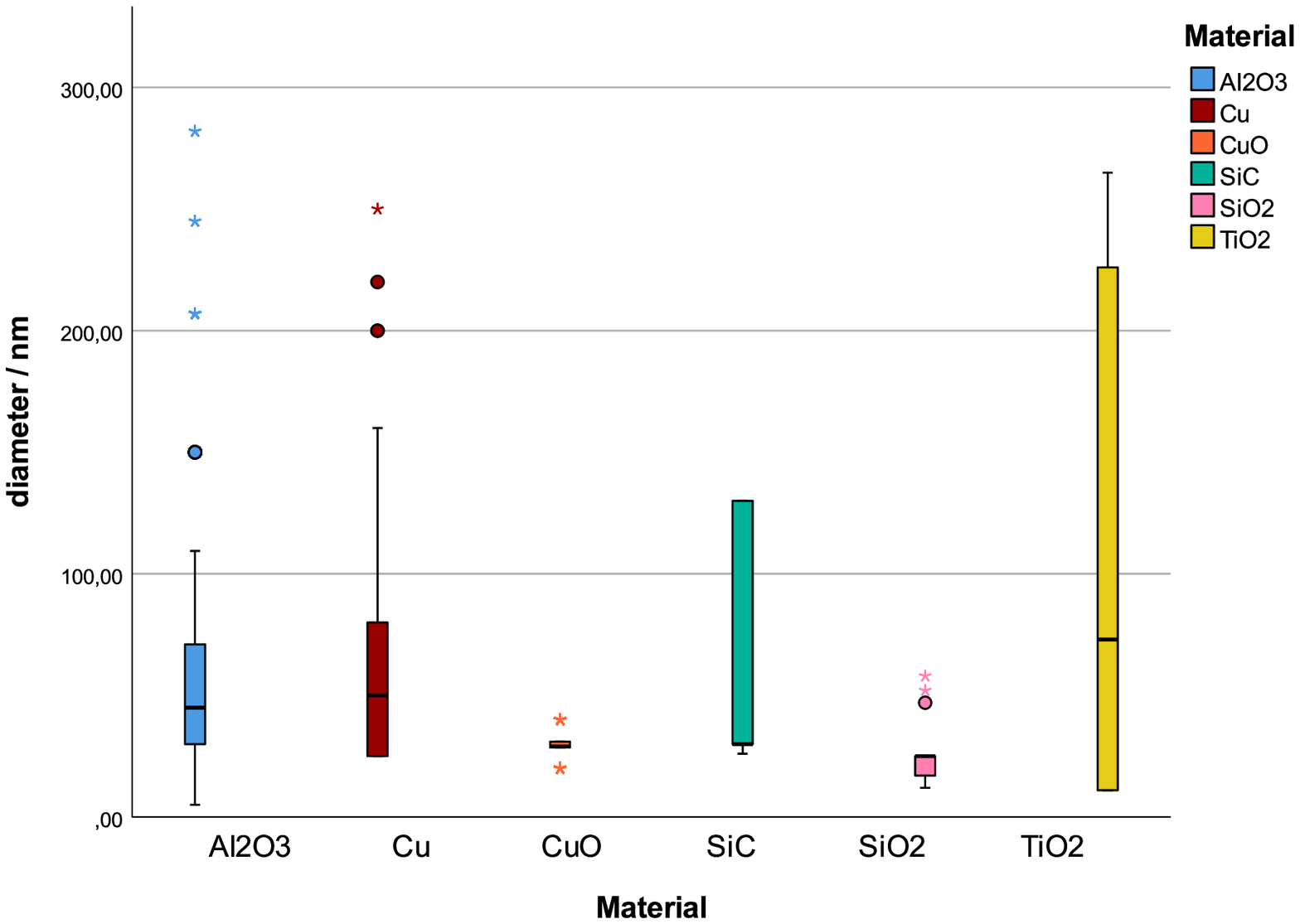} \\
    \end{tabular}
    \caption{Boxplott diagrams showing the distribution of the data for the analyzed materials with respect to the thermal conductivity $k_\text{eff}/k_\text{f}$, concentration  $\varphi$, the temperature $T$, and the nanoparticle size $d$.}
    \label{fig:box2}
\end{figure}

The following table \ref{tab:descriptive}a-d show descriptive statistics of the data with respect to thermal conductivity $k_\text{eff}/k_\text{f}$, the concentration  $\varphi$, the temperature $T$, and the nanoparticle size $d$. Figures \ref{fig:box1}--\ref{fig:box2} show their corresponding histograms.

\begin{table} [ht]
    \centering
    \caption{Mean, minimum, maximum values and percentiles of the parameters $k_\text{eff}/k_\text{f},\  \varphi, \  T, \  d$ \ for all data points and the single materials.}
    \label{tab:descriptive} 
    \end{table}

a) Thermal Conductivity $k_\text{eff}/k_\text{f}$ \\

\begin{tabular}  {ll ccccccc }
\hline
            &    & all data & $Al_2O_3$ & $TiO_2$ & $CuO$  & $Cu$  & $SiO_2$   & $SiC$   \\ \hline
mean value  &    & 1.084    & 1.078     & 1.070   & 1.081  & 1.111  & 1.024    & 1.086 \\
minimum     &    & 0.592    & 0.968     & 0.941   & 1.000  & 1.005  & 0.990    & 1.010 \\
maximum     &    & 1.483    & 1.325     & 1.332   & 1.350  & 1.483  & 1.061    & 1.290 \\
percentiles & 01 & 0.970    & 0.982     & 0.972   & 1.000  & 1.005  & 0.990    & 1.010 \\
            & 05 & 0.990    & 1.002     & 1.006   & 1.0015 & 1.008  & 1.002    & 1.011 \\
            & 10 & 1.000    & 1.015     & 1.014   & 1.017  & 1.021  & 1.007    & 1.017 \\
            & 25 & 1.024    & 1.030     & 1.025   & 1.033  & 1.056  & 1.016    & 1.040 \\
            & 50 & 1.060    & 1.063     & 1.045   & 1.056  & 1.106  & 1.025    & 1.075 \\
            & 75 & 1.120    & 1.104     & 1.098   & 1.096  & 1.149  & 1.032    & 1.111 \\ 
            & 90 & 1.220    & 1.172     & 1.154   & 1.201  & 1.205  & 1.040    & 1.208 \\
            & 95 & 1.280    & 1.236     & 1.230   & 1.247  & 1.255  & 1.045    & 1.227 \\
            & 99 & 1.373    & 1.299     & 1.331   & 1.349  &        &          &  \\\hline
\end{tabular}

\vspace{0.5cm}

{b) Concentration $\varphi$ ($\cdot 10^{-2}$)} \\

\begin{tabular} {ll ccccccc }   
\hline
            &    & all data & $Al_2O_3$ & $TiO_2$  & $CuO$  & $Cu$  & $SiO_2$   & $SiC$ \\ \hline
mean value  &    & 1.047    & 1.874    & 1.016    & 1.946   & 0.635  & 0.923    & 1.191 \\
minimum     &    & 0.001    & 0.003    & 0.001    & 0.015   & 0.001  & 0.09     & 0.01 \\
maximum     &    & 18.0     & 18.0     & 11.2      & 14.0    & 3.0    & 4.0      & 7.5  \\
percentiles & 01 & 0.002    & 0.003    & 0.001    & 0.05    & 0.001  & 0.09     & 0.01 \\
            & 05 & 0.008    & 0.01     & 0.002    & 0.05    & 0.001  & 0.11     & 0.04 \\
            & 10 & 0.02     & 0.025    & 0.003    & 0.1     & 0.023  & 0.13     & 0.04 \\ 
            & 25 & 0.044     & 0.10     & 0.01     & 0.2     & 0.045  & 0.25     & 0.10 \\
            & 50 & 0.22     & 1.0      & 0.2     & 0.5     & 0.165  & 0.75     & 0.80 \\
            & 75 & 1.0      & 2.0      & 1.0     & 3.0     & 1.0    & 1.31     & 1.75 \\ 
            & 90 & 3.0      & 4.0      & 2.54      & 6.0     & 1.30   & 2.0      & 4.0 \\
            & 95 & 4.0      & 6.5      & 5.54      & 7.7     & 2.25   & 2.0      & 4.06 \\
            & 99 & 11.2       & 15.3     & 11.2      & 13.9    &        &          &  \\ \hline
\end{tabular}

\vspace{0.5cm}

{c) Temperature $T$} \\

\begin{tabular}  {ll ccccccc }
\hline
            &    & all data & $Al_2O_3$ & $TiO_2$  & $CuO$  & $Cu$      & $SiO_2$    & $SiC$  \\ \hline
mean value  &    & 307.1     & 305.8   & 312.3    & 313.0    & 302.0    & 307.2      & 299.6  \\
minimum     &    & 277       & 283     & 288      & 293      & 293      & 290        & 277    \\
maximum     &    & 358       & 353     & 353      & 353      & 329      & 333        & 343    \\
percentiles & 01 & 283       & 283     & 288      & 293      & 293      & 290        & 277    \\
            & 05 & 288       & 293     & 293      & 294      & 293      & 293        & 277    \\
            & 10 & 293       & 293     & 293      & 298      & 298      & 293        & 277    \\ 
            & 25 & 298       & 298     & 298      & 302      & 298      & 298        & 283    \\
            & 50 & 303       & 300     & 313    & 309      & 298      & 305.5      & 293    \\
            & 75 & 313.3       & 313     & 323      & 323      & 302.3    & 318        & 313    \\
            & 90 & 328       & 324     & 333      & 336      & 316.5    & 323        & 323    \\
            & 95 & 333       & 333     & 342.2      & 349      & 323      & 323        & 343    \\
            & 99 & 348       & 344     & 353      & 353      &       &            &      \\\hline
\end{tabular}

\vspace{0.5cm}

{d) Nanoparticle size $d$}  \\

\begin{tabular}  {ll ccccccc }
\hline
            &    & all data & $Al_2O_3$ & $TiO_2$  & $CuO$  & $Cu$  & $SiO_2$ & $SiC$  \\ \hline
mean value  &    & 81.3     & 61.1      & 99.4    & 29.7   & 47.8  & 23.0   & 95.7 \\
minimum     &    & 0.34     & 5         & 11       & 20     & 25    & 12     & 26   \\
maximum     &    & 600      & 282       & 265      & 40     & 160   & 58     & 600  \\
percentiles & 01 & 0.34     & 05        & 11       & 20     & 25    & 12     & 26   \\
            & 05 & 03       & 05        & 11       & 20     & 25    & 12     & 26  \\
            & 10 & 10       & 11        & 11       & 20     & 25    & 15     & 26.4  \\ 
            & 25 & 15       & 30        & 11       & 28.6   & 25    & 17     & 30   \\
            & 50 & 30       & 45        & 73       & 29     & 50    & 25     & 30  \\
            & 75 & 73     & 71        & 226      & 31     & 57.5  & 25     & 130  \\
            & 90 & 265      & 150       & 265      & 40     & 80    & 25     & 130   \\
            & 95 & 550      & 207       & 265      & 40     & 80    & 25     & 600  \\
            & 99 & 550      & 245       & 265      & 40     &       &        &    \\\hline
\end{tabular}

\newpage

\newpage

\section{Linear Regression}

The following table \ref{tab:linear} a-g \ show the results for the statistical analysis done with SPSS. Next to the general dataset, statistics were done on six single materials. 

\begin{table}[h]
    \centering
    \caption{Results for linear regression as described in section 2.1. The table shows $R, R^2$, sum of squares, the coefficients with corresponding errors, 95\% interval and $\beta$-values.}
    \label{tab:linear}
\end{table}

{a) General model} \\

\begin{tabular} {c c c c c }
\hline
$R$     & $R^2$          & corrected $R^2$ & standard error & sum of squares    \\  \hline
0.538   & 0.289          & 0.288           & 0.078          & 4.095              \\ \hline
\end{tabular}

\vspace{0.5cm}

\begin{tabular} {c c c c c c}
\hline
& \multicolumn{3}{l}{}                       & \multicolumn{2}{c}{95\% Interval} \\ \hline
      & coefficient & error        & $\beta$              & lower bound       & upper bound   \\ \hline
$C_0$          & 1.031       & 0.003        &                & 1.025             & 1.037         \\
$C_\varphi$    & 1.812       & 0.097        & 0.391          & 1.622             & 2.002         \\
$C_T$          & 0.506       & 0.040        & 0.267          & 0.429             & 0.584         \\
$C_S$          & 0.092       & 0.006        & 0.342          & 0.081             & 0.103    \\ \hline

\end{tabular}

\vspace{0.5cm}

{b) Alumina $Al_2O_3$} \\

\begin{tabular} {c c c c c}
\hline
$R$     & $R^2$          & corrected $R^2$ & standard error & sum of squares    \\ \hline
0.727   & 0.529          & 0.526           & 0.046          & 1.114                     \\ \hline
\end{tabular}

\vspace{0.5cm}

\begin{tabular} {c c c c c c}
\hline
    &   \multicolumn{3}{c}{}                       & \multicolumn{2}{c}{95\% Interval} \\ \hline
      & coefficient & error        & $\beta$              & lower bound       & upper bound   \\ \hline
$C_0$          & 1.025       & 0.004        &                & 1.017             & 1.032         \\
$C_\varphi$    & 1.748        & 0.079       & 0.715          & 1.594             & 1.903         \\
$C_T$          & 0.311       & 0.048        & 0.209          & 0.216             & 0.406         \\
$C_S$          & 0.164       & 0.043        & 0.123          & 0.080             & 0.248  \\ \hline

\end{tabular}

\vspace{0.5cm}

{c) Titania $TiO_2$} \\

\begin{tabular}  {c c c c c}
\hline
$R$     & $R^2$          & corrected $R^2$ & standard error & sum of squares       \\ \hline
0.870   & 0.756               & 0.752        & 0.034          & 0.644                \\ \hline
\end{tabular}

\vspace{0.5cm}

\begin{tabular} {c c c c c c}
\hline
    &             \multicolumn{3}{c}{}                      & \multicolumn{2}{c}{95\% Interval} \\ \hline
      & coefficient & error        & $\beta$              & lower bound       & upper bound   \\  \hline
$C_0$          & 1.018       & 0.005        &                & 1.007             & 1.028         \\
$C_\varphi$    & 1.694       & 0.111        & 0.579          & 1.476             & 1.913         \\
$C_T$          & 0.629       & 0.046        & 0.512          & 0.538             & 0.721         \\
$C_S$          & -0.186      & 0.073        & -0.097         & -0.331            & -0.041 \\ \hline

\end{tabular}

\newpage

{d) Copper oxide $CuO$} \\

\begin{tabular}  {c c c c c}
\hline
$R$     & $R^2$          & corrected $R^2$ & standard error & sum of squares          \\ \hline
0.499     & 0.249            & 0.227        & 0.064          & 0.141                    \\ \hline
\end{tabular} 

 \vspace{0.5cm}
 
  \begin{tabular} {c c c c c c}
  \hline
    &               \multicolumn{3}{c}{}                  & \multicolumn{2}{c}{95\% Interval} \\ \hline
      & coefficient & error        & $\beta$              & lower bound       & upper bound   \\  \hline
$C_0$          & 1.051       & 0.027        &                & 0.997             & 1.105         \\
$C_\varphi$    & 1.452       & 0.251        & 0.544          & 0.954             & 1.949         \\
$C_T$          & 0.271       & 0.134        & 0.199          & 0.005             & 0.537         \\
$C_S$          & -0.473      & 0.754        & -0.057         & -1.968            & 1.023        \\ \hline

\end{tabular}

\vspace{0.5cm}

{e) Copper $Cu$} \\

\begin{tabular} {c c c c c}
\hline
$R$     & $R^2$          & corrected $R^2$ & standard error & sum of squares       \\ \hline
0.668   & 0.446         & 0.427            & 0.059          & 0.25                \\ \hline
\end{tabular} 

\vspace{0.5cm}

\begin{tabular}{c c c c c c}
\hline
    &       \multicolumn{3}{c}{}                      & \multicolumn{2}{c}{95\% Interval} \\ \hline
      & coefficient     & error      & $\beta$        & lower bound       & upper bound   \\ \hline
$C_0$          & 1.059   & 0.029     &                & 1.002             & 1.116         \\
$C_\varphi$    & 7.458   & 1.098     & 0.724          & 5.277             & 9.638         \\
$C_T$          & -0.24   & 0.237     & -0.086         & -0.712            & 0.231         \\
$C_S$          & 0.476   & 0.739     & 0.073          & -0.992            & 1.944        \\ \hline

\end{tabular}

\vspace{0.5cm}

{f) Silica $SiO_2$} \\

\begin{tabular}  {c c c c c}
\hline
$R$     & $R^2$          & corrected $R^2$ & standard error & sum of squares      \\ \hline
0.541   & 0.292       & 0.266        & 0.011          & 0.004                   \\ \hline
 \end{tabular} 
 
 \vspace{0.5cm}
 
 \begin{tabular} {c c c c c c}
 \hline
    &          \multicolumn{3}{c}{}                    & \multicolumn{2}{c}{95\% Interval} \\ \hline
      & coefficient & error        & $\beta$              & lower bound       & upper bound   \\ \hline
$C_0$          & 0.994       & 0.005        &                & 0.983             & 1.005         \\
$C_\varphi$    & 0.608       & 0.170        & 0.351          & 0.269             & 0.947         \\
$C_T$          & 0.104       & 0.032        & 0.311          & 0.040            & 0.168         \\
$C_S$          & 0.415       & 0.085        & 0.469          & 0.245             & 0.585 \\ \hline

\end{tabular}

\vspace{0.5cm}

{g) Silicon carbide $SiC$} \\

\begin{tabular} {c c c c c}
\hline
$R$     & $R^2$          & corrected $R^2$ & standard error & sum of squares           \\  \hline
0.837    & 0.700         & 0.682        & 0.036          & 0.150              \\ \hline
\end{tabular} 

\vspace{0.5cm}

\begin{tabular} {c c c c c c}
\hline
    &   \multicolumn{3}{c}{}                  & \multicolumn{2}{c}{95\% Interval} \\ \hline
      & coefficient & error        & $\beta$              & lower bound       & upper bound   \\ \hline
$C_0$          & 1.082       & 0.014        &                & 1.053             & 1.111         \\
$C_\varphi$    & 2.965       & 0.366        & 0.702          & 2.229             & 3.701         \\
$C_T$          & -0.018      & 0.081        & -0.017          & -0.180             & 0.145         \\
$C_S$          & -1.246      & 0.440        & -0.247          & -2.131             & -0.362 \\ \hline

\end{tabular}

\subsection{Restriction to $\varphi \leq 0.02$}

\begin{figure} [!h]
    \centering
    \includegraphics[width=\textwidth]{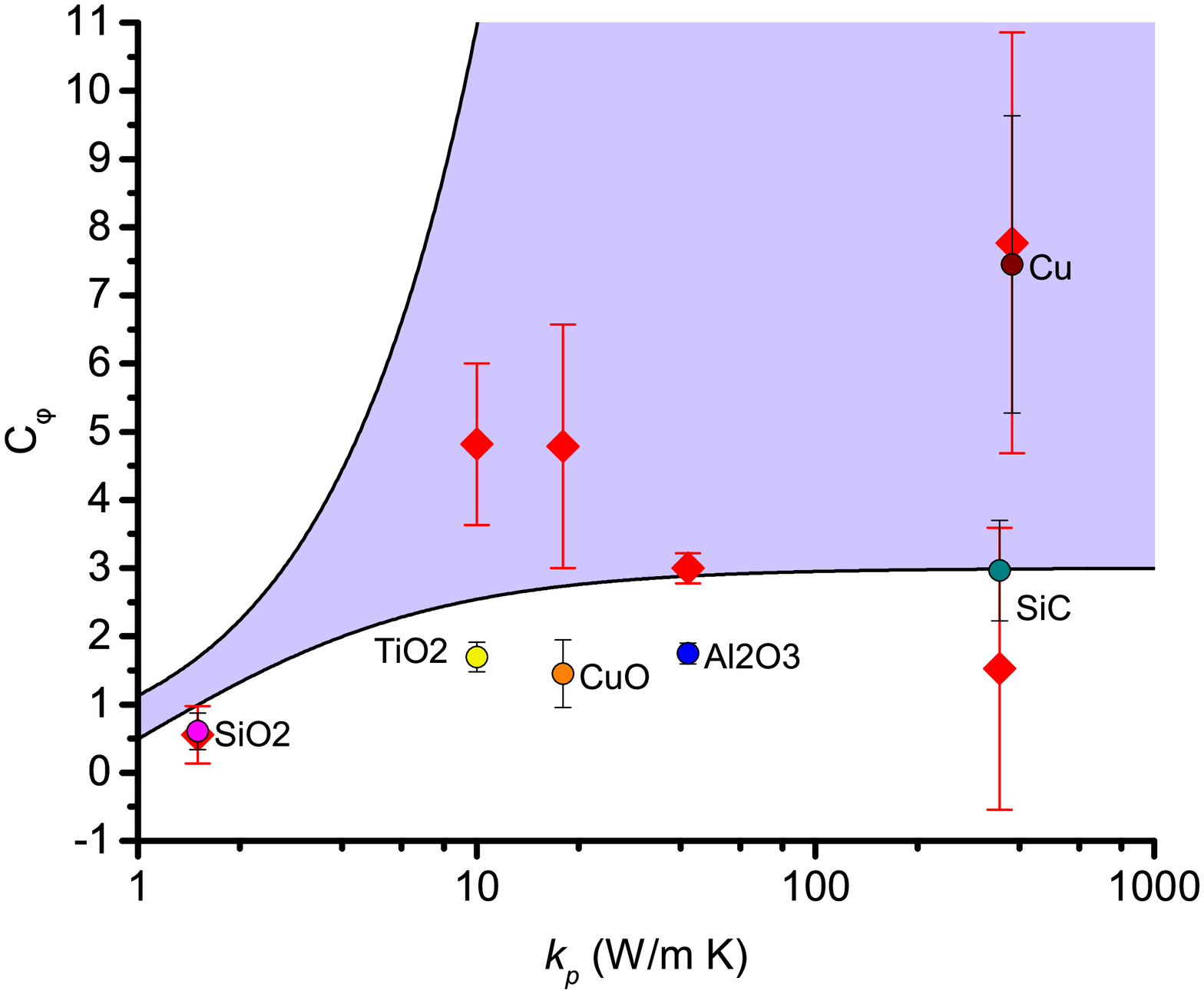}
    \caption{Thermal conductivity \(k_\text{p}\) of each material versus model parameter $C_\varphi$ with corresponding 95\% confidence-interval (error-bars) for linear regression. Red diamonds (including 95\% error-bars) display the change in parameter $C_\varphi$ with concentration $\varphi \leq 0.02$.  The linearized HS-bounds are displayed as solid black line. The lower bound is given by the linearized Maxwell equation (4), whereas the upper HS-bound was calculated by linearizing the right-hand-side of eq.~(2).}
    \label{fig:phi2}
\end{figure}

As the linear regression only works for small concentrations, the data have been restricted to concentrations $\varphi \leq 0.02$. The changes in the $C_\varphi$ parameter in comparison to no restriction is displayed in fig.~\ref{fig:phi2} with red diamonds.\\
Major changes only occurred for alumina $Al_2O_3$, titania $TiO_2$ and copper oxide $CuO$ with an increased $C_\varphi$, which is now fitting or exceeding Maxwell's prediction. The change in the case of $SiC$ is not in the uncertainties of the non-restricted regression, but still fits Maxwell's prediction. The changes in respect to the other materials, $SiO_2$ and $CNT$ lie within their corresponding uncertainties. Higher concentrations decrease the (relative) performance of $Al_2O_3$, $TiO_2$ and $CuO$, possibly due to unfavourable agglomeration effects.

\subsection{Use of surfactants}

\begin{figure} 
    \centering
    \begin{tabular} {c}
    \includegraphics[width=0.8\textwidth]{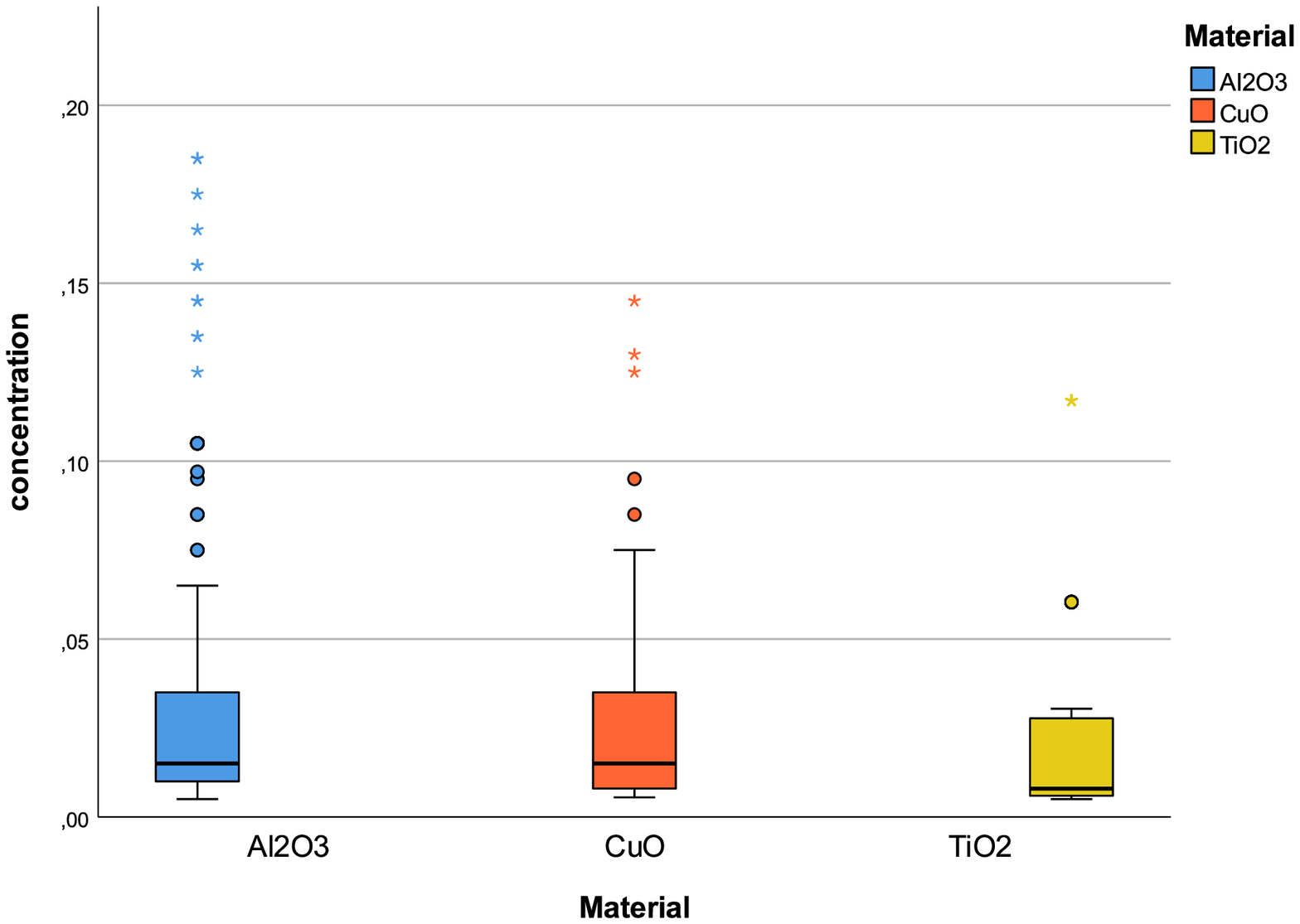} \\
    \includegraphics[width=0.8\textwidth]{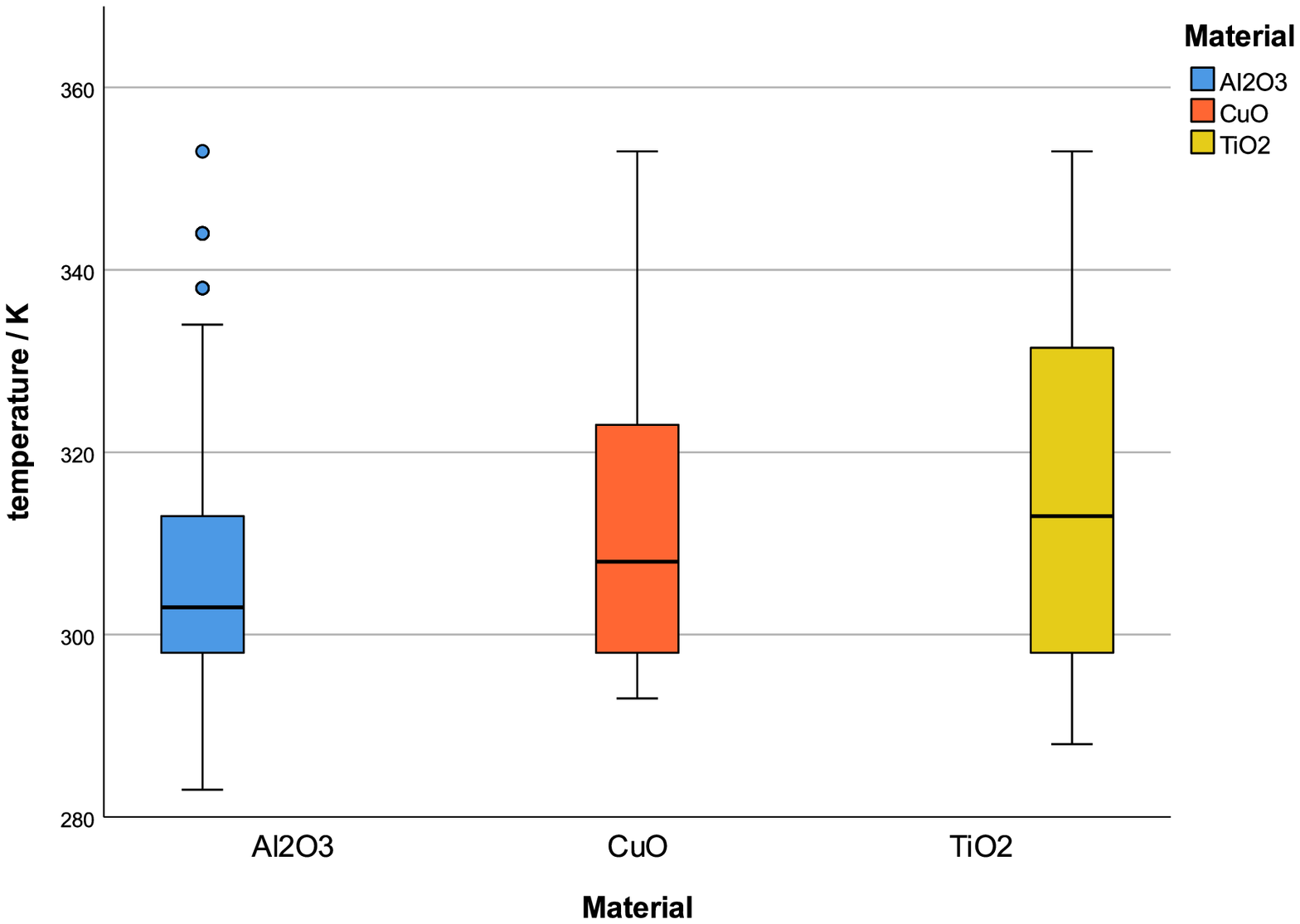} \\ 
    \includegraphics[width=0.8\textwidth]{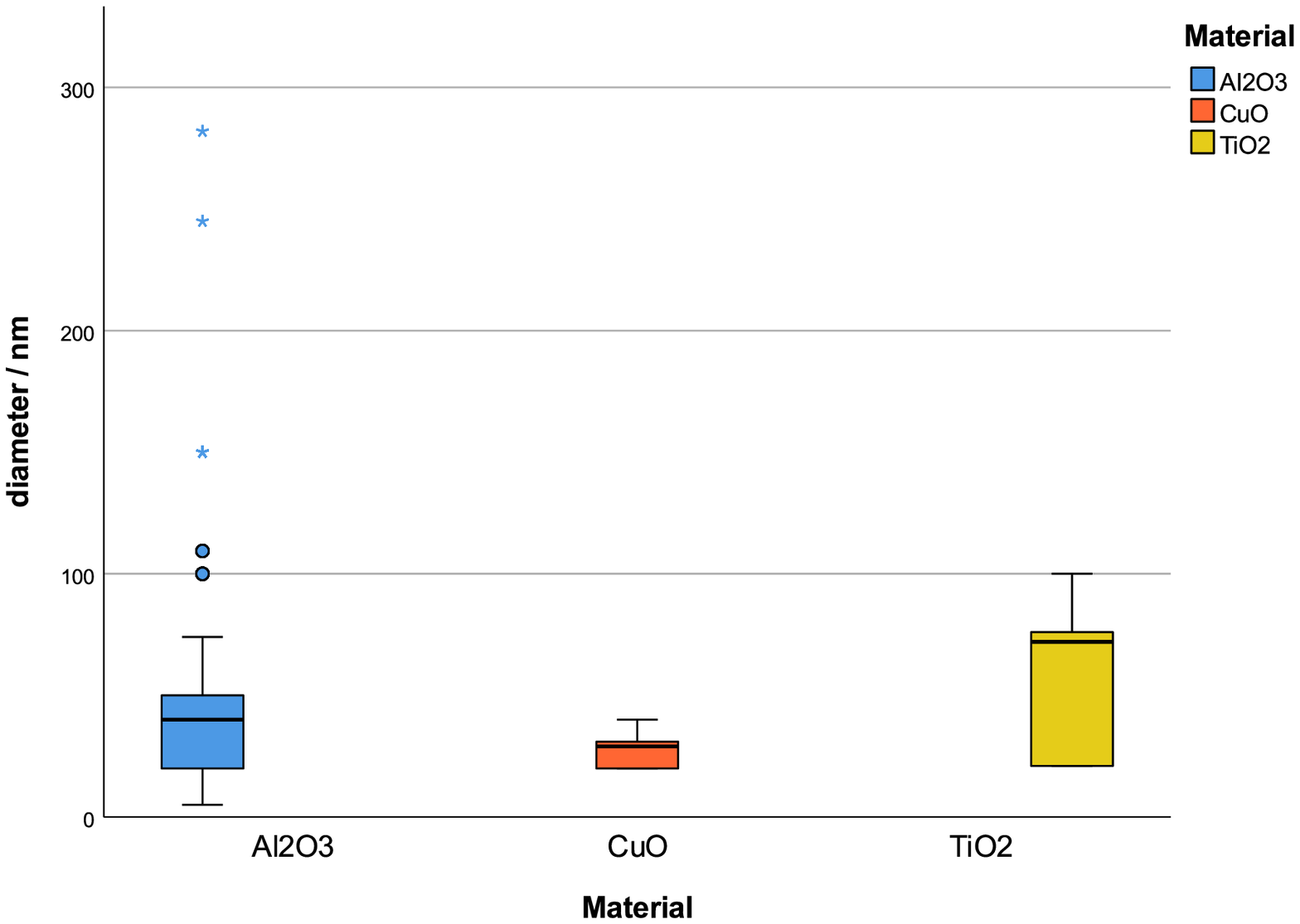} \\
    \end{tabular}
    \caption{Histograms showing the distribution of the data with exclusion of surfactants for $Al_2O_3$, $CuO$, and $TiO_2$ with respect to the  concentration  $\varphi$, the temperature $T$, and the nanoparticle size $d$.}
    \label{fig:H2O_box_surf}
\end{figure}

As the use of surfactants changes the resulting thermal conductivity, we filtered the data in which surfactants were explicitly used. This corresponds to 37\% (618 data points) of the data, which were excluded. An analysis of the different materials with the linear regression was only possible for $Al_2O_3$, $TiO_2$, and $CuO$. In the case of $SiO_2$ and $SiC$ no surfactants were used and in the case of $Cu$ only 10 data points were without the use of a surfactant, so that no linear regression is applicable. The changes in the descriptive statistics is shown in table~\ref{tab:descr-S} a-c. The results in the linear regression are shown in table~\ref{tab:Surfactants} a-c and figure~\ref{fig:H2O_box_surf}.

\begin{table} [ht]
    \centering
    \caption{Mean, minimum and maximum values of the parameters $\varphi, \  T, \  d$ \ for the single materials $Al_2O_3$, $TiO_2$, and $CuO$ with exclusion of surfactants and in comparison to with surfactant as taken from table~\ref{tab:descriptive}.}
    \label{tab:descr-S} 
    \end{table}
    
{a) Concentration $\varphi$ ($\cdot 10^{-2}$)} \\

\begin{tabular} {ll ccc |ccc }   
\hline
& &  \multicolumn{3}{c |}{Without Surfactants} & \multicolumn{3}{c}{With Surfactants}  \\ \hline
            &   & $Al_2O_3$ & $TiO_2$  & $CuO$   & $Al_2O_3$ & $TiO_2$  & $CuO$  \\ \hline
mean value  &   & 2.136    & 1.912    & 2.180    & 1.874     & 1.016    & 1.946  \\
minimum     &   & 0.003    & 0.002    & 0.05     & 0.003     & 0.001    & 0.015\\
maximum     &   & 18.0     & 11.2      & 14.0     & 18.0      & 11.2      & 14.0 \\\hline
\end{tabular}

\vspace{0.5cm}

{b) Temperature $T$} \\

\begin{tabular}  {ll ccc | ccc }
\hline
& &  \multicolumn{3}{c |}{Without Surfactants} & \multicolumn{3}{c}{With Surfactants}  \\ \hline
            &    & $Al_2O_3$ & $TiO_2$  & $CuO$  & $Al_2O_3$ & $TiO_2$  & $CuO$ \\ \hline
mean value  &    & 306.2   & 314.7    & 312.4    & 305.8     & 312.3    & 313.0 \\
minimum     &    & 283     & 288      & 293      & 283       & 288      & 293 \\
maximum     &    & 353     & 353      & 353      & 353       & 353      & 353 \\\hline
\end{tabular}

\vspace{0.5cm}

{c) Nanoparticle size $d$}  \\

\begin{tabular}  {ll ccc | ccc }
\hline
& &  \multicolumn{3}{c |}{Without Surfactants} & \multicolumn{3}{c}{With Surfactants}  \\ \hline
            &    & $Al_2O_3$ & $TiO_2$  & $CuO$ & $Al_2O_3$ & $TiO_2$  & $CuO$\\ \hline
mean value  &    & 47.9      & 58.4    & 28.3   & 61.1      & 99.4    & 29.7\\
minimum     &    & 5         & 21       & 20    & 5         & 11       & 20 \\
maximum     &    & 282       & 100      & 40    & 282       & 265      & 40 \\\hline
\end{tabular}

\newpage

\vspace{0.5cm}

\begin{table}[h]
    \centering
    \caption{Results for linear regression as described in section 2.1 of the paper with the exclusion of surfactants. The table shows $R, R^2$, sum of squares, the number of data points $N$, the coefficients with corresponding errors, 95\% interval and $\beta$-values for $Al_2O_3$, $TiO_2$ and $CuO$.}
    \label{tab:Surfactants}
\end{table}

{a) Alumina $Al_2O_3$} \\

\begin{tabular} {c c c c c c}
\hline
$R$     & $R^2$      & corrected $R^2$ & standard error & sum of squares & $N$   \\ \hline
0.747   & 0.558      & 0.554           & 0.047          & 1.094        & 405         \\ \hline
\end{tabular}

\vspace{0.5cm}

\begin{tabular} {c c c c c c}
\hline
    &   \multicolumn{3}{c}{}                       & \multicolumn{2}{c}{95\% Interval} \\ \hline
      & coefficient & error        & $\beta$              & lower bound       & upper bound   \\ \hline
$C_0$          & 1.018       & 0.004        &                & 1.010             & 1.027         \\
$C_\varphi$    & 1.788        & 0.082       & 0.735          & 1.626             & 1.950         \\
$C_T$          & 0.340       & 0.052        & 0.224          & 0.238             & 0.442         \\
$C_S$          & 0.245       & 0.047        & 0.174          & 0.152             & 0.338  \\ \hline

\end{tabular}

\vspace{0.5cm}

{b) Titania $TiO_2$} \\

\begin{tabular}  {c c c c c c}
\hline
$R$     & $R^2$          & corrected $R^2$ & standard error & sum of squares   & $N$    \\ \hline
0.945   & 0.893               & 0.889        & 0.030          & 0.576        &  83       \\ \hline
\end{tabular}

\vspace{0.5cm}

\begin{tabular} {c c c c c c}
\hline
    &             \multicolumn{3}{c}{}                      & \multicolumn{2}{c}{95\% Interval} \\ \hline
      & coefficient & error        & $\beta$              & lower bound       & upper bound   \\  \hline
$C_0$          & 0.936       & 0.009        &                & 0.919             & 0.954   \\
$C_\varphi$    & 1.985       & 0.105        & 0.725          & 1.775            & 2.194   \\
$C_T$          & 0.826       & 0.051        & 0.626          & 0.724             & 0.928     \\
$C_S$          & 1.931       & 0.211        & 0.367          & 1.511             & 2.351 \\ \hline

\end{tabular}

\vspace{0.5cm}

{c) Copper oxide $CuO$} \\

\begin{tabular}  {c c c c c c}
\hline
$R$     & $R^2$          & corrected $R^2$ & standard error & sum of squares & $N$      \\ \hline
0.485     & 0.235           & 0.210        & 0.068          & 0.126    & 94       \\ \hline
\end{tabular} 

 \vspace{0.5cm}

  \begin{tabular} {c c c c c c}
  \hline
    &               \multicolumn{3}{c}{}                  & \multicolumn{2}{c}{95\% Interval} \\ \hline
      & coefficient & error        & $\beta$              & lower bound     & upper bound   \\  \hline
$C_0$          & 1.063       & 0.034        &                & 0.996             & 1.131         \\
$C_\varphi$    & 1.396       & 0.272        & 0.517          & 0.855             & 1.937         \\
$C_T$          & 0.266       & 0.150        & 0.192          & -0.033             & 0.565         \\
$C_S$          & -0.725      & 0.924        & -0.080         & -2.560            & 1.110        \\ \hline

\end{tabular}

\vspace{0.5cm}

As to be seen in the tables, the changes due to the use of surfactants in the case of $Al_2O_3$ and $CuO$ lie within their uncertainties of the regressions without restrictions. In the case of $TiO_2$, the $C_\varphi$ increases and is nearly in agreement with Maxwell's boundaries. The $C_\text{t}$ also increases beyond the uncertainties of the regression without restrictions. Different to the other materials, the $C_0$ coefficient significantly decrases below the ideal $C_0=1$. Additionally, the $C_\text{S}$ coefficient becomes significant showing a strong size-dependency.

\subsection{Restrictions on Copper data set}

The regression with all data in the case of copper resulted in a high $C_0$ coefficient. Out of all data sets, only the study from Liu [62] used no surfactants. Exclusion of this data set results in a significantly increased regression, which was used in the paper. The regression with all data points is again displayed in the table~\ref{tab:Cu}.

\begin{table}[h]
    \centering
    \caption{Results for linear regression as described in section 2.1 with all data points (a) and without the study of Liu (b), see also table~\ref{tab:linear}. The table shows $R, R^2$, sum of squares, the number of data points $N$, the coefficients with corresponding errors, 95\% interval and $\beta$-values.}
    \label{tab:Cu}
\end{table}

{a) All data points (with Liu)} \\

\begin{tabular} {c c c c c c}
\hline
$R$     & $R^2$      & corrected $R^2$ & standard error & sum of squares & $N$   \\ \hline
0.624   & 0.389      & 0.371           & 0.061          & 0.234        & 102         \\ \hline
\end{tabular}

\vspace{0.5cm}

\begin{tabular} {c c c c c c}
\hline
    &             \multicolumn{3}{c}{}                      & \multicolumn{2}{c}{95\% Interval} \\ \hline
      &   coefficient &       error        & $\beta$     & lower bound       & upper bound   \\  \hline
$C_0$          & 1.098       & 0.021        &                & 1.056             & 1.140   \\
$C_\varphi$    & 6.303       & 0.973        & 0.602          & 4.373             & 8.233   \\
$C_T$          & -0.385      & 0.235        & -0.135         &-0.851             & 0.082     \\
$C_S$          & -0.432      & 0.579        & -0.071         & -1.582            & 0.718 \\ \hline

\end{tabular}

\vspace{0.5cm} 

{b) Without Liu} \\

\begin{tabular} {c c c c c}
\hline
$R$     & $R^2$          & corrected $R^2$ & standard error & sum of squares       \\ \hline
0.668   & 0.446         & 0.427            & 0.059          & 0.25                \\ \hline
\end{tabular} 

\vspace{0.5cm}

\begin{tabular}{c c c c c c}
\hline
    &       \multicolumn{3}{c}{}                      & \multicolumn{2}{c}{95\% Interval} \\ \hline
      & coefficient     & error      & $\beta$        & lower bound       & upper bound   \\ \hline
$C_0$          & 1.059   & 0.029     &                & 1.002             & 1.116         \\
$C_\varphi$    & 7.458   & 1.098     & 0.724          & 5.277             & 9.638         \\
$C_T$          & -0.24   & 0.237     & -0.086         & -0.712            & 0.231         \\
$C_S$          & 0.476   & 0.739     & 0.073          & -0.992            & 1.944        \\ \hline

\end{tabular}

\vspace{0.5cm}

The corrected correlation coefficient increases from $R^2=0.371$ to $R^2=0.427$ with the exclusion of Liu. Additionally, the $C_0$ coefficient decrased, while the $C_\varphi$ coefficient increased. 

\newpage

\section{Nonlinear regression}

The following table \ref{tab:nonlinear} a-g \ show the results for the statistical analysis done with SPSS. Next to the general dataset, statistics were done on six single materials. The nonlinear regression is described as:
\begin{equation}
    k^*(\varphi,S,T)= 1 + C_\varphi \,\varphi + C_\text{T}\, \Delta T + C_\text{S}\, \Delta S
    \label{eq:nonlin}
\end{equation}
The starting values for the coefficients were for all data and the single materials: 
$C_\varphi=2,\  C_\text{T}=0.5, \ C_\text{S}=0.1$.\\

\begin{figure} [ht]
    \centering
    \includegraphics[width=0.8\textwidth]{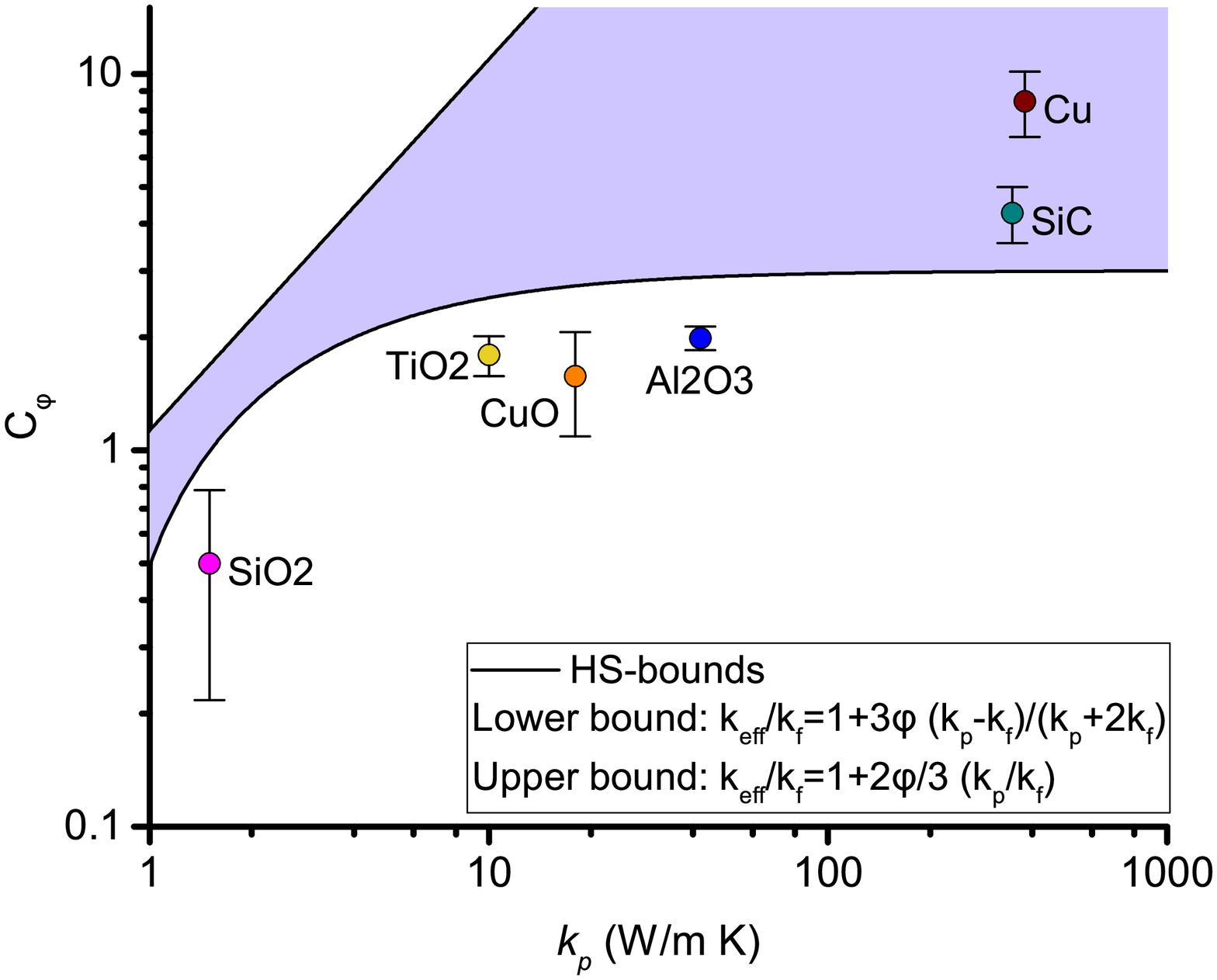}
    \caption{Thermal conductivity \(k_\text{p}\) of each material versus model parameter $C_\varphi$ with corresponding 95\% confidence-interval (error-bars) for the non-linear regression. The linearized HS-bounds are displayed as solid black line. The lower bound is given by the linearized Maxwell equation (4), whereas the upper HS-bound was calculated by linearizing the right-hand-side of eq.~(2).}
    \label{fig:C}
\end{figure}

\begin{figure} [ht]
    \centering
    \includegraphics[width=0.8\textwidth]{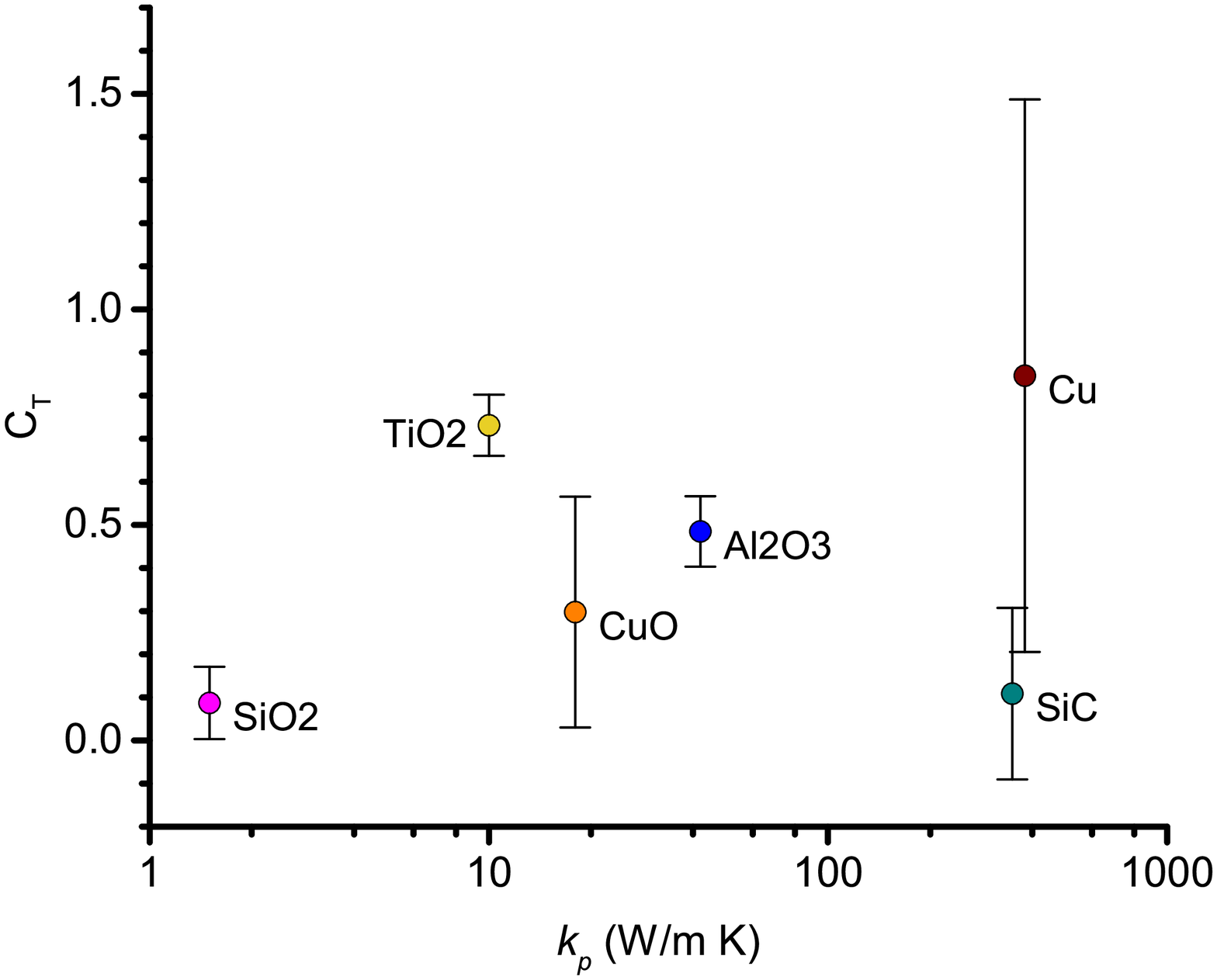}
    \caption{Thermal conductivity \(k_\text{p}\) of each material versus model parameter $C_T$ with corresponding 95\% confidence-interval (error-bars) for the non-linear regression.}
    \label{fig:T}
\end{figure}

\begin{figure} [ht]
    \centering
    \includegraphics[width=0.8\textwidth]{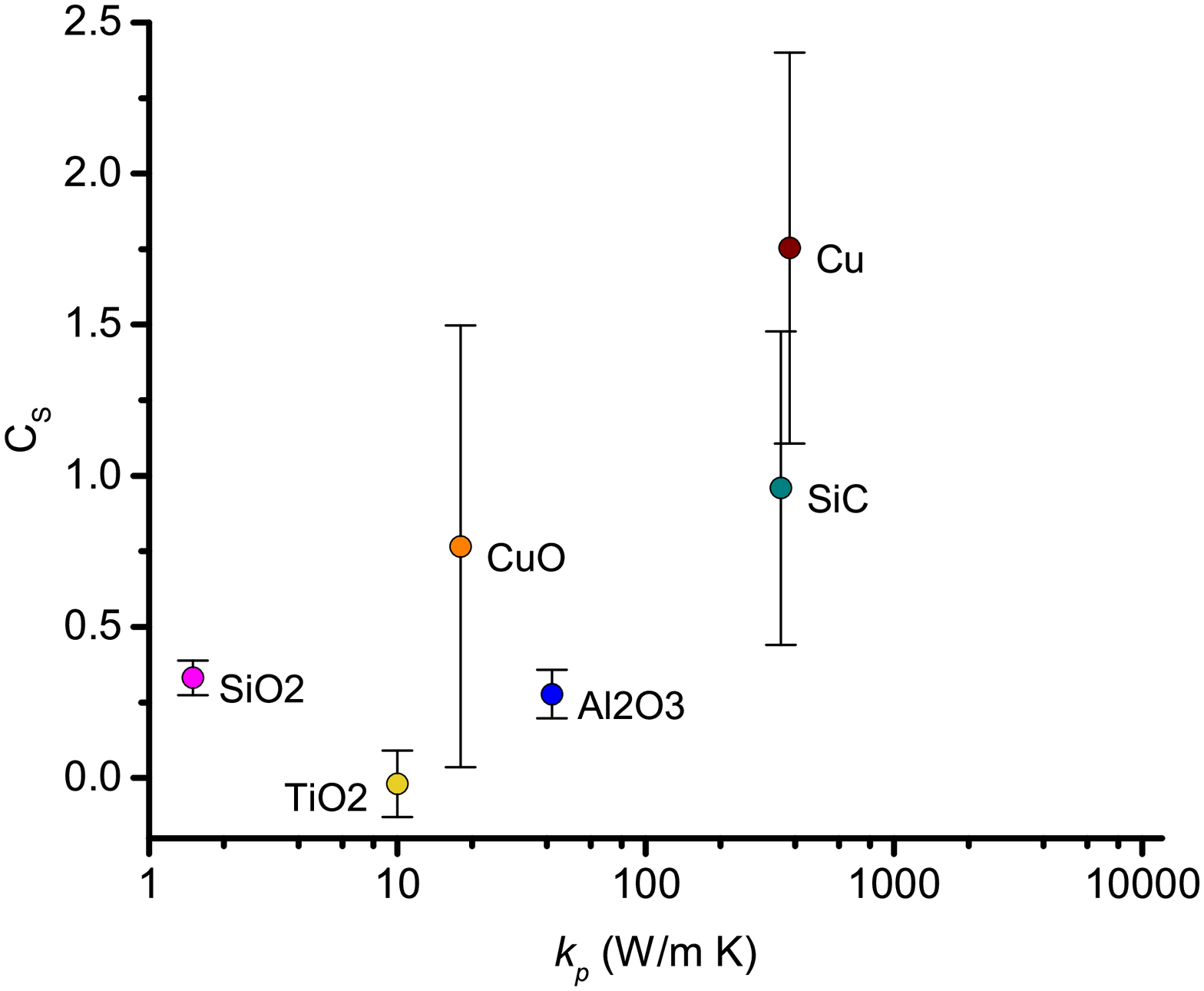}
    \caption{Thermal conductivity \(k_\text{p}\) of each material versus model parameter $C_S$ with corresponding 95\% confidence-interval (error-bars) for the non-linear regression.}
    \label{fig:S}
\end{figure}

The figures \ref{fig:C}, \ref{fig:T}, and \ref{fig:S} show the coefficients as a function of the thermal conductivity of the particle material, similar to the figures with the linear regression.

\begin{table}[h]
    \centering
    \caption{Results for nonlinear regression as described above. The table shows $R^2$, the coefficients with corresponding errors and 95\% interval.}
    \label{tab:nonlinear}
\end{table}

{a) General model} \\

$R^2=0.243$ \\

\begin{tabular} {c c c c c}
\hline
       \multicolumn{3}{l}{}            & \multicolumn{2}{c}{95\% Interval} \\ \hline
               & coefficient   & error        & lower bound       & upper bound  \\  \hline
$C_\varphi$   & 2.208        & 0.092          & 2.028             & 2.387    \\
$C_T$          & 0.778       & 0.030          & 0.718             & 0.837       \\
$C_S$          & 0.107       & 0.006          & 0.097             & 0.118 \\ \hline

\end{tabular}

\vspace{0.5cm}

{b) Alumina $Al_2O_3$} \\

$R^2=0.486$ \\

\begin{tabular} {c c c c c}
\hline
    \multicolumn{3}{c}{}              & \multicolumn{2}{c}{95\% Interval} \\ \hline
             & coefficient   & error        & lower bound       & upper bound  \\ 
$C_\varphi$   & 1.988        & 0.072        & 1.845             & 2.130    \\
$C_T$          & 0.485       & 0.042         & 0.403             & 0.568       \\
$C_S$          & 0.276       & 0.041          & 0.196             & 0.356 \\ \hline

\end{tabular}

\vspace{0.5cm}

{c) Titania $TiO_2$} \\

$R^2=0.741$ \\

\begin{tabular} {c c c c c}
\hline
   \multicolumn{3}{c}{}         & \multicolumn{2}{c}{95\% Interval} \\ \hline
           & coefficient   & error        & lower bound       & upper bound  \\ 
$C_\varphi$    & 1.791      & 0.110          & 1.574             & 2.008    \\
$C_T$          & 0.731      & 0.036          & 0.660             & 0.802       \\
$C_S$          & -0.020     & 0.056          & -0.130            & 0.090 \\ \hline

\end{tabular}

\vspace{2.5cm}

{d) Copper oxide $CuO$} \\

$R^2=0.223$ \\

\begin{tabular} {c c c c c}
\hline
     \multicolumn{3}{c}{}          & \multicolumn{2}{c}{95\% Interval} \\ \hline
           & coefficient   & error        & lower bound       & upper bound  \\ \hline
$C_\varphi$   & 1.575       & 0.245        & 1.089             & 2.060   \\
$C_T$          & 0.298       & 0.135         & 0.030             & 0.566       \\
$C_S$          & 0.766       & 0.369          & 0.034             & 1.497 \\ \hline

\end{tabular}

\vspace{0.5cm}

{e)  Copper $Cu$} \\

$R=0.244$ \\

\begin{tabular} {c c c c c}
\hline
      \multicolumn{3}{c}{}          & \multicolumn{2}{c}{95\% Interval} \\ \hline
           & coefficient   & error        & lower bound       & upper bound  \\ 
$C_\varphi$   & 9.089       & 0.776         & 7.547             & 10.630    \\
$C_T$         & 0.017       & 0.206         & -0.392            & 0.425       \\
$C_S$         & 1.906       & 0.263          & 1.384            & 2.428 \\ \hline

\end{tabular}

\vspace{0.5cm} 

{f) Silica $SiO_2$} \\

$R^2=0.281$ \\

\begin{tabular} {c c c c c}
\hline
  \multicolumn{3}{c}{}           & \multicolumn{2}{c}{95\% Interval} \\ \hline
           & coefficient   & error        & lower bound       & upper bound  \\ \hline
$C_\varphi$   & 0.500       & 0.142        & 0.218             & 0.783    \\
$C_T$          & 0.087       & 0.029         & 0.030             & 0.144       \\
$C_S$          & 0.331        & 0.043          & 0.246             & 0.415 \\ \hline

\end{tabular}

\vspace{0.5cm}

{g) Silicon carbide $SiC$} \\

$R^2=0.618$ \\

\begin{tabular}  {c c c c c}
\hline
    \multicolumn{3}{c}{}          & \multicolumn{2}{c}{95\% Interval} \\ \hline
           & coefficient   & error        & lower bound       & upper bound  \\ \hline
$C_\varphi$   & 4.276       & 0.360         & 3.552             & 4.999    \\
$C_T$          & 0.109       & 0.099         & -0.089             & 0.308       \\
$C_S$          & 0.959        & 0.257          & 0.442             & 1.477 \\ \hline

\end{tabular}

\newpage

\subsection{Variation in concentration term}

To proof that the use of the linear regression in terms of concentration is valid, we analyzed the data set with quadratic and cubic terms for the concentration parameter as shown in regressions (\ref{eq:quadr}) and (\ref{eq:cub}). Results are shown in table~\ref{tab:Quad-Cub} a-b.\\
\begin{equation}
    k^*(\varphi,S,T)= 1 + C_\varphi \,\varphi + C_{\varphi,2} \,\varphi^2 + C_\text{T}\, \Delta T + C_\text{S}\, \Delta S
    \label{eq:quadr}
\end{equation}

\begin{equation}
    k^*(\varphi,S,T)= 1 + C_\varphi \,\varphi + C_{\varphi,2} \,\varphi^2 + C_{\varphi,3} \,\varphi^3 + C_\text{T}\, \Delta T + C_\text{S}\, \Delta S
    \label{eq:cub}
\end{equation}

The starting values for the coefficients $C_\varphi,C_\text{T}, C_\text{S}$ were the same as in regression (\ref{eq:nonlin}, while the starting values of the coefficients $C_{\varphi,2},C_{\varphi,3}$ were set to:
$C_{\varphi,2}=1,\  C_{\varphi,3}=1$.\\

\begin{table}[h]
    \centering
    \caption{Results for nonlinear regression as described in regressions (\ref{eq:quadr}) and (\ref{eq:cub}). The table shows $R^2$, the coefficients with corresponding errors, and 95\% interval}
    \label{tab:Quad-Cub}
\end{table}

{a) Quadratic model} \\

$R^2=0.264$ \\

\begin{tabular} {c c c c c}
\hline
       \multicolumn{3}{l}{}            & \multicolumn{2}{c}{95\% Interval} \\ \hline
               & coefficient   & error        & lower bound       & upper bound  \\  \hline
$C_\varphi$   & 3.660        & 0.188           & 3.292             & 4.028    \\
$C_{\varphi,2}$   & -14.921        & 1.694           & -18.243             & -11.599    \\
$C_T$          & 0.702       & 0.031          & 0.642             & 0.763       \\
$C_S$          & 0.106       & 0.005          & 0.096             & 0.117 \\ \hline

\end{tabular}

\vspace{0.5cm}

{b) Cubic model} \\

$R^2=0.297$ \\

\begin{tabular} {c c c c c}
\hline
       \multicolumn{3}{l}{}            & \multicolumn{2}{c}{95\% Interval} \\ \hline
               & coefficient   & error        & lower bound       & upper bound  \\  \hline
$C_\varphi$   & 5.255        & 0.299           & 4.669             & 5.841   \\
$C_{\varphi,2}$   & -62.083        & 7.131           & -76.069             & -48.097    \\
$C_{\varphi,3}$   & 247.097        & 36.319           & 175.861             & 318.332    \\
$C_T$          & 0.661       & 0.031          & 0.600             & 0.721       \\
$C_S$          & 0.105       & 0.005          & 0.095             & 0.116 \\ \hline

\end{tabular}
\vspace{0.5cm}

With regard to the correlation coefficient changing only some percent, the use of the linear term for the concentration is appropiate.

